\documentclass[manuscript]{aastex63}

\usepackage{threeparttable}

\shorttitle{different origins of asymmetric substructures}
\shortauthors{Xu et al.}

\graphicspath{{./}{figures/}}

\begin{document}

\title{ {Exploring asymmetric substructures of the outer disk based on the conjugate angle of the radial action}}

\author[0000-0002-2459-3483]{Y.~Xu} 
\affiliation{CAS Key Laboratory of Optical Astronomy, National Astronomical Observatories, Chinese Academy of Sciences, Beijing, 100101, China}
\email{xuyan@nao.cas.cn}

\author[0000-0002-1802-6917]{C.~Liu}
\affiliation{Key Laboratory of Space Astronomy and Technology, 
National Astronomical Observatories, 
Chinese Academy of Sciences, 
Beijing 100101, PR China;}

\author[0000-0001-5017-7021]{Z. ~Li}
\affiliation{Department of Astronomy, School of Physics and Astronomy, Shanghai Jiao Tong University, 800 Dongchuan Road, Shanghai 200240, China; } 
\affiliation{Key Laboratory for Particle Astrophysics and Cosmology (MOE)/Shanghai Key Laboratory for Particle Physics and Cosmology, Shanghai 200240, China}

\author[0000-0003-3347-7596]{H.~Tian}
\affiliation{Key Laboratory of Space Astronomy and Technology, 
National Astronomical Observatories, 
Chinese Academy of Sciences, 
Beijing 100101, PR China;}

\author[0000-0002-5469-5149]{Sarah A. Bird} 
\affiliation{College of Science, China Three Gorges University, Yichang 443002, People's Republic of China} 
\affiliation{Center for Astronomy and Space Sciences, China Three Gorges University, Yichang 443002, People's Republic of China}

\author[0000-0001-8348-0983]{H.~J.~Newberg}
\affiliation{Department of Physics, Applied Physics and Astronomy, Rensselaer Polytechnic Institute, Troy, NY 12180, USA}

\author[0000-0001-8382-6323]{S.~Shao} 
\affiliation{Key Laboratory for Computational Astrophysics, National Astronomical Observatories, Chinese Academy of Sciences, Beĳing 100101, China}

\author[0000-0001-9073-9914]{L.~C.~Deng} 
\affiliation{CAS Key Laboratory of Optical Astronomy, National Astronomical Observatories, Chinese Academy of Sciences, Beijing, 100101, China}

\begin{abstract}
{We use the conjugate angle of radial action ($\theta_R$), the best representation of the orbital phase, to explore  the ``mid-plane, north branch, south branch" and ``Monoceros area" disk structures that were previously revealed in the  LAMOST K giants \citep{2020ApJ...905....6X}. The former three substructures,  identified by their 3D kinematical distributions, have been shown to be projections of the phase
space spiral (resulting from nonequilibrium phase mixing). In this work, we find that all of these substructures associated with the phase spiral show high aggregation in conjugate angle phase space, indicating that
the clumping in conjugate angle space is a feature of ongoing, incomplete phase mixing. 
We do not find the $Z-V_Z$ phase spiral located in the ``Monoceros area", but we do find a very highly concentrated substructure in the quadrant of conjugate angle space with the orbital phase from the apocenter to the guiding radius. The existence of the clump in conjugate angle space 
provides a complementary way to connect the ``Monoceros area" with the direct response to a perturbation from a significant gravitationally interactive event. 
Using test particle simulations, we show that these features are analogous to disturbances caused by the impact of the
last passage of the Sagittarius dwarf spheroidal galaxy.}

\end{abstract}


\section{Introduction} \label{sec:intro}
Data releases from large sky surveys such as \textit{Gaia} \citep{2016A&A...595A...1G}, Large Sky Area Multi-Object Fibre Spectroscopic Telescope
 \citep[LAMOST;][]{2012RAA....12.1197C, 2012RAA....12..723Z, 2012RAA....12..735D}, Sloan Digital Sky Survey \citep[SDSS;][]{2000AJ....120.1579Y}, Sloan
Extension for Galactic Understanding and Exploration \citep[SEGUE;][]{2009AJ....137.4377Y}, Radial Velocity Experiment \citep[RAVE;][]{2006AJ....132.1645S}, Galactic Archaeology with HERMES \citep[GALAH;][]{2015MNRAS.449.2604D} and APOGEE \citep{2017AJ....154...94M}, reveal vast evidence of the nonequilibrium kinematics of the Milky Way disk. Many recent studies use the above mentioned surveys to show the bulk motion and substructures of the disk stars of the Milky Way in different parameter spaces in order to explore the properties and origins of these features. 

Studies reveal the asymmetry of the density of the disk in the solar neighborhood \citep{2012ApJ...750L..41W, 2003ApJ...588..824Y, 2019MNRAS.482.1417B},  the over-density of Monoceros ring-like structures \citep{2002ApJ...569..245N}, and the oscillation of the disk \citep{2015ApJ...801..105X}.  Other studies map the spatial  distribution of  velocity substructure in the disk \citep{2010ApJ...716....1B} or study in-plane velocity space moving groups \citep{1998AJ....115.2384D, 2005A&A...430..165F, 2018MNRAS.479L.108K, 2019MNRAS.489.4962K, 2019MNRAS.488.3324F}. The disk stars are also studied in $Z-V_Z$ and other phase space projections \citep{2018Natur.561..360A, 2018MNRAS.478.3809S,2018MNRAS.481.1501B,2019MNRAS.486.1167B, 2019ApJ...872L...1C, 2019MNRAS.485.3134L,2019MNRAS.488.3324F, 2019MNRAS.485L.104M,2021MNRAS.504.3168B}. 

Besides studying configuration and velocity phase space,  \citet{2019MNRAS.484.3154S} point out that action-angle space represents the information of the entire orbit, not just instantaneous position and velocity.
 \citet{2019MNRAS.484.3291T} study the distribution of stars in the solar neighborhood using action-angle space. Some studies show the relationship of the action to  metallicity and age \citep{2018ApJ...867...31B, 2019ApJ...878...21T, 2019MNRAS.486.1167B, 2019ApJ...880..134G}. Using Hipparcos data, \citet{2019MNRAS.484.3154S} find that conjugate angles contain additional information. Such studies prove that action-angle space is a convenient way to describe orbital properties and find stars with similar dynamical features. 

\citet{2021ApJ...911..107L} shows that using the guiding center radius instead of  Galactocentric radius makes the phase spiral and kinematic features easier to explore. This result reveals that the mixing of orbital phase can blur the kinematic information.

In this work, we classify the orbital phase of LAMOST K giants with the conjugate angle of radial action ($\theta_R$) and study the properties of the ``mid-plane",  ``north branch", ``south branch" and ``Monoceros area"  \citep{2020ApJ...905....6X} kinematic substructures in conjugate angle of radial action space.  
 
 The ``mid-plane", ``north branch" and ``south branch" are projections of the phase spiral onto the $R-Z$ plane \citep{2020ApJ...905....6X}. The ``south branch" may be connected with the overdensity of the ``south middle structure" in the south of the disk \citep{2015ApJ...801..105X, 2018MNRAS.478.3367W}. 
 
 The ``Monoceros area" may be associated with the Monoceros ring and other anti-center stellar streams. The Monoceros ring was found in the SDSS equatorial stripe \citep{2002ApJ...569..245N}. Follow-up works trace the density, metallicity, population and age of the candidate stars in Monoceros \citep{2003MNRAS.340L..21I,2013ApJ...777...91Y,2003ApJ...594L.119C,2012ApJ...757..151L} and associated anti-center stellar streams, such as the Anticenter Stream and Eastern Banded Structure
 \citep[ACS, EBS;][]{2011ApJ...738...98G}, A13,  and the overdensity in the direction of the Triangulum and Andromeda galaxies  \citep[Tri-And;][]{2004ApJ...615..738M,2004ApJ...615..732R,2007ApJ...668L.123M,2018MNRAS.473..647D,2015MNRAS.452..676P,2014ApJ...793...62S,2018MNRAS.473.2428D}. Initially, the origin of the Monoceros ring was explained by dwarf galaxy debris \citep{2005ApJ...626..128P}. More recently, evidence shows the Monoceros ring more likely belong to the disk \citep{2020MNRAS.492L..61L, 2018Natur.555..334B}.  In this work, we show  new evidence that the  ``Monoceros area"  likely originates from a disturbance caused by the last impact of the Sagittarius dwarf spheroidal galaxy (Sgr dSph). 
 
 The paper is organized as follows.
   The sample and the analyzed quantities are introduced in Section \ref{sec:sample}.
In Section \ref{sec:orbitalphase} we describe the orbital phase of our data using the conjugate angle of radial action ($\theta_R$) and proceed to separate our data using $\theta_R$ for further investigation.
   The kinematic and chemical properties of the stars in each $\theta_R$ bin are illustrated in Section \ref{sec:properties}.
   In Section  \ref{sec:asymmetric}, we define asymmetric kinematic substructures in each $\theta_R$ bin as high relative fractions of stars and find that they are related with known kinematic substructures that previous literature explains as likely products from the last impact of the Sgr dSph.
   In Section \ref{sec:testparticle}, we use a Milky Way-like test particle simulation with a Sgr dSph impact to the disk and find that the Sgr dSph impact can produce the asymmetric features in $\theta_R$ orbital phase space that are similar to observations.
   In Section \ref{comparison}, we compare the asymmetric kinematic features of the test particle simulations to the observed Milky Way disk substructures.
   We provide a discussion in Section \ref{discussion} and summarize in Section \ref{summary}.

 \section{sample}\label{sec:sample}
We use a sample of  429,500 LAMOST K giants with the same distance and cylindrical velocity estimation as we previously used in \citet{2020ApJ...905....6X}. In this work, we calculate the azimuthal, radial and vertical actions ($J_\phi, J_R, J_Z$), conjugate angles ($\theta_\phi, \theta_R, \theta_Z$) and frequencies ($\Omega_\phi, \Omega_R, \Omega_Z$) \citep{2008gady.book.....B} making use of the software package $\tt {galpy}$ \citep{2015ApJS..216...29B} that uses the galpy potential $\tt MWPotential2014$ and Staeckel approximation \citep{2012MNRAS.426.1324B}.  Actions and frequencies are conserved quantities that describe the orbital properties of the stars. The azimuthal action ($J_\phi$) describes the amount of rotation around the Galactic center. The radial action ($J_R$) describes the radial extent of in-plane motion. The vertical action ($J_Z$) describes vertical oscillation. The conjugate angles describe the orbital phase. The angle $\theta_\phi$ describes the azimuth of a star’s guiding center. The angles $\theta_R$ and $\theta_Z$ are the radial and vertical phases.  We describe $\theta_R$ in more detail in the next section. The frequencies $\Omega_\phi$, $\Omega_R$, and $\Omega_Z$ are the rotational, the epicyclic, and the vertical frequencies, respectively.   The action is calculated assuming the Sun's radius at the Galactic plane and speed are ($R_0, V_{\phi,0}$$)=($8 kpc, 220 km/s). In this case, for a star of ($R, V_{\phi}$$)=($8 kpc, 220 km/s) with a circular orbit at the Galactic plane, the actions are equal to $J_\phi=1$, $J_R=0$, $J_Z=0$.

\section{Separation of the orbital phase with $\theta_R$}\label{sec:orbitalphase}
\begin{figure}
\centering
\includegraphics[width=9cm]{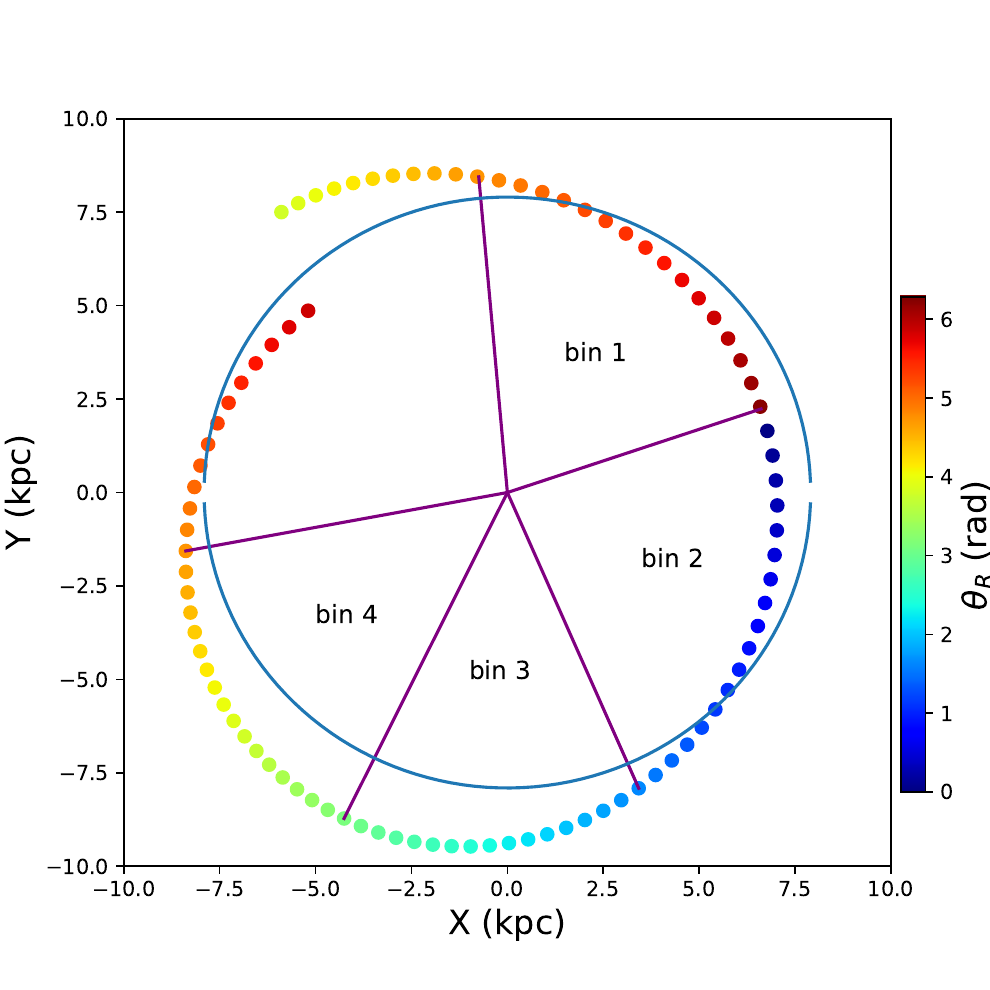}
\caption{Representative orbit of a star in the $X-Y$ plane. The star orbits clockwise. The colors indicate the $\theta_R$, orbital phase. The blue circle shows the guiding center radius of the star. The bins are separated by $\theta_R$ in the range of [3$\pi$/2, 2$\pi$] (bin 1), [0,$\pi$/2] (bin 2), [$\pi$/2,$\pi$] (bin 3), and [$\pi$,3$\pi$/2] (bin 4).}
\label{sampleAngleR}
\end{figure}

The conjugate angle of radial action, $\theta_R$, is the in-plane phase of a star around its epicycle \citep{2019MNRAS.484.3154S}.  Figure~\ref{sampleAngleR} shows how  $\theta_R$ changes along a stellar orbit.  When $\theta_R=0$ or $2\pi$, the star is at pericenter. When $\theta_R=\pi$, the star is at apocenter. With a conjugate angle of $\theta_R=\pi/2$, $3\pi/2$, the star is near the location of the guiding center radius. For orbital motion from pericenter to apocenter, $\theta_R$ ranges from 0 to $\pi$. For orbital motion from apocenter to pericenter, $\theta_R$ ranges from $\pi$ to 2$\pi$.  

We separate the orbital phase into four bins based on $\theta_R$. In the first bin, $\theta_R$ ranges from 3$\pi$/2 to 2$\pi$. These stars move in the direction from the guiding center radius toward pericenter. In the second bin, $\theta_R$ ranges from 0 to $\pi$/2. These stars move away from pericenter toward the guiding center radius.  In the third bin, $\theta_R$ ranges from $\pi$/2 to $\pi$. These stars move from the guiding center radius toward apocenter. In the fourth bin, $\theta_R$ ranges from $\pi$ to 3$\pi$/2. These stars move away from apocenter toward the guiding center radius. These bins are labeled in Figure~\ref{sampleAngleR}.

 \section{Properties of stars in the four angular regions} \label{sec:properties}
\begin{figure}
\centering
\includegraphics[width=20cm]{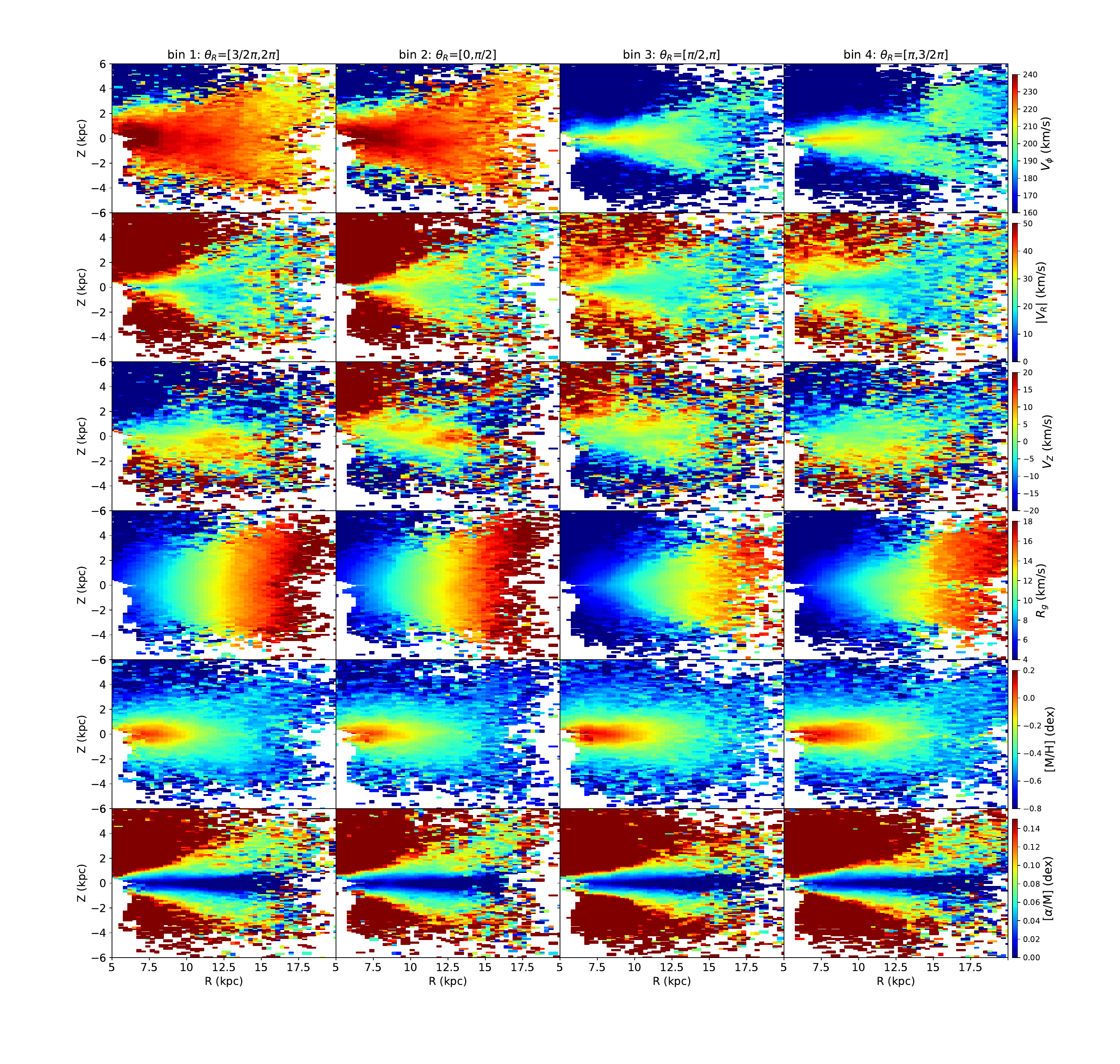}
\caption{Distributions of $V_\phi$, $|V_R|$, $V_Z$, guiding center radius ($R_g$), [M/H], and [$\alpha$/M], in four $\theta_R$ bins for LAMOST K giants in the $R-Z$ map. We make the maps by dividing the stars into bins of ($\Delta R$, $\Delta Z$)=(0.2, 0.2) kpc and present the medians of each property of the stars as represented by the colorbar. In the second row, $V_R$ is positive in the second and third bin, and it is negative in the first bin and the fourth bin; in each panel of the second row, the absolute value of the $V_R$ distribution is presented to make it easier to compare stars in adjacent bins. }
\label{velocityfeh}
\end{figure}

\subsection{$V_\phi$ and $R_g$}
Figure~\ref{velocityfeh} shows the kinematical and chemical distributions of our LAMOST K giant sample as functions of $(R, Z)$. We make the maps by dividing the stars into bins of ($\Delta R$, $\Delta Z$)=(0.2, 0.2) kpc, the colorbars indicate the medians of each stellar property. Each column of Figure~\ref{velocityfeh} presents the stars separated into one of the four $\theta_R$ bins described in Section 3.  In Figure~\ref{velocityfeh},  stars in a given $(R, Z)$ in bin 1 (bin 2) are leaving (approaching)  pericenter, and stars of bins 3 (bin 4) are leaving (approaching)  apocenter. As seen in the first row of Figure~\ref{velocityfeh}, the stars in bins 1 and 2 have higher values of $V_\phi$ than  in bins 3 and 4 at the same $R$. Also, the stars in bins 1 and 2  in the fourth row of Figure~\ref{velocityfeh} have larger guiding center radii ($R_g$) than the stars of bins 3 and 4  at the same $R$.   In the region  $12<R<13$ kpc and $-2<Z<2$ kpc, the median $R_g$  is 13.27, 13.22, 11.52, and 11.22 kpc for stars in bin 1 to bin 4, respectively. These results support the expectation that in a given volume, stars near pericenter will have higher velocities and larger guiding center radii than those that are closer to apocenter.

\subsection{Metallicity and [$\alpha$/M]}
The stars in  bins 1 and 2 are closer to pericenter. The stars in bins 3 and 4 are closer to apocenter. Therefore, in some $R$ range the stars in bins 1 and 2 will have the same $R_g$ as the stars in bins 3 and 4 within a larger $R$ range. Because of this, the stars with $\mathrm{[M/H]}>-0.1$ dex (the red part in the plots of the fifth row of Figure~\ref{velocityfeh}) in bins 1 and 2 show smaller scale length than stars with $\mathrm{[M/H]}>-0.1$ dex in bins 3 and 4. Also the stars with  $[\alpha/\mathrm{M}]<0.1$ dex (the blue and green part in the plots of the sixth row of Figure~\ref{velocityfeh}) in bins 1 and 2 show a wider flare than the stars with $[\alpha/$M$]<0.1$ dex of bins 3 and 4 at the same $R$. Here we make the assumption that the metallicities of stars are correlated with their guiding radii in order to explain our observations.

\subsection{$V_R$}
$V_R$ is defined to be positive going outwards. The stars in bin 1 are approaching pericenter. The stars in bin 4 are receding from apocenter. In both cases, we expect $V_R$ to be negative as is observed in bins 1 and 4 in the second row of Figure~\ref{velocityfeh}.
The stars in bin 2 are receding from pericenter. The stars of bin 3 are approaching apocenter. As expected, the $V_R$ of stars in bins 2 and 3 of the second row of Figure~\ref{velocityfeh}  are positive. The  absolute value of median $V_R$  plotted in the second row of Figure~\ref{velocityfeh}, so that it is clear that the median $V_R$ in bins 1 and 2  and bins 3 and 4 differ only by a sign flip.

\subsection{$V_Z$}
The third row of Figure~\ref{velocityfeh} shows  the $V_Z$ distribution in each bin. In the first and second bins, the mid-plane in the range $10<R<15$ kpc and $-1<Z<1$ kpc shows a feature with positive median $V_Z$, which is consistent with the feature of the warp \citep{2020NatAs...4..590P}.  

\begin{figure}
\centering
\includegraphics[width=11cm]{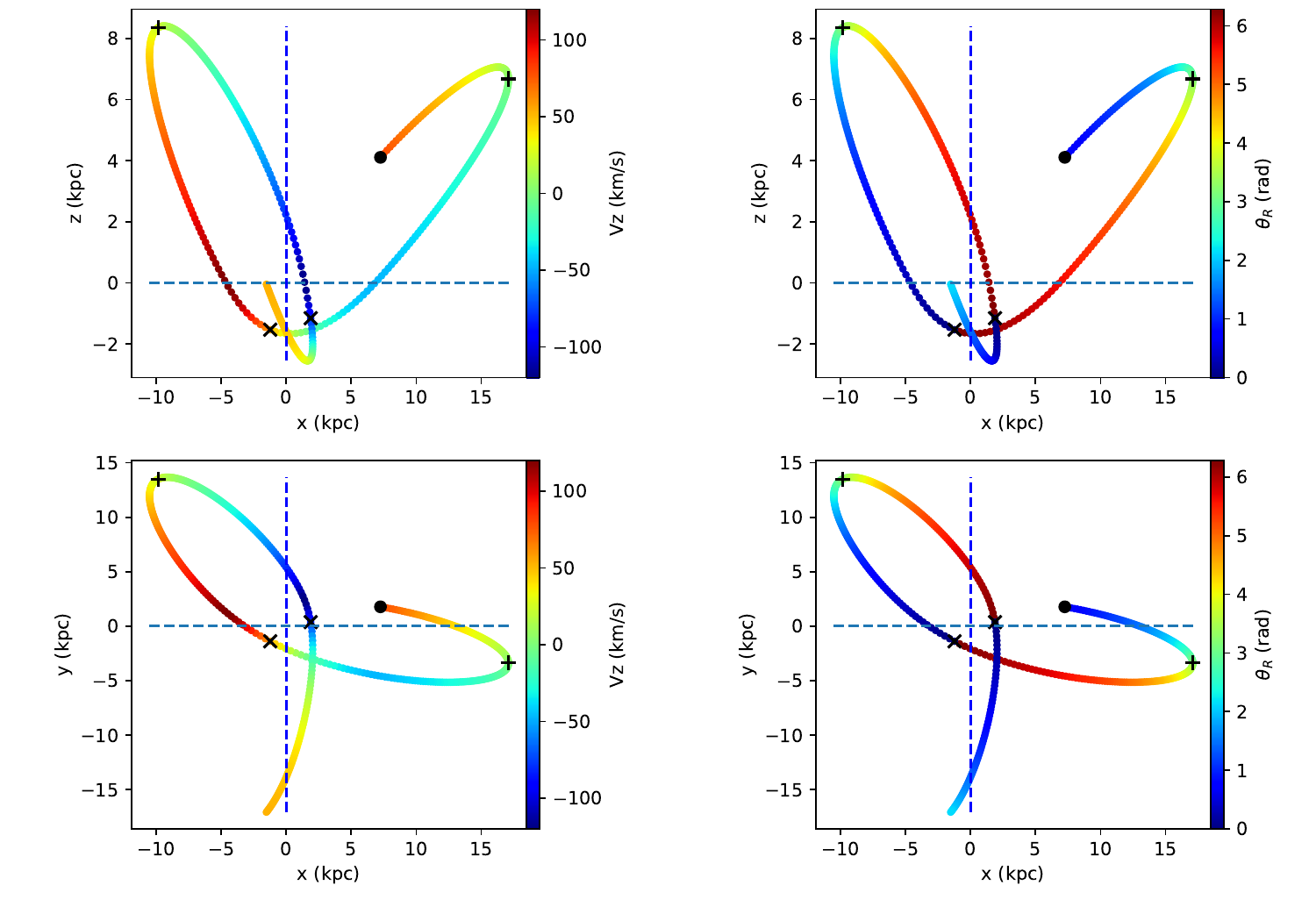}
\caption{Orbit of one typical star that reaches high Galactic height, integrated from $t=0$ to 0.5 Gyr. The black dot is the current location of this star with $(R, Z)=(7.14, 4.12)$ kpc, $\theta_R=0.66$ (rad), and $V_Z=117.06$ (km/s).  The plus signs label the location of $\theta_R=\pi$. The crosses label the location of $\theta_R=0,2\pi$. Upper left panel: the orbit in the $X-Z$ plane is color coded by $V_Z$.  Upper right panel: the orbit in the $X-Z$ plane is color coded by $\theta_R$. Lower left panel: the orbit in the $X-Y$ plane is color coded by $V_Z$. Lower right panel: the orbit in the $X-Y$ plane is color coded by $\theta_R$.}
\label{orbitsample1}
\end{figure}

\begin{figure}
\centering
\includegraphics[width=11cm]{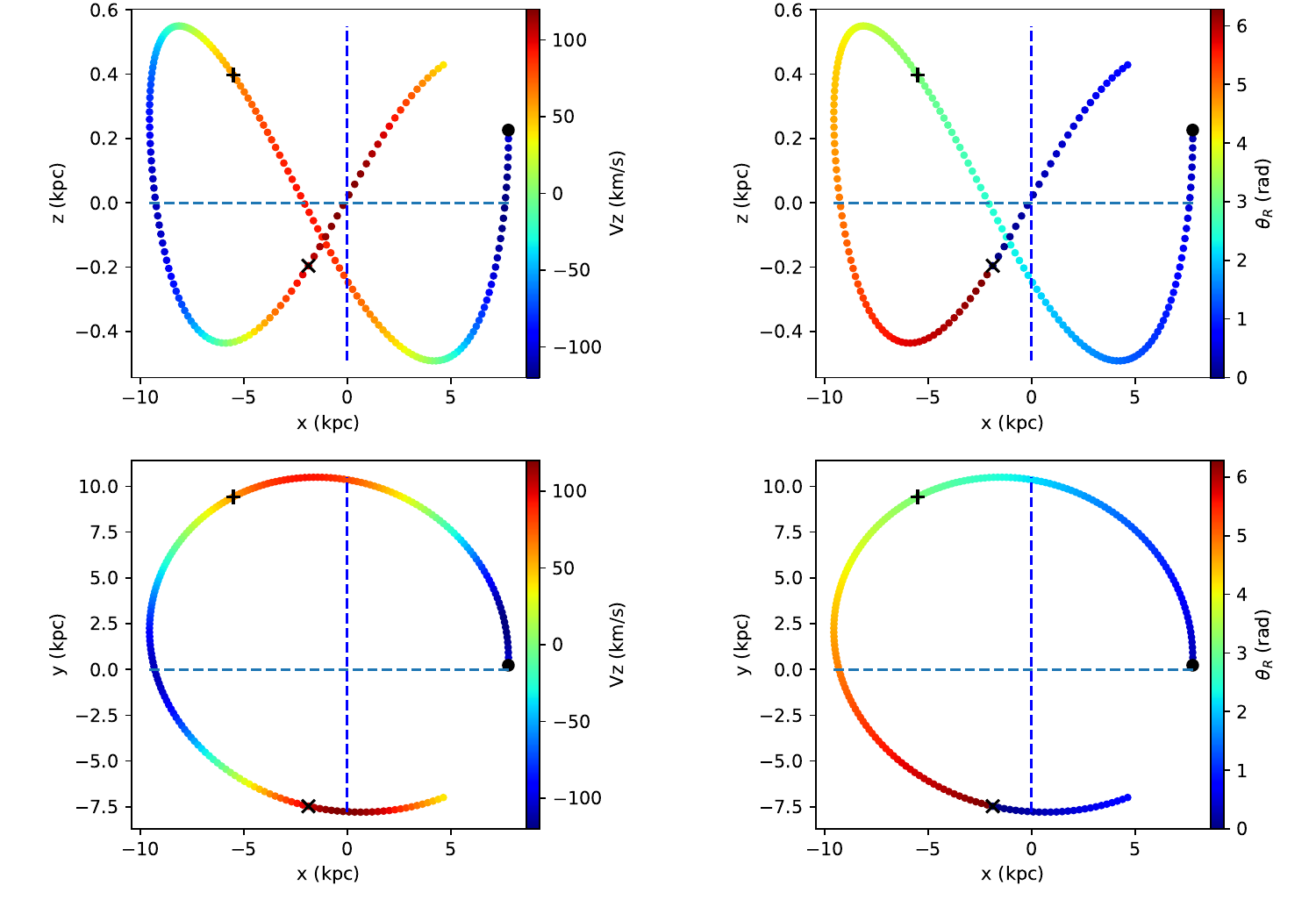}
\caption{ Orbit of one typical star that can only reach to low Galactic height, integrated from $t=0$ to 0.2 Gyr. The black dot is the current location of this star with $(R, Z)=(7.44, 0.23)$ kpc, $\theta_R=0.26$ (rad), and $V_Z=-25.79$ (km/s). The plus signs label the location of $\theta_R=\pi$. The crosses label the location of $\theta_R=0,2\pi$. Upper left panel: the orbit in the $X-Z$ plane is color coded by $V_Z$.  Upper right panel: the orbit in the $X-Z$ plane is color coded by $\theta_R$. Lower left panel: the orbit in the $X-Y$ plane is color coded by $V_Z$. Lower right panel: the orbit in the $X-Y$ plane is color coded by $\theta_R$.}
\label{orbitsample2}
\end{figure}

The $V_Z$ distribution shows a reverse breathing mode, especially in the range of $R<10$ kpc, $|Z|>2$ kpc. In bins 1 and 4, $V_Z>0$ km/s when $Z>2$ kpc and $V_Z<0$ km/s when $Z<-2$ kpc. In bins 2 and 3, the situation is reversed,  $V_Z<0$ km/s when $Z>2$ kpc and  $V_Z>0$ km/s when $Z<-2$ kpc.  The pattern is not visible at low $Z$. This phenomenon is related to the fact that the Milky Way's velocity ellipsoid is more tilted with increasing distance from the Galactic plane.  From Figures 15 and 16 of \citet{2019MNRAS.486.1167B}, the tilted velocity ellipsoid for stars above or below the mid-plane produces the $V_R$ or $\theta_R$ quadrupole signature in the $Z-V_Z$ plane. Correspondingly, the $V_Z$ distribution of stars above or below the mid-plane is also related to $\theta_R$. The stars near the mid-plane do not show this relation.

 To exhibit the difference between the two kinds of orbits, 
we chose two stars in the same $\theta_R$ bin ($\theta_R\in[0,\pi/2]$) at similar $R$, but with high and low Galactic height; the current locations of the two stars are $(R, Z)=(7.14, 4.12)$ and (7.44, 0.23) kpc, respectively.  We explore the relationship between $\theta_R$ and $V_Z$ for two kinds of orbits in Figures~\ref{orbitsample1} and \ref{orbitsample2}. 

Figure~\ref{orbitsample1} shows a typical orbit for a star that has large orbital inclination and can therefore reach large heights from the Galactic mid-plane.  In Figure~\ref{orbitsample1}, the pericenter is labeled by the cross and the apocenter is labeled by the plus. For this kind of elliptical  orbit, the apocenter is usually near the location with maximum Galactic height. The pericenter is usually near the Galactic center. When the $\theta_R$ is in the range of [0,$\pi$/2], the star is   moving away from pericenter. The motion is upward and decelerating when the star reaches high Galactic height with $Z>0$ kpc, as is the case for the orbit shown in Figure~\ref{orbitsample1}. Likewise the motion is downward and decelelerating as the star is far from the mid-plane with $Z<0$ kpc. In the range of $0<\theta_R<\pi/2$, for this kind of orbit, $V_Z>0$ km/s when $Z>0$ kpc, and $V_Z<0$ km/s when $Z<0$ kpc. 

 Figure~\ref{orbitsample2} shows one orbit of  a star that can only reach low heights from the Galactic mid-plane. This orbit has a smaller orbital inclination and makes more oscillations across the mid-plane during one orbital period.  When $\theta_R$ is in the range of $[0,\pi/2]$, the star also is moving away from the pericenter just like the star in Figure~\ref{orbitsample1}, but this star does not necessarily tend to move upward and decelerate because the apogalacticon and perigalacticon are not well correlated with distance from the plane. 
 In the case of Figure~\ref{orbitsample2}, from the initial position,  the star first moves downward across the mid-plane of the disk  in the range of $\theta_R<\pi/2$. Then it moves upward and finally reaches the apocenter that is labeled by the plus sign ($\theta_R=\pi$).  The star moves up and down in the disk between apogalacticon and perigalacticon due to the more rapid oscillations for this type of orbit. 

In the other bins ($\theta_R\in[\pi/2,2\pi]$), the situation is similar; stars  whose orbits extend to large heights from the Galactic mid-plane tend to have a monotonic relationship between $V_Z$ and  $\theta_R$.   
This is the reason that median $V_Z$ reaches extremely positive or negative values far from the disk mid-plane, as seen in the panels of the third row of Figure~\ref{velocityfeh}. 
This trend also explains why median $V_Z$ has opposite signs at negative and positive $Z$. 

\section{The asymmetric substructures } \label{sec:asymmetric}
\begin{figure}
\centering
\includegraphics[width=8cm]{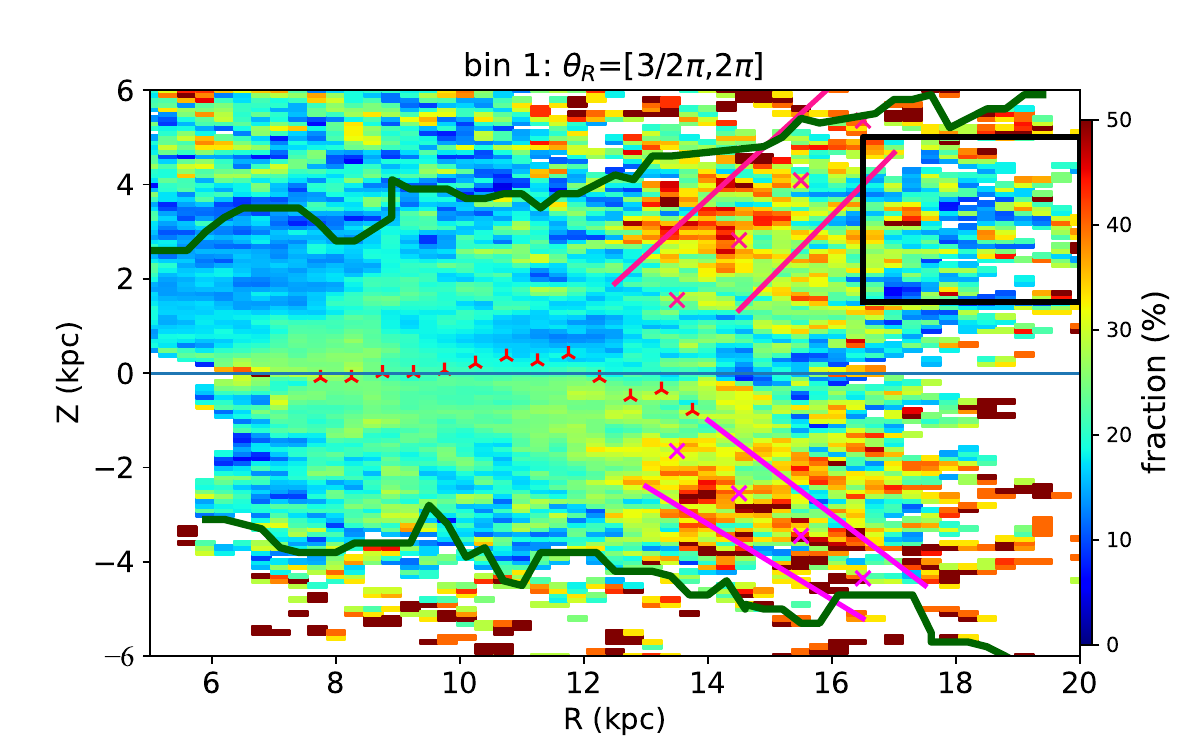}
\includegraphics[width=8cm]{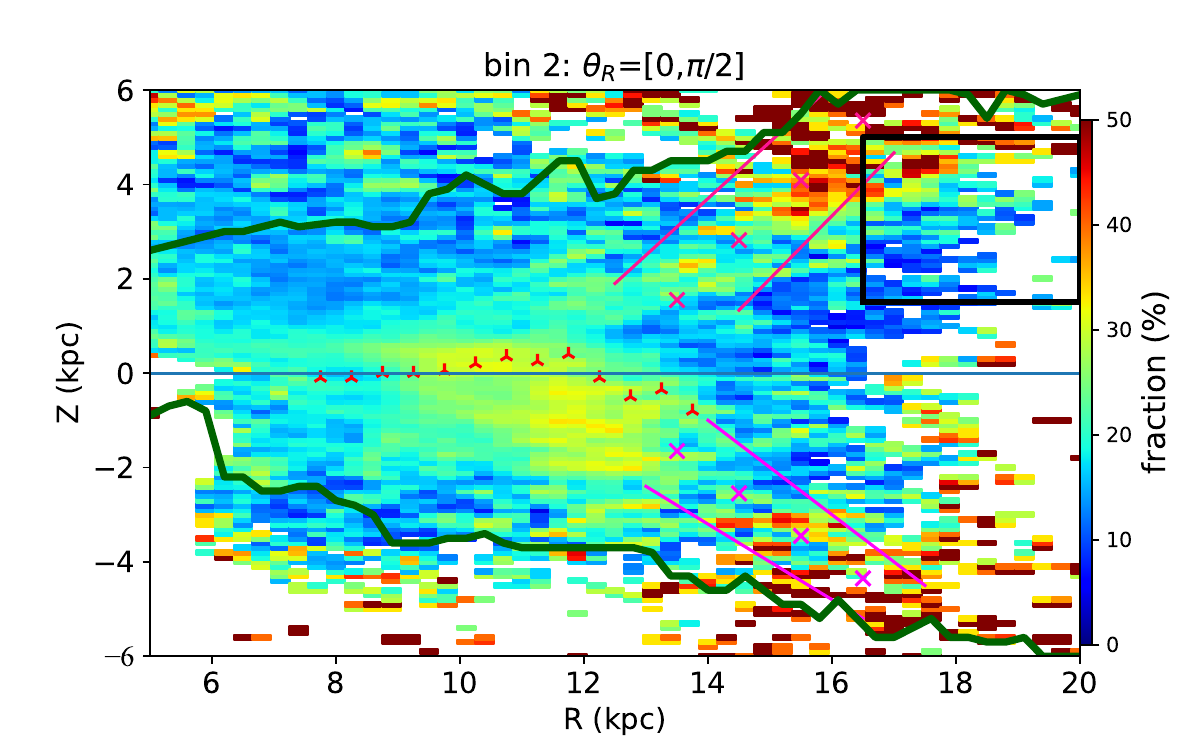}

\includegraphics[width=8cm]{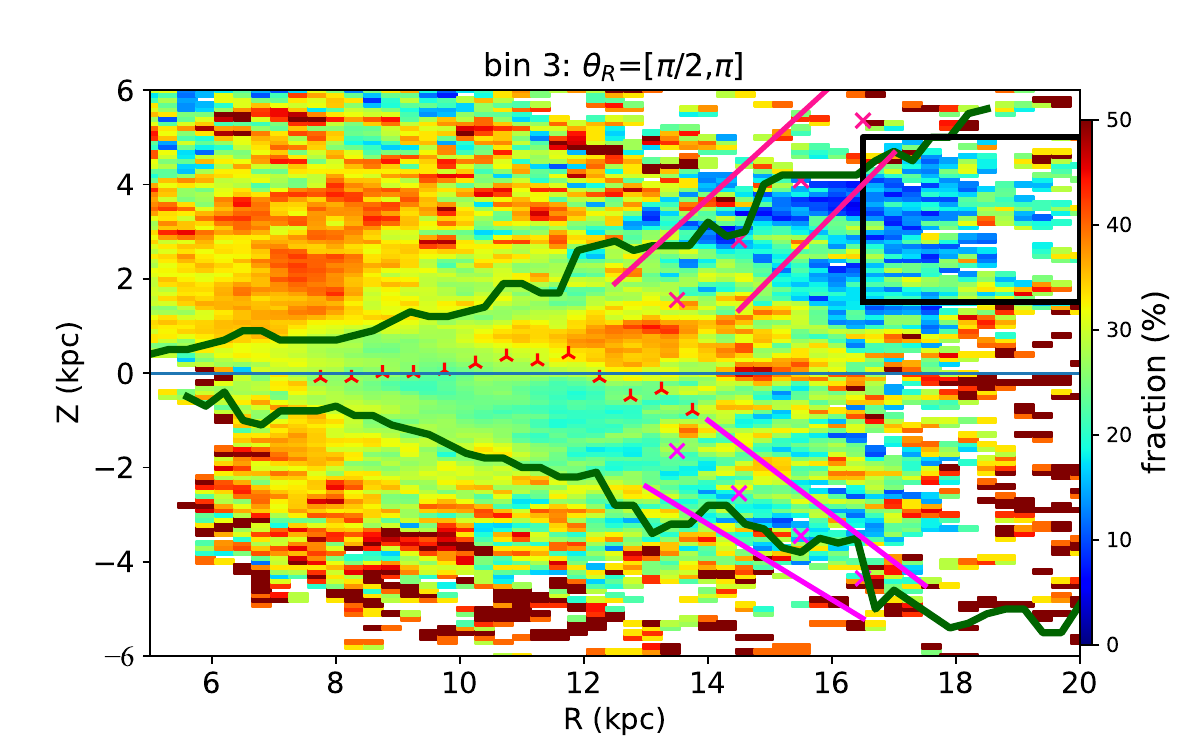}
\includegraphics[width=8cm]{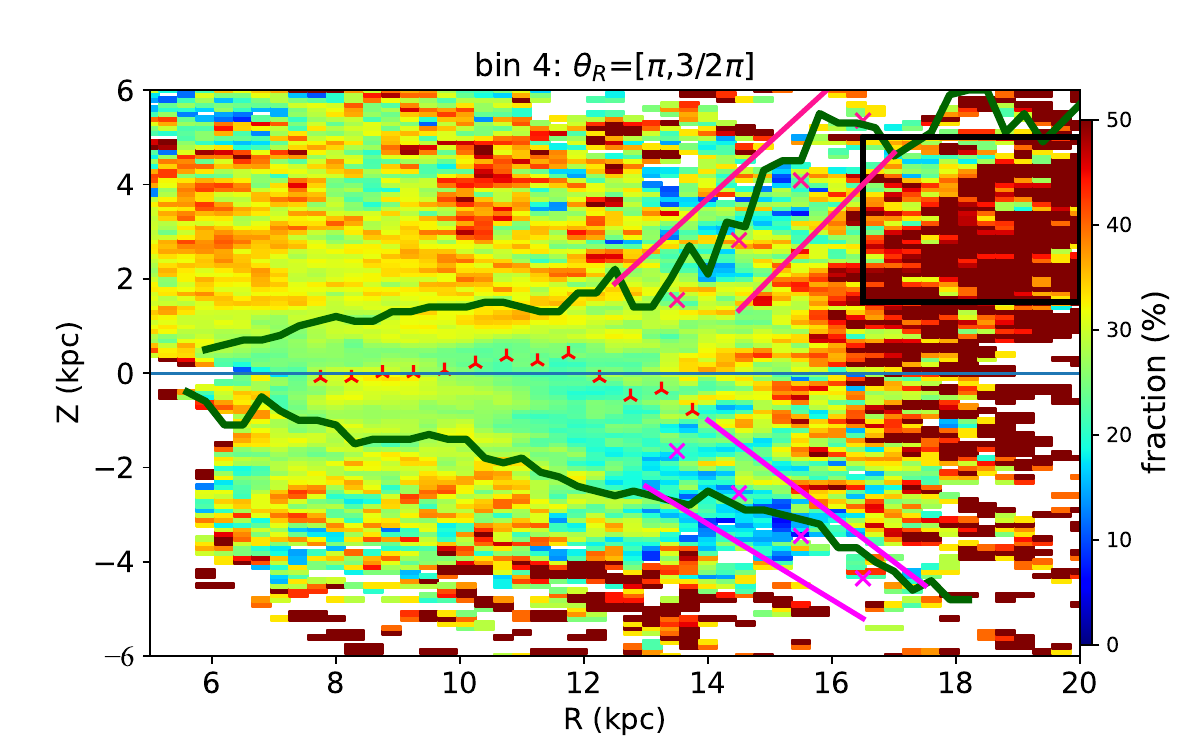}

\caption{This picture shows the relative fraction of stars in each $\theta_R$ bin, where the relative fraction is defined as the number of stars in a particular  $\theta_R$ bin divided by the total number of stars at a particular region of $(R, Z)$.  The boundaries and peaks of the ``north branch" and ``south branch" are labeled  with pink lines and crosses, respectively. The oscillation of the ``mid-plane" is labeled with triangles. The position of  ``Monoceros area" is labeled  by the black rectangle \citep[see Figure 8 of][]{2020ApJ...905....6X}. The wavy dark green curves label the boundary area at which median  $V_\phi$  is 190 km/s. }
\label{numAngleR}
\end{figure}

\begin{deluxetable*}{l*{7}{c}}
\tablewidth{0pt}
\label{Table:fraction}
\tiny
\tabletypesize{\footnotesize}
\renewcommand{\arraystretch}{1.1}
\tablecaption{Median and standard deviation (std) of the fraction of  stars in each $\theta_R$ bin for the full sample and for the  kinematic substructures of \citet{2020ApJ...905....6X}.}
\tablehead{
\colhead{} &
\colhead{median (std)  } &
\colhead{median (std)  } & 
\colhead{median (std)  } &
\colhead{median (std)  } &
\colhead{} &
\colhead{} \\
\colhead{} &
\colhead{ fraction (\%) of } &
\colhead{ fraction (\%) of } & 
\colhead{ fraction (\%) of } &
\colhead{ fraction (\%) of } &
\colhead{} &
\colhead{} \\\colhead{} &
\colhead{total sample of this $\theta_R$ bin} &
\colhead{ north branch} & 
\colhead{south branch} &
\colhead{ Monoceros area} &
\colhead{} &
\colhead{} }
\startdata
$\theta_R=[3/2\pi,2\pi]$ (bin 1)  & 22.2 (7.6)  & 29.4 (10) & 33.3 (11.7) & 20 (10.4) \\
$\theta_R=[0,1/2\pi]$ (bin 2) & 20.1 (7.3)  & 30.7 (13.1)  & 25 (12.1) & 16.1 (8.7) \\
$\theta_R=[\pi/2,\pi]$ (bin 3) & 29.8 (8.7) & 20.9 (10.5) & 24.2 (9.2) & 18.2 (7.9) \\
$\theta_R=[\pi,3/2\pi]$ (bin 4) & 31.1 (9.1) & 25 (9.1) & 20 (8.1) & 50 (17.9) \\
 \enddata
\end{deluxetable*}  

In this section, we study how  ``north branch", ``south branch" and ``Monoceros area", discovered as disk kinematic substructures in $R-Z$ space \citep{2020ApJ...905....6X},  are distributed in orbital phase space. 

Figure~\ref{numAngleR} shows the relative fraction of stars in each $\theta_R$ bin, as a function of position in the $(R, Z)$ plane. 
 Table \ref{Table:fraction} shows the median fraction and the standard deviation of the distribution of fractions within each bin. The median fractions of bin 1 and bin 2 are 22.2\% and 20\%. The median fractions of bin 3 and bin 4 are 29.8\% and 31.1\%. The median fractions of bins 3 and 4 are higher than bins 1 and 2. This may be caused by selection effects. The bin 1 and 2 stars are distributed from guiding center radius to pericenter with higher $V_\phi$; that means that these stars spend little time in these bins and thus have a lower sampling rate. The bin 3 and 4 stars are distributed from the guiding center radius to apocenter with lower $V_\phi$; that means that the stars spend a longer time near apocenter and thus have a higher sampling rate. 

 If the disk of the Galaxy is in an equilibrium state,  we expect the relative fractional distribution in each $\theta_R$ bin to be a smooth distribution in $(R, Z)$. This is not the case, as seen in Figure~\ref{numAngleR}; we find many asymmetric substructures.

\subsection{Number fraction distribution of substructures associated with the phase spirals}

To guide the eyes, the boundaries of the high $V_\phi$ (median $V_\phi>$190 km/s) region in the first row of the Figure~\ref{velocityfeh} is labeled with  dark green curves on each panel of Figure~\ref{numAngleR}. The kinematic features of ``mid-plane", ``north branch",  ``south branch", and ``Monoceros area" found in the first paper \citep{2020ApJ...905....6X} are also labeled in Figure~\ref{numAngleR}. 

The ``mid-plane" is identified as having the smallest standard deviation of $V_\phi$, $V_Z$  in Figures 5 and 8 of \citet{2020ApJ...905....6X}. The ``north branch" and  ``south branch" identified in Figures 4 and 8 of \citet{2020ApJ...905....6X} are high $V_\phi$ features compared with the $V_\phi$ distribution in the adjacent area. In Figure~\ref{numAngleR}, the ``mid-plane" is labeled with pink triangles. The boundaries and peak lines of the ``north branch" and ``south branch" are labeled with pink lines and crosses, respectively. 
The ``Monoceros area"  in Figure 4 of \citet{2020ApJ...905....6X} is a low $V_\phi$ structure. In  Figure~\ref{numAngleR}, the boundary of ``Monoceros area" is indicated b a black rectangle. 

The median fraction of each bin in Figure~\ref{numAngleR}, the median fraction within the areas of the above kinematic substructures, and the standard deviation of the distribution of those fractions are listed in Table \ref{Table:fraction}.  We find that stars belonging to each kinematic feature are not evenly distributed in $\theta_R$.

In the ``north branch"region  in Figure~\ref{numAngleR},  there is a higher number fraction in bins 1 and 2 than  is found in the nearby area. In Table \ref{Table:fraction}, the median fractions of the ``north branch" in bin 1 and bin 2 are  29.4\% and 30.7\%, which is higher than the median fraction of the total sample of these bins (22.2\%, 20\%). In  bins 3 and 4, the median fractions (20.9\%, 25\%) in the same  ``north branch" region are lower than the median fraction of the total sample of bins 3 and 4 (29.8\%, 31.1\%). 

Similarly, in the region of ``south branch" in  Figure~\ref{numAngleR}, there is a higher number fraction than nearby area. In Table \ref{Table:fraction}, the median fractions of the ``south branch" in bin 1 and bin 2 are 25.4\% and 33.3\% that is higher than the median fraction of the total sample of bin 1 and bin 2 (22.2\%, 20\%). In bin 3 and bin 4, at the location of the ``south branch", there are obvious valleys in the median fraction (24.2\%, 20\%) that are lower than the median fraction of the total sample (29.8\%, 31.1\%) in those same bins. 

In bin 2 panel of Figure~\ref{numAngleR}, there is an obvious are with a high fraction (larger than 30\%) of stars in the range of $10<R<14$ kpc, $-3<Z<0.5$ kpc. This high fraction (in bin 2) substructure roughly follows the same trend as the ``mid-plane" labeled by triangles.  Inside this substructure, there is high fraction in the range of $10<R<12$ kpc, $0<Z<0.5$ kpc and high fraction in the range of $11<R<14$ kpc, $-2.5<Z<0$ kpc.  This is consistent with the trend of the ``mid-plane" shifting north before $R<12$ kpc and shifting south at larger distances.

From a comparison between the kinematic structures of \citet{2020ApJ...905....6X} and the number fraction distributions in Figure~\ref{numAngleR}, we see they are related. From  \citet{2020ApJ...905....6X}, we know the ``north branch" and ``south branch" substructures are projections of the $V_\phi$ phase spiral in an $R-Z$ map. The oscillation of the ``mid-plane" is consistent with the centroid of the phase spiral. In recent work \citep{2018Natur.561..360A, 2018MNRAS.481.1501B, 2021MNRAS.504.3168B, 2022MNRAS.515.5951T}, the phase spiral is found to most likely be the result of a perturbation caused by the last impact of Sgr dSph. In bins 1 and 2 of Figure~\ref{numAngleR},  there are high fractions at the oscillation of the ``mid-plane", ``north branch" and ``south branch". The inference is that the perturbation caused by the last impact of Sgr dSph also caused stellar asymmetric distributions  in conjugate angle space. 

\subsection{Number fraction distribution of substructure associated with $V_R$ ripples }

\begin{figure}
\centering
\includegraphics[width=8cm]{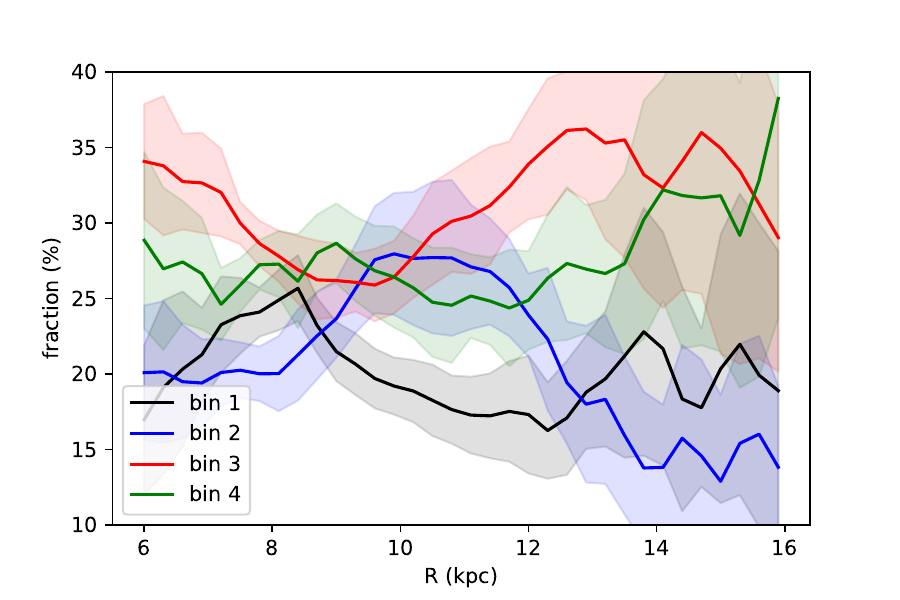}
\caption{ Number fraction of stars in the range of $0<Z<1$ kpc. The black curve is the median number fraction of the stars of bin 1, and blue curve is the median number fraction of the stars in bin 2.  The  red curve is the median number fraction of the stars in bin 3, and the green curve is the median number fraction of the stars in bin 4. The shadow shows the standard deviation of the number fraction distribution.}
\label{fractionbin1234}
\end{figure}

In Figure~\ref{numAngleR}, in bins 2 and 3, in the range of $10<R<13$ kpc and $0<Z<1$ kpc, the number fraction is higher than that of the adjacent area.  In bins 1 and 4, there are correspondingly low number fractions. The number fractions within $0<Z<1$ kpc in each $\theta_R$ bin data set are plotted in Figure~\ref{fractionbin1234}. The number fractions in bins 2 and  3 with $10<R<13$ kpc, $0<Z<1$ kpc can reach up to 28\% and 35\%, respectively. The dips in the number fraction for the same range of $(R, Z)$ for bins 1 and 4 are around 18\% and 25\%, respectively. From Figure~\ref{velocityfeh}, we know that the $V_R$ of bins 2 and 3 is positive, and the $V_R$ of bins 1 and 4 is negative. In this $R$, $Z$ range ($10<R<13$ kpc, $0<Z<1$ kpc), there are more stars in bins 2 and 3 than in bins 1 and 4. Thus, the median $V_R$ of the total sample of this $(R, Z)$ range is positive. The same positive $V_R$ substructure is detected by many previous works. There is a positive $V_R$ substructure beyond $R\sim9$ kpc found in a sample of LAMOST red clump stars \citep{2017RAA....17..114T,2020MNRAS.491.2104W} that was explained by a perturbation due to the rotating bar or an  effect due to the  spiral arms or a minor merger. This is also consistent with the positive $V_R$ ripple  detected in a sample of  LAMOST O, B stars \citep{2019ApJ...872L...1C} that is likely associated with disturbances due to the passage of the Sgr dSph \citep{2019MNRAS.486.1167B}.

\subsection{ Number fraction distribution of the ``Monoceros area" sustructure}

In the range  $R>16$ kpc, $Z>0$ kpc in bin 4 of Figure~\ref{numAngleR},  in the black rectangle of the ``Monoceros area", there is an overdense area with a number fraction of over 50\%.   In the same  ``Monoceros area" of bins 1, 2 and 3, the number fractions are 20.0\%, 16.1\% and 18.2\%, respectively. This substructure dominates in bin 4 where the orbital phase is receding from apocenter.

We illustrate that the kinematic feature in bin 4 of Figure~\ref{velocityfeh} is consistent with the  ``Monoceros area" kinematic feature observed in the full sample of Xu et al. (2020).
Table 2 summarizes  the chemical and kinematic properties of stars in the ``Monoceros area".  The stars in ``Monoceros area" have relatively low $V_\phi$ and disk-like metallicity. The  metallicity and $\alpha$ abundance of the substructure are similar to that of metal-poor thin disk stars.  This is consistent with the result of \citet{2020MNRAS.492L..61L} who analyzed the metallicity of Monoceros with APOGEE data. The median $J_R$ is relatively low. That means that the orbits of the stars are relatively round, which is consistent with result of \citet{2021ApJ...910...46L}. The metallicity and kinematics of the stars in the ``Monoceros area" show that the stars from this substructure are consistent with the disk component. 

We gain additional insight by separating the data by orbital phase. 
 In Figure~\ref{velocityfeh}, in bin 4 of the first row of the $V_\phi$ distribution, we see a discontinuity in the  distribution of $V_\phi$ at $R=15$ kpc, $Z>1$ kpc that is at the edge of ``Monoceros area". This shows that the flared disk is not a smooth structure.

\begin{deluxetable*}{l*{7}{c}}
\tablewidth{0pt}
\tabletypesize{\footnotesize}
\renewcommand{\arraystretch}{1.1}
\tablecaption{Properties of stars in the  ``Monoceros area"}
\tablehead{
\colhead{} &
\colhead{``Monoceros area" } &
\colhead{``Monoceros area" } & 
\colhead{} &
\colhead{} \\
\colhead{} &
\colhead{total sample} &
\colhead{stars with $\theta_R=[\pi,3/2\pi]$ (bin 4)} & 
\colhead{} &
\colhead{} }

\startdata
median [M/H] (dex) & $-0.56$  & $-0.53$  \\
 median $[\alpha$/H] (dex)& 0.085 & 0.082   \\
median $J_R$ &0.022 & 0.023  \\
median $V_\phi$ (km/s)& 195.5 & 193  \\
median $V_R$ (km/s)& $-12.8$ & $-19.9$  \\
median $V_Z$ (km/s)& $-7$ & $-9.5$  \\
 \enddata
\label{Table:summary}
\end{deluxetable*} 
 
 \citet{2020ApJ...905....6X} show that the phase spiral vanishes when $R>15$ kpc. Forming a phase spiral takes a longer time at larger $R$, due to the longer dynamical time scale. The ``Monoceros area" at $R>16$ kpc,  is not on the  $Z-V_Z$ phase space spiral in Figure 7 of \citet{2020ApJ...905....6X}.  In this work, we find that the ``Monoceros area" is highly centralized in the radial action conjugate angle, just like the ``north branch'' and ``south branch". In Section 5.1, we mentioned that the substructures ``north branch" and ``south branch" are projections of phase sprials. The phase spirals are likely produced by the impact of Sgr dSph as explained in many recent works \citep{2018Natur.561..360A, 2012MNRAS.426.1324B,2019MNRAS.486.1167B,2019MNRAS.485.3134L}. Because the ``Monoceros area" shows the same features in conjugate angle space of radial action  as the kinematic substructures associated with the last Sgr dSph impact in conjugate angle space of radial action, there is  a possibility that the ``Monoceros area" may also be associated with a gravitational disturbance such as occurred during the Sgr dSph impact. We shows in Section 7 that the ``Monoceros area" could be a perturbed section of the disk that has not had time to wind up due to the longer dynamical time of larger radius. 

 We will illustrate the possibility of a dwarf galaxy impact producing highly concentrated substructures in conjugate angle space using a test particle simulation in next section.
 
\subsection{ Number fraction distribution beyond the disk }
In the first row of Figure~\ref{velocityfeh}, there is rapid transition from stars with $V_\phi>210$ km/s to stars with $V_\phi<190$ km/s in bins 1 and 2. And there is rapid transition from stars with $V_\phi>190$ km/s to stars with $V_\phi<170$ km/s in bins 3 and 4. In Figure~\ref{numAngleR}, we show the transition at $V_\phi$=190 km/s. Beyond the rapid transition, the stars are dominated by a combination of thick disk stars and halo stars. In this work, we discuss the relationship between the disturbance and the clumps in radial action conjugate angle space  for disk stars.  But the discussion does not relate to the clumps beyond the rapid transtion, such as the region where $6<R<8$ kpc and $2<Z<2$ kpc in bin 3 of Figure~\ref{numAngleR}; the response of thick disk stars and halo stars to a perturbation such as the impact of Sgr dSph needs further study.

\section{test particle simulation}\label{sec:testparticle}
In order to compare the clustering of disk stars in the conjugate angle of radial action with expectations for a dwarf galaxy perturbation,
we made a test particle simulation.  The initial conditions of the test particle simulation are the same as outlined in  Section 9.1 of \citet{2020ApJ...905....6X}. The disk is simulated with the $\tt MWPotential2014$ potential of galpy \citep{2015ApJS..216...29B} with a Galaxy  virial mass of 0.8$\times10^{12} M_\odot$. The Sun is located at $(X, Y, Z)=(-8, 0, 0)$ kpc. The disk stars rotate clockwise, looking down on the $X-Y$ plane. The azimuth angle $\phi$ is defined as $\phi=0$ in the direction from the Galactic center to the Sun and increases in the clockwise direction. A dwarf galaxy that is represented by a point mass of  $2\times10^{10} M_\odot$ passes through the disk perpendicularly and just once \citep{2018MNRAS.481.1501B}. The orbit of the dwarf galaxy is not influenced by the Galactic gravity as it passes. The dwarf galaxy appears at the position $(X, Y, Z)=(-15, 0, 10)$ kpc, passes through the disk, and finally arrives at $(X, Y, Z)=(-15, 0, -10)$ kpc at which point the influence of the dwarf galaxy vanishes. The test particle simulation starts integrating  at $t=-500$ Myr. The dwarf galaxy appears at $Z=10$ kpc at $t=-66$ Myr. The time duration of the impact is 66 Myr \citep{2018MNRAS.481.1501B}. The completion of the impact occurres at $t=0.0$ Myr, when the satelltie is at $Z=-10$ kpc.

\subsection{Character of influenced stars after impact}
In Section \ref{sec:asymmetric}, we suggested the possibility that a dwarf galaxy impact may produce highly concentrated substructure in conjugate angle space. We want to know if this possibility can be corroborated with a test particle simulation. 
In this subsection we show how the distribution in the conjugate angle space of radial action changes due to the disturbance caused by the passage of a dwarf galaxy. In particular, we examine selected stars located near the projection point of the dwarf galaxy at 15 Myr after the start of the impact. 

\begin{figure}
\centering
\includegraphics[width=7cm]{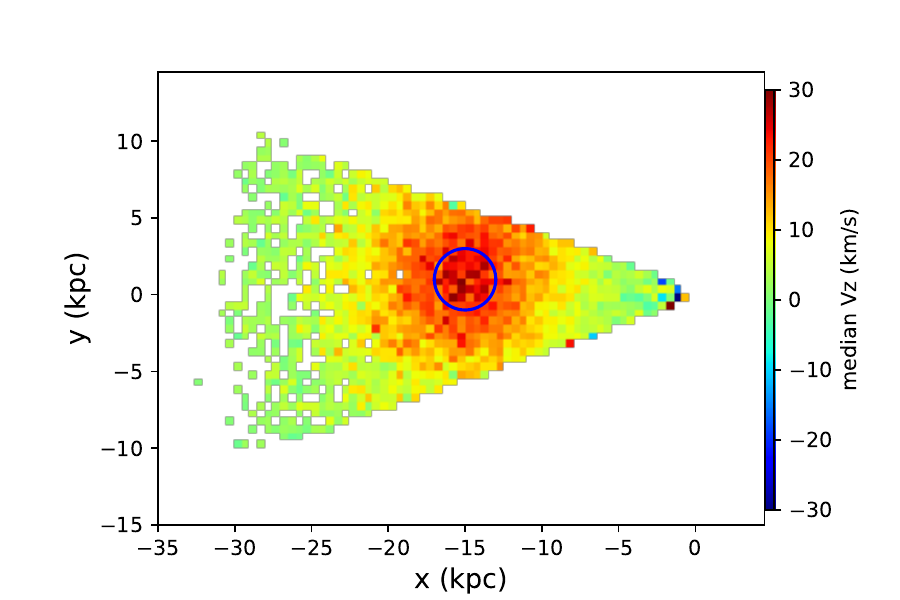}
\caption{This plot shows that the median $V_Z$ distribution of disk stars in the wedge is influenced by the impact dwarf galaxy, 15 Myr after the beginning of the impact. Once the dwarf galaxy appears at $(X,Y,Z)=(-15,0,10)$ kpc, the stars in the disk are perturbed towards positive $V_Z$ around the position $(X, Y)=(-15,0)$ kpc.  These influenced stars rotate clockwise.  The blue circle highlights the stars with the largest positive $V_Z$ in a radius of 2 kpc centered on $(X, Y)=(-15,1)$ kpc. }
\label{85sample}
\end{figure}

At the beginning of the impact, the influenced disk stars are pulled up and dragged towards the dwarf galaxy; the influence accumulates with time.
Figure~\ref{85sample} shows the median $V_Z$ distribution of disk stars in a wedge in the $X-Y$ plane with $-20^\circ<\phi<20^\circ$ at 15 Myr after the beginning of the impact. The blue circle in Figure~\ref{85sample} shows the 2 kpc radius around the  point $(X, Y)=(-15, 1)$ kpc, highlighting the stars with largest positive $V_Z$. These stars that are in the blue circle at 15 Myr after the beginning of the impact in Figure~\ref{85sample} are selected to study the effect that the passage of a dwarf galaxy has on the distribution of $\theta_R$ at the end of the impact. 

\begin{figure}
\centering
\includegraphics[width=12cm]{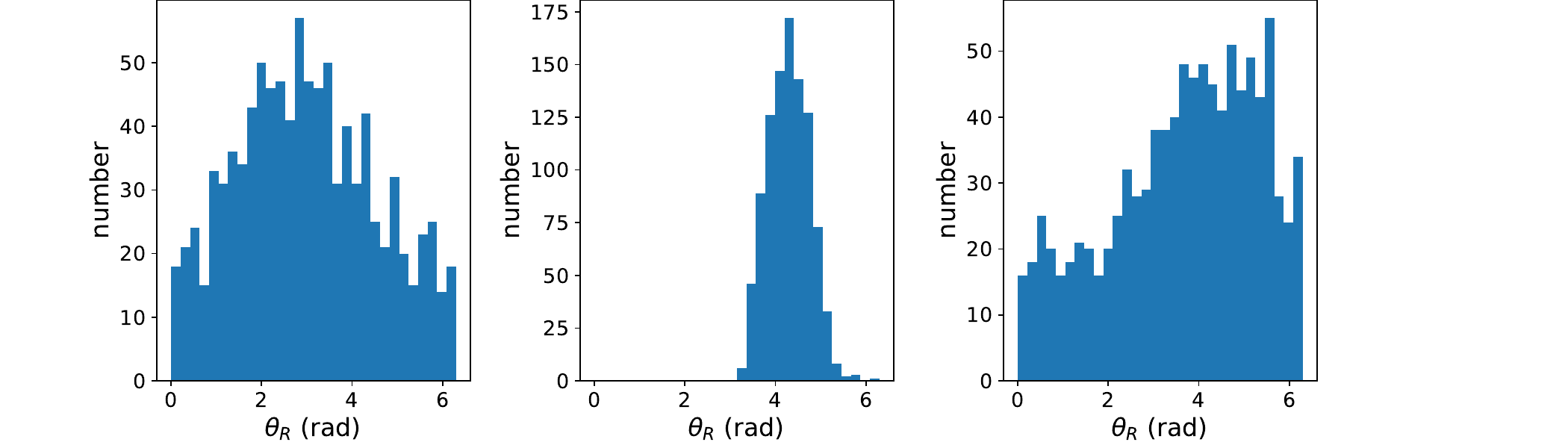}
\caption{This plot shows the $\theta_R$ distributions of stars in the blue circle of Figure~\ref{85sample}.  The left panel shows the $\theta_R$ distribution of these stars at the start of the impact ($t=-66$ Myr). The middle panel shows the $\theta_R$ distributon of these same stars at the end of  the impact ($t=0$ Myr). The right panel shows the $\theta_R$ distribution that these same stars in the left panel would have at the end of the ``impact" if the simulation is run without the  passage of a dwarf galaxy.}
\label{85and155AngleR}
\end{figure}

The $\theta_R$ distribution of stars within the blue circle in Figure~\ref{85sample} at the start and end of the impact is shown in Figure~\ref{85and155AngleR}. From the left and middle panel of Figure~\ref{85and155AngleR},  the standard deviation of the $\theta_R$ distribution at the start and end of the impact changes from 1.64 to 0.45 radians. The decreasing standard deviation in the distribution of $\theta_R$ for the test particle simulation is caused by the influence of the passage of the dwarf galaxy. The stars within the blue circle have a similar location in the disk and they are dragged in a similar way by the perturbation of the dwarf galaxy. When the perturbation is strong enough, it can wipe away the orbital phase difference of these stars. 

 The right panel of Figure~\ref{85and155AngleR} shows the $\theta_R$ distribution of the same stars at the same time as the middle panel, if the simulation  is run without  the disturbance of the passage of the dwarf galaxy. The standard deviation of the $\theta_R$ distribution is 1.55 radians, which is similar to the standard deviation in the left panel of Figure~\ref{85and155AngleR}. 

\subsection{Substructures found within the test particle simulation}
In Section \ref{sec:asymmetric}, the ``north branch", ``south branch" and ``Monoceros area" show the same features in the conjugate angle space of radial action. If the substructures are all produced by a gravitational disturbance such as the Sgr dSph impact, we seek to explain why the ``north branch" and ``south branch" regions show a strong  high $V_\phi$ spiral while the ``Monoceros area" shows a low $V_\phi$ substructure.  To seek answers, we turn to our test particle simulation to search for clues to the
explanation.  In the previous subsection, we focused on the test particle simulation stars that were disturbed by the dwarf galaxy  and studied the change in the distribution of these stars in conjugate angle space during the impact. In this subsection, we search for a portion of the test particle simulation after the impact that exhibits similar phase space features to those that we  observe in our LAMOST K giant sample.  We then trace back  the locations and orbital properties of the stars in those features at the beginning of the impact.

 \subsubsection{Selection of simulation that is similar to observations}

In Figure~\ref{numAngleR}, the most significant substructures are the ``north branch", ``south branch" and ``Monoceros area". The ``north branch" and ``south branch" are  projections on the $R-Z$ map of a series of high $V_\phi$ phase spirals \citep{2020ApJ...905....6X}. The ``Monoceros area" is a structure with small velocity dispersion and relatively low $V_\phi$.  We are therefore looking for a wedge of disk stars in the simulation with both a high $V_\phi$ phase spiral (similar to the ``north branch" and ``south branch") and a substructure with low $V_\phi$ and small velocity dispersion at larger Galactic radius (similar to the ``Monoceros area").  

\begin{figure}
\centering
\includegraphics[width=8cm]{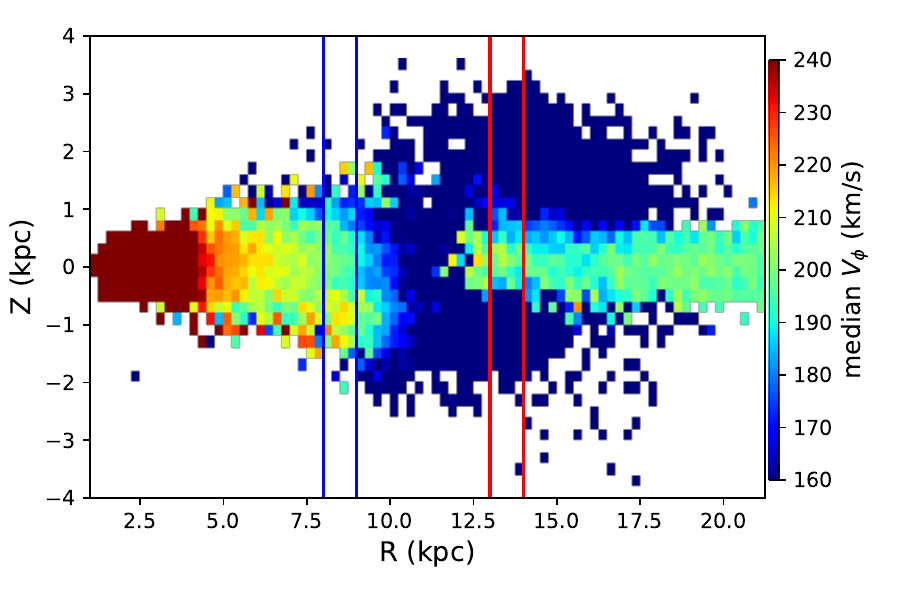}
\includegraphics[width=8cm]{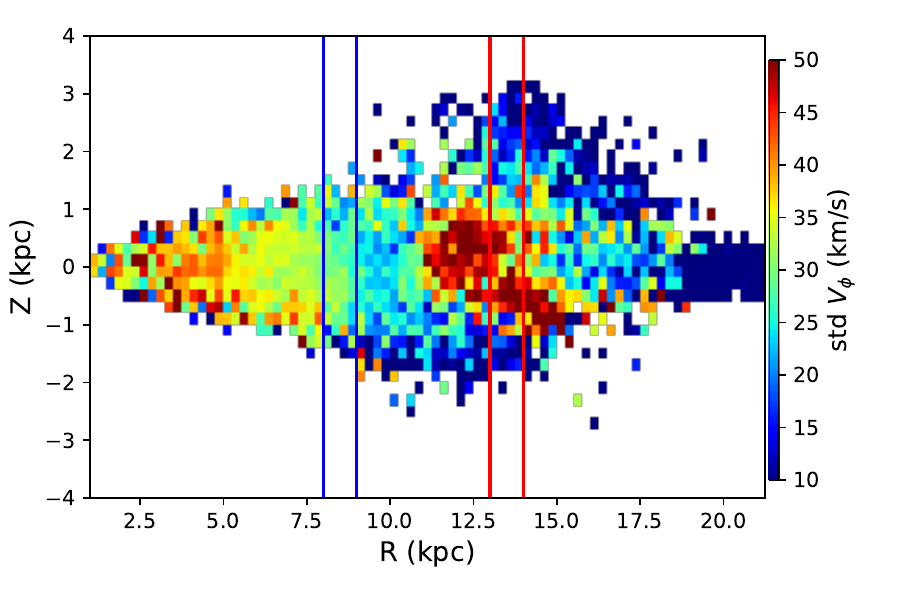}
\caption{The left panel shows the median $V_\phi$ distribution in the $R-Z$ map of the simulation bodies with $270^\circ-20^\circ<\phi<270^\circ+20^\circ$ at 200 Myr after the impact. The right panel shows the distribution of the standard deviation in $V_\phi$ for the same sample slice. The slices with $8<R<9$ kpc and $13<R<14$ kpc are selected for further analysis. The  $8<R<9$ kpc slice is named ``SampleR8", the $13<R<14$ kpc slice is named ``SampleR13". }
\label{RZhess180_270}
\end{figure}

Figure~\ref{RZhess180_270} shows the chosen wedge.  It shows the $V_\phi$ distribution and its standard deviation in an $R-Z$ map of a slice with  $270^\circ-20^\circ<\phi<270^\circ+20^\circ$ at 200 Myr after the impact. We choose this snapshot because it happens that this wedge shows an obvious high $V_\phi$ phase spiral at smaller Galactocentric radius and a low $V_\phi$ substructure with small velocity dispersion at larger Galactiocentric radius.
 
 \begin{figure}
\centering
\includegraphics[width=6cm]{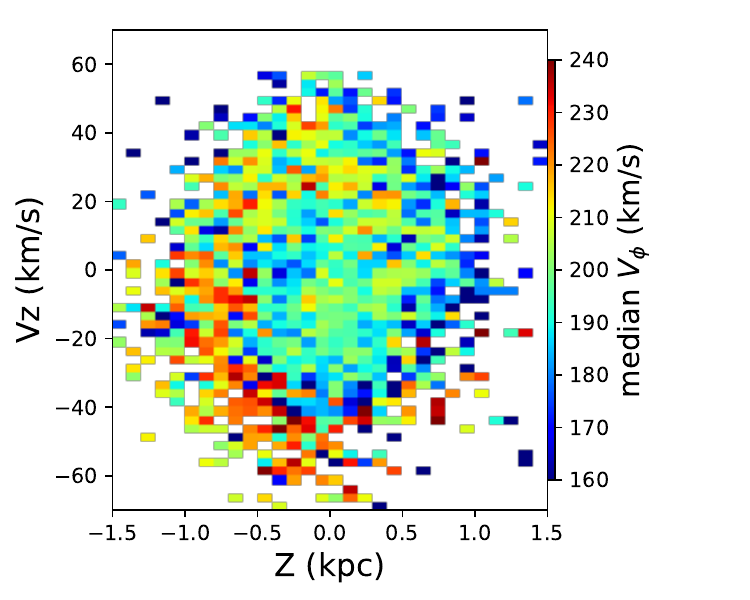}
\includegraphics[width=6cm]{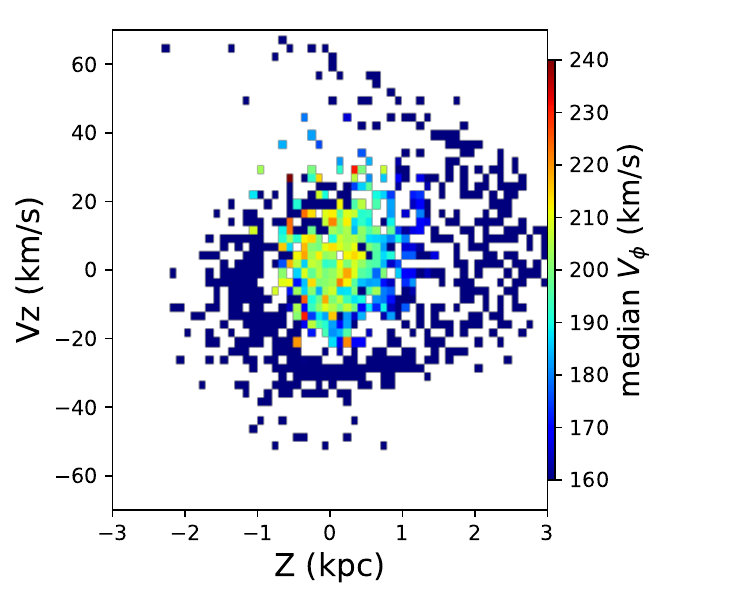}
\caption{The left and right panel shows the $V_\phi$ phase spiral in $Z-V_Z$ phase space for stars of ``SampleR8" and ``SampleR13" in Figure~\ref{RZhess180_270}.  The left panel shows a  high $V_\phi$ spiral and the right panel shows a low $V_\phi$ spiral.}
\label{PS180_270}
\end{figure}

 Two slices are cut on Figure~\ref{RZhess180_270} in the range of $8<R<9$ and $13<R<14$ kpc to sample the areas with  high and low $V_\phi$  substructures, repectively.
 We refer to the stars in the slice with $8<R<9$ kpc, $270^\circ-20^\circ<\phi<270^\circ+20^\circ$ at 200 Myr after the impact as ``SampleR8",
and the stars in the slice with $13<R<14$ kpc, $270^\circ-20^\circ<\phi<270^\circ+20^\circ$  at 200 Myr after the impact   as ``SampleR13", as shown in Figure~\ref{RZhess180_270}.
  Figure~\ref{PS180_270} shows the $V_\phi$ distribution in $Z-V_Z$ phase space of stars in the two slices. There is a high $V_\phi$ phase spiral in  ``SampleR8". There is a low $V_\phi$ phase spiral in ``SampleR13". 

\begin{figure}
\centering
\includegraphics[width=18cm]{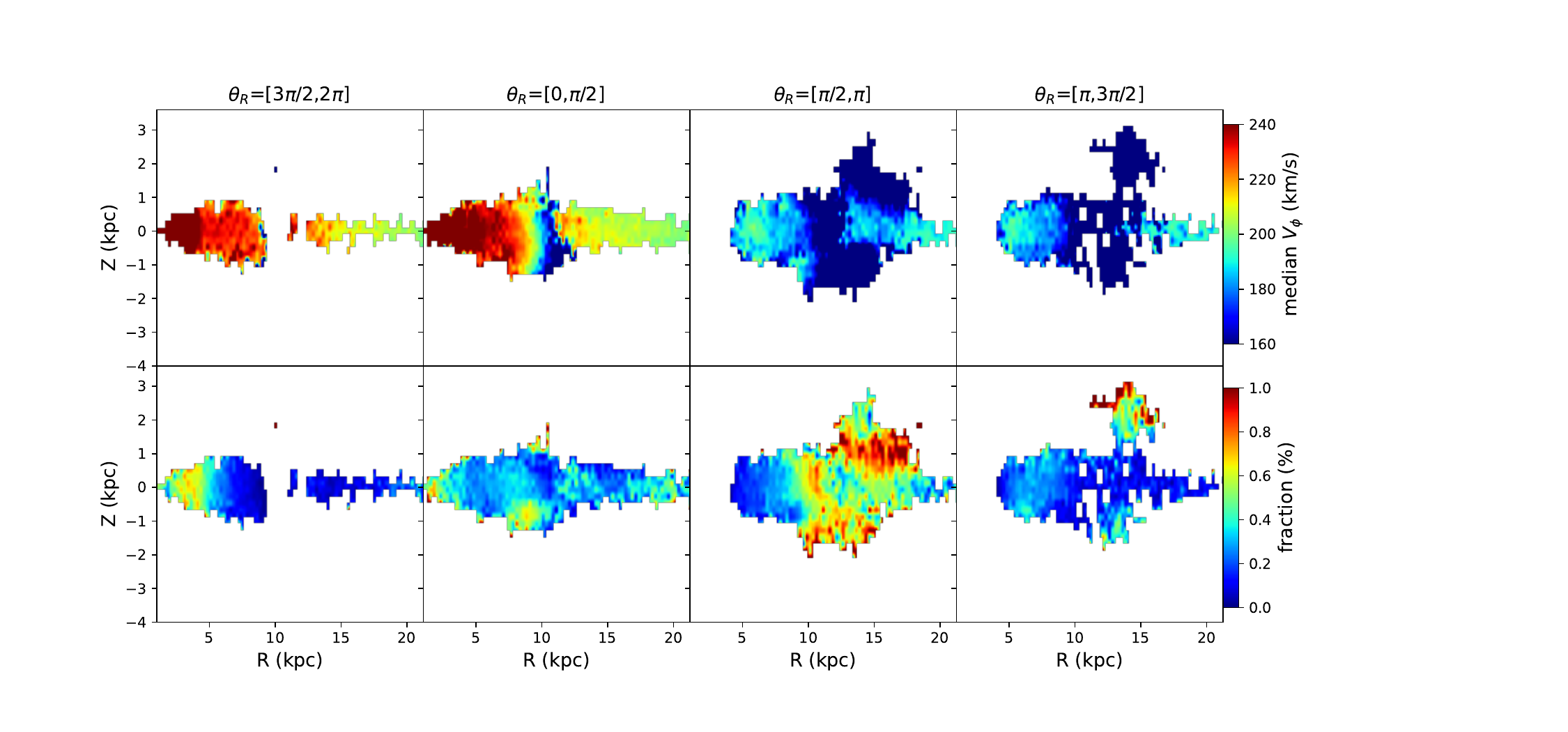}
\caption{ The panels show the $V_\phi$ distribution and number fraction of the total sample for the same wedge as shown in Figure~\ref{RZhess180_270}, now divided into four bins in $\theta_R$.}
\label{RZhess180_270_AngleR}
\end{figure}

 The sample shown in Figure~\ref{RZhess180_270} is divided into the same 4 bins in $\theta_R$ as described in Section \ref{sec:sample}. 
Figure~\ref{RZhess180_270_AngleR} shows the median $V_\phi$ and  proportional distribution in the $R-Z$ map of the 4 subsamples. The number fraction here is defined the same way as the fraction of total sample in Section \ref{sec:asymmetric}. From Figure~\ref{RZhess180_270}, we see a low $V_\phi$ substructure with small standard deviation in $V_\phi$ at $12<R<14$ kpc, $Z<-1.5$ kpc and $13<R<16$ kpc and $Z>1.5$ kpc. 
 In Figure~\ref{RZhess180_270_AngleR}, we see that the low $V_\phi$ substructures in the range  $12<R<14$ kpc, $Z<-1.5$ kpc and $13<R<16$ kpc and $Z>1.5$ kpc mainly belong to bin 3, where $\theta_R$ $\in[\pi/2, \pi]$. The fraction of the substructures reaches 80\% in this bin.
 
 \begin{figure}
\centering
\includegraphics[width=5cm]{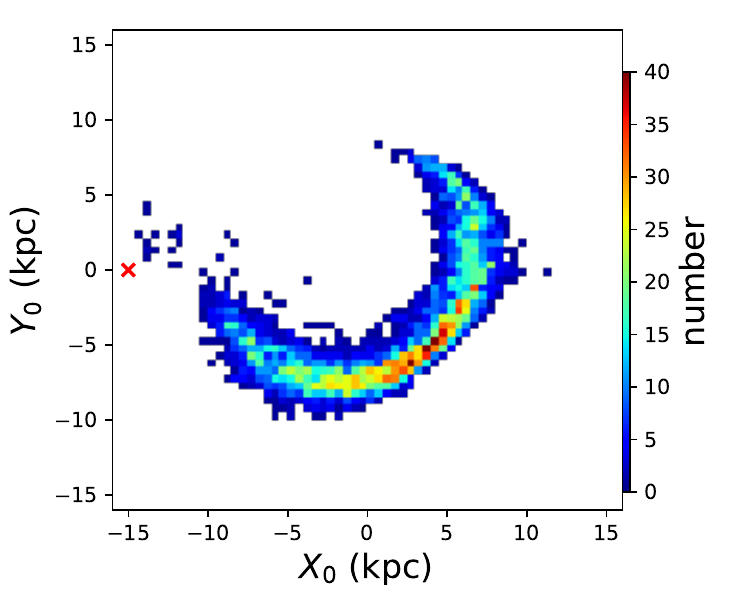}
\includegraphics[width=6cm]{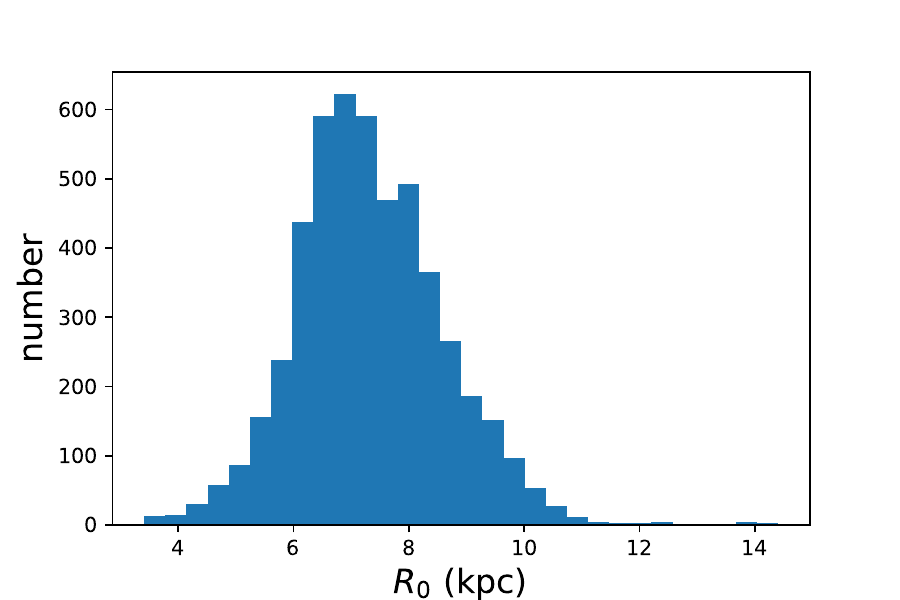}
\includegraphics[width=6cm]{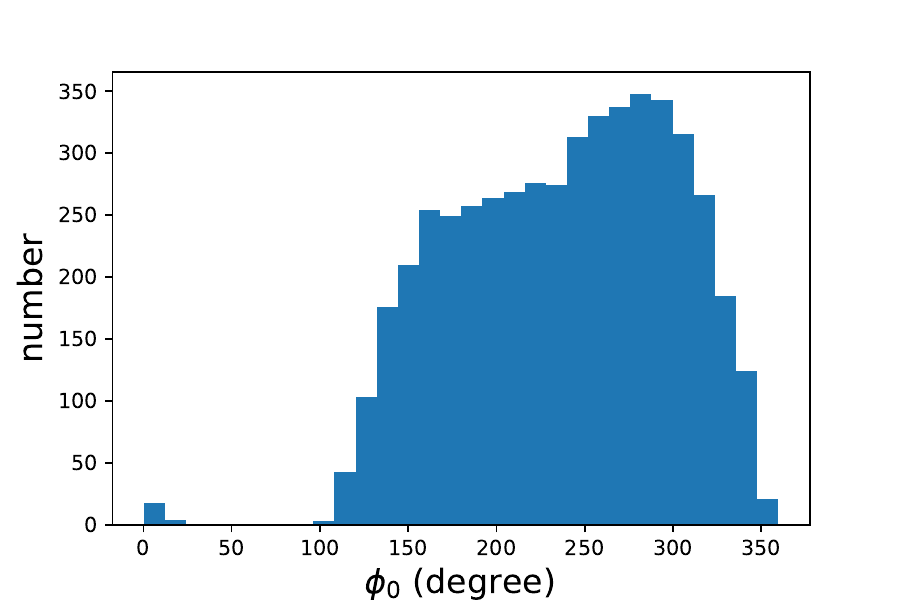}
\caption{The plots show the spatial distribution at the beginning of the impact for the stars of ``SampleR8" shown in Figure~\ref{RZhess180_270}.  The first panel shows the the number counts of these stars in the $X_0-Y_0$ plane. The red cross shows the projection point of the dwarf galaxy passage through the disk in the $X_0-Y_0$ plane. The second panel shows a histogram of the $R_0$ distribution. The third panel shows a histogram of the $\phi_0$ distribution.}
\label{Rphi180_270_7}
\end{figure}

\begin{figure}
\centering
\includegraphics[width=5cm]{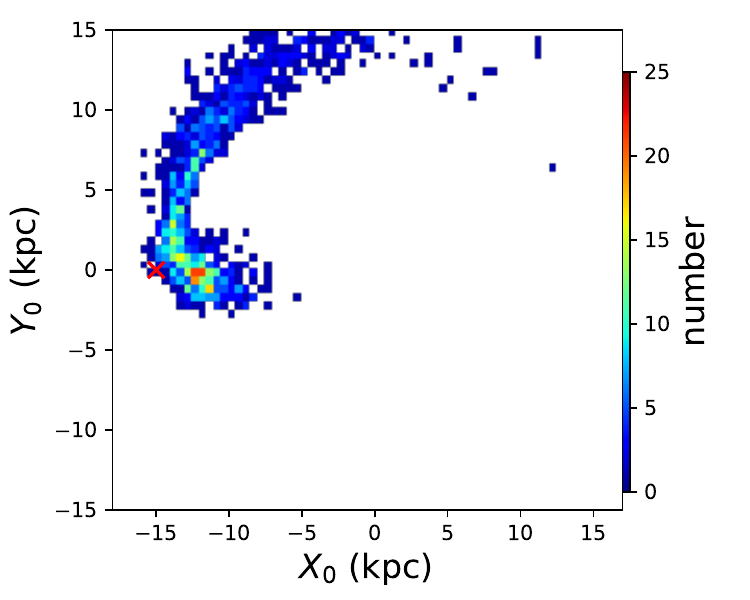}
\includegraphics[width=6cm]{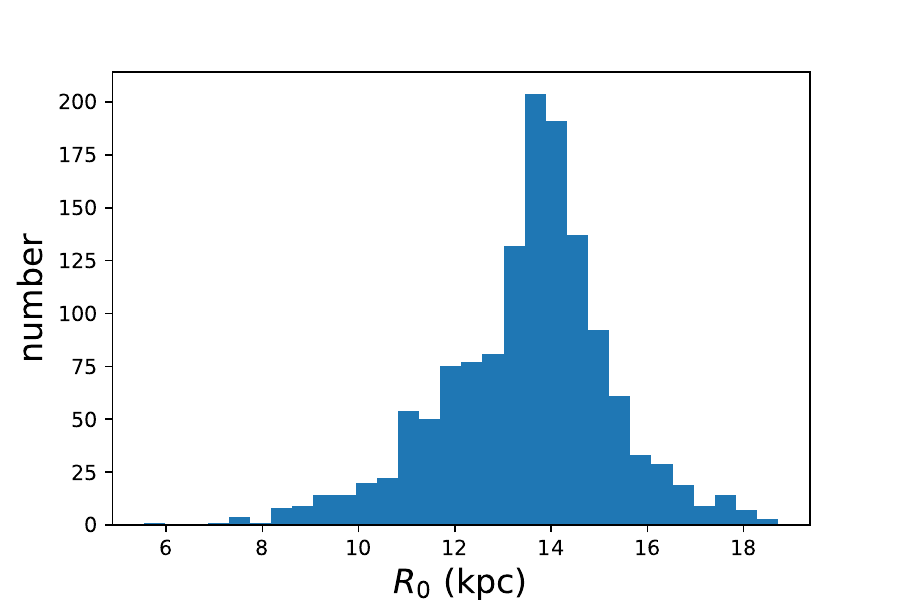}
\includegraphics[width=6cm]{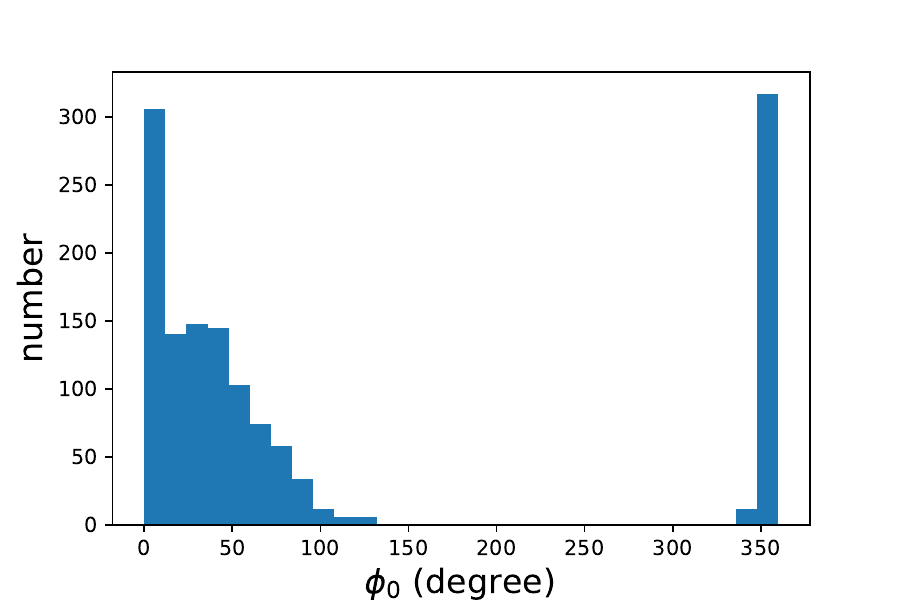}
\caption{The plots show the spatial distribution at the beginning of the impact  for the stars of ``SampleR13" shown in Figure~\ref{RZhess180_270}. The first panel shows the the number counts of these stars in $X_0-Y_0$ plane. The red cross shows the projection point of the dwarf galaxy passage through the disk in $X_0-Y_0$ plane. The second panel shows a histogram of the $R_0$ distribution. The third panel shows a histogram of the $\phi_0$ distribution.}
\label{Rphi180_270_12}
\end{figure}

Note that in the \citet{2020ApJ...905....6X} test particle simulation, the phase spiral appears at $8-9$ kpc from 120 Myr to 360 Myr after the impact in our test particle simulation.  The duration of the phase spiral is shorter than the suggested time in many other works, especially  in results of $N$-body simulations, e.g. \citet{2018MNRAS.481..286L}, because self-gravity is not included in our test particle simulation. Since we use only a toy model, we can approximate the effect of the impact only in the most simplified way,  the impact time is not accurate. 

\subsubsection{Spatial distribution of ``SampleR8" and ``SampleR13" at the beginning of the impact} 

We trace back the orbits of the stars of ``SampleR8" and ``SampleR13" to find the characteristics of the stars in the two slices at the time of impact,  relative to the projection point. Figure~\ref{Rphi180_270_7} shows the location of the stars of ``SampleR8" at the beginning of the impact. Here $R_0$ and $\phi_0$  describe the location of stars at the beginning of the simulation. These stars are distributed in the range  $4<R_0<10$ kpc; the peak is at 6.5 kpc. They occupy a wide $\phi_0$ range, $130^\circ<\phi_0<360^\circ$. At the beginning of the impact, the dwarf galaxy is at $(R_0, Z_0, \phi_0)=(15$  kpc, 10 kpc, $ 0^\circ)$. The disk rotates along the direction of increasing $\phi_0$. So, during the impact, these stars were rotating towards the dwarf galaxy.

Figure~\ref{Rphi180_270_12} shows the location of stars of ``SampleR13" at the beginning of the impact. The $R_0$ range is $8<R_0<18$ kpc and the peak is 14 kpc. The $\phi_0$ range is $0^\circ<\phi_0<100^\circ$. The $\phi_0$ distribution of stars of slice ``SampleR13"  is more concentrated than that of the stars of ``SampleR8". 
During the impact, the stars of ``SampleR13" were moving away from the passage of dwarf galaxy. These stars are decelerated during the impact.

\subsubsection{The change in the integral invariants of ``SampleR8" during the impact}
From subsection 6.2.2, we know the locations of ``SampleR8" and ``SampleR13" are different relative to the projection point of passage of dwarf galaxy of are different at the beginning of the impact.  We also wonder how the integral invariants change due to the impact. The integral invariants of stars describe their orbital properties. We care about angular momentum ($L_Z$), pericentric radius ($r_\mathrm{peri}$) and apocentric radius ($r_\mathrm{apo}$) that determine the eccentricity of the orbit, and vertical action ($J_Z$) and frequency of the vertical action ($\Omega_Z$) that determine the orbit of star in $Z-V_Z$ space.

\begin{figure}
\centering
\includegraphics[width=18cm]{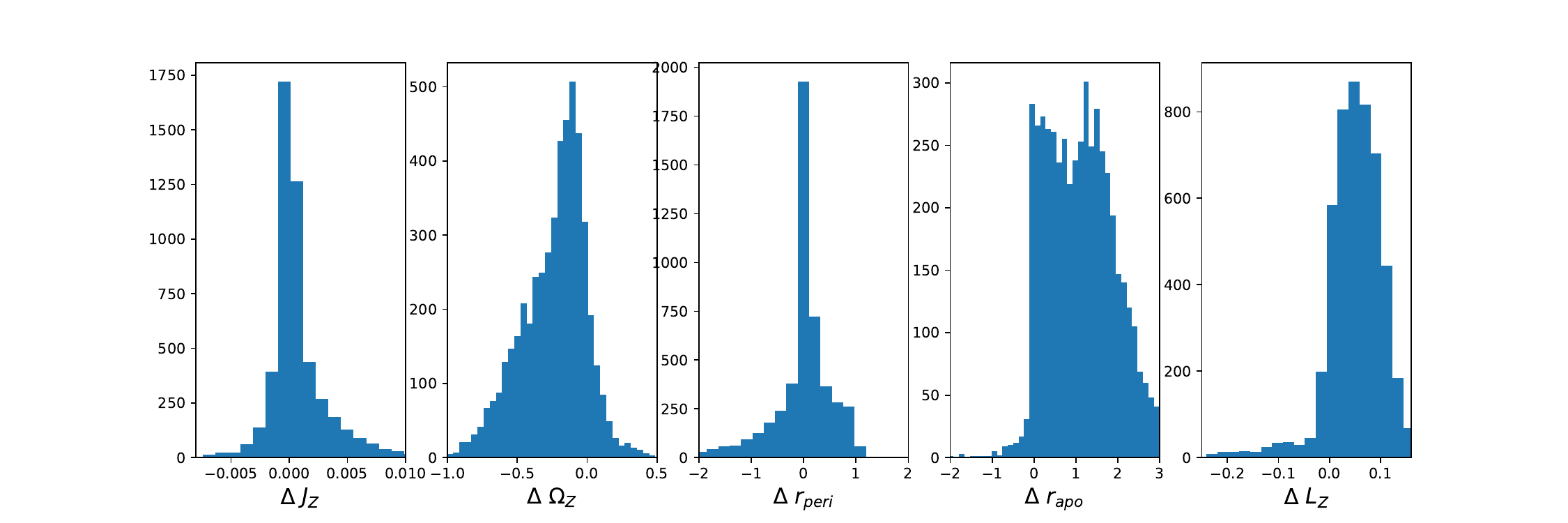}
\caption{Histogram of the amount of change in the integral invariants (vertical action $J_Z$, frequency of vertical action $\Omega_Z$, pericentric radius $r_\mathrm{peri}$, apocentric radius $r_\mathrm{apo}$, angular momentum $L_Z$) before and after the impact for the stars  of ``SampleR8" from Figure~\ref{RZhess180_270}. }
\label{histR8}
\end{figure}

  Figure~\ref{histR8} shows a histogram of the change in the integral invariants of these stars before and after the impact of ``SampleR8" ( $J_Z$,  $\Omega_Z$,  $r_\mathrm{peri}$,  $r_\mathrm{apo}$,  $L_Z$).  From Figure~\ref{histR8}, we see that the angular momentum increases with the gravitational pull of the Sgr dSph. The orbital radius gets larger due to the increasing angular momentum.  The apocentric radius of most of these stars shows an obvious increase. On the other hand, the vertical frequency of most of these stars  decreases. The full maps of the changes in the integral invariants are shown in Figures~\ref{Lz} to ~\ref{rap} of the Appendix. Comparing Figures~\ref{Rphi180_270_7} \&~\ref{histR8} to Figure~\ref{Rphi180_270_12} \&~\ref{histR12}, we know that the amount of change in the integral invariants is related to the relative position of the stars to the passage of dwarf galaxy during the impact.

\begin{figure}
\centering
\includegraphics[width=8cm]{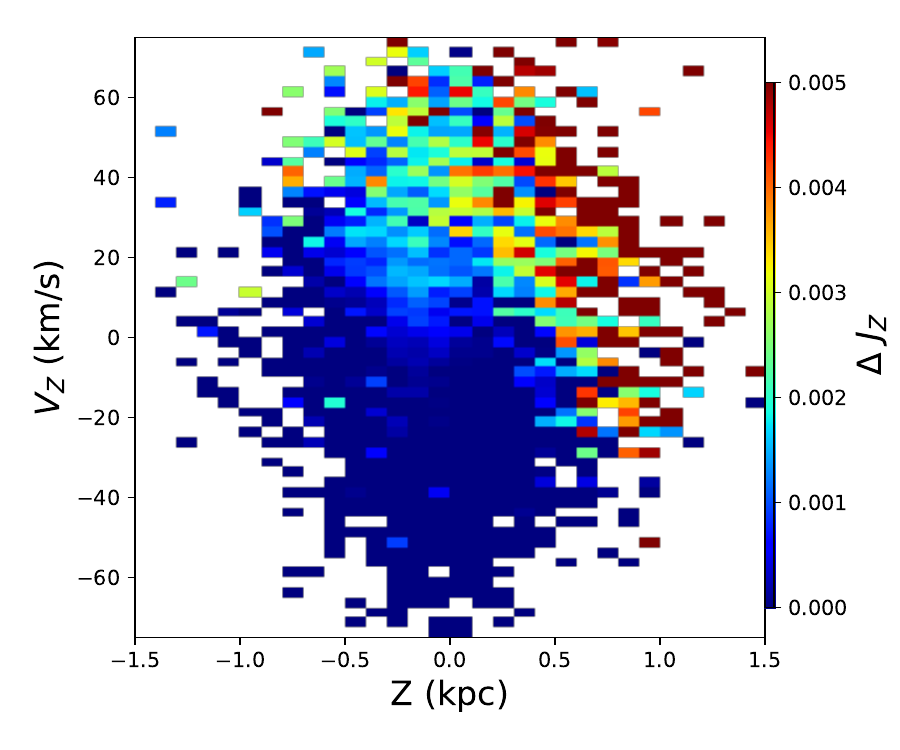}
\includegraphics[width=8cm]{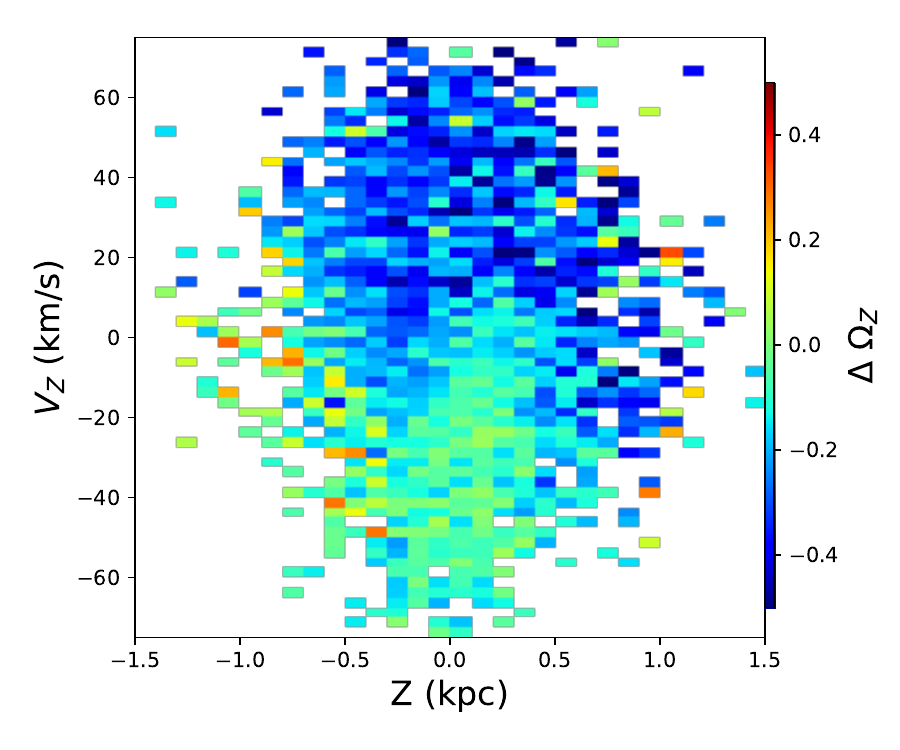}
\includegraphics[width=8cm]{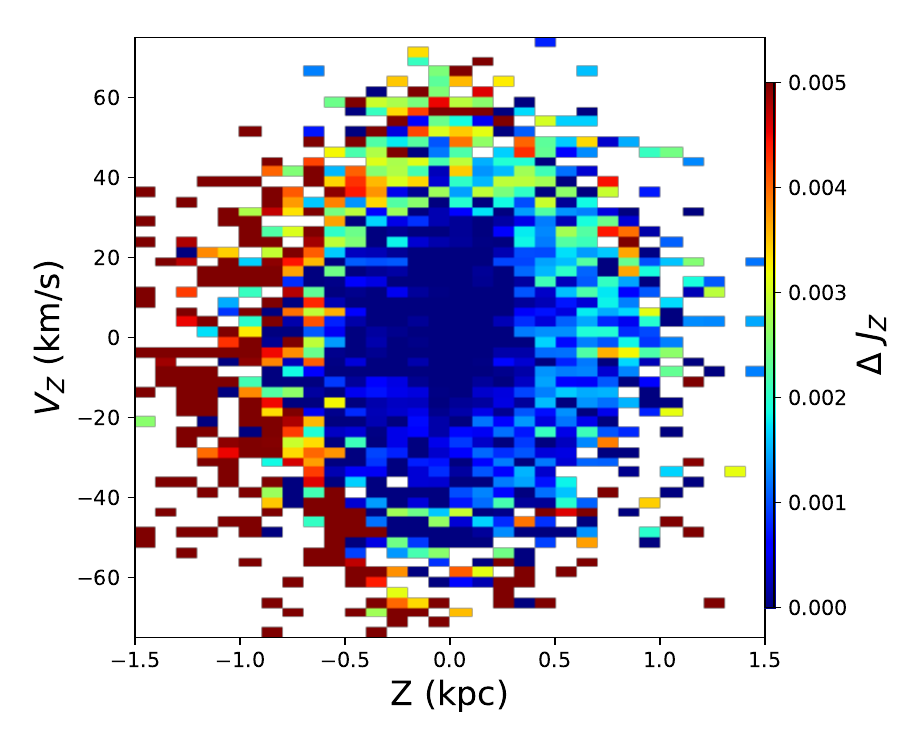}
\includegraphics[width=8cm]{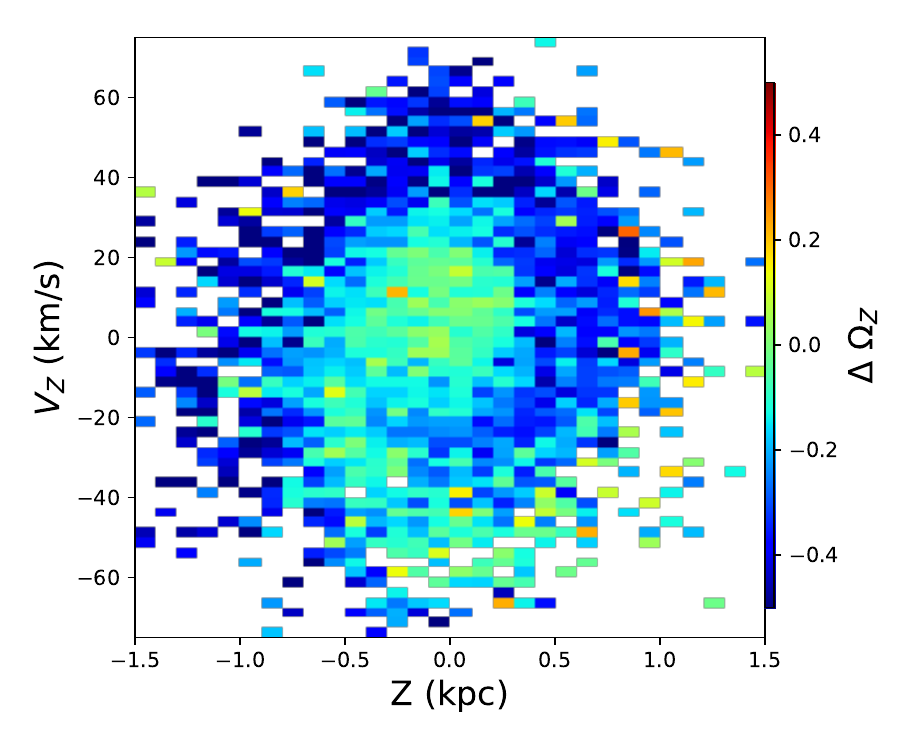}
\caption{The two upper panels show the distribution of stars of ``SampleR8" at the beginning of the impact in the $Z-V_Z$ phase space, color coded by median of amount of change in $J_Z$ and $\Omega_Z$ during the impact.  The two lower panels show the distribution of stars in ``SampleR8" at 200 Myr after  the impact in the $Z-V_Z$ phase space color, coded by median of amount of change in $J_Z$ and $\Omega_Z$ during the impact. The amount of change is defined similarly for $J_Z$ and $\Omega_Z$; $\Delta \Omega_Z=\Omega_Z$(at the beginning of the impact)$-\Omega_Z$(at the end of the impact). }
\label{ZVzR8}
\end{figure}

\begin{figure}
\centering
\includegraphics[width=6cm]{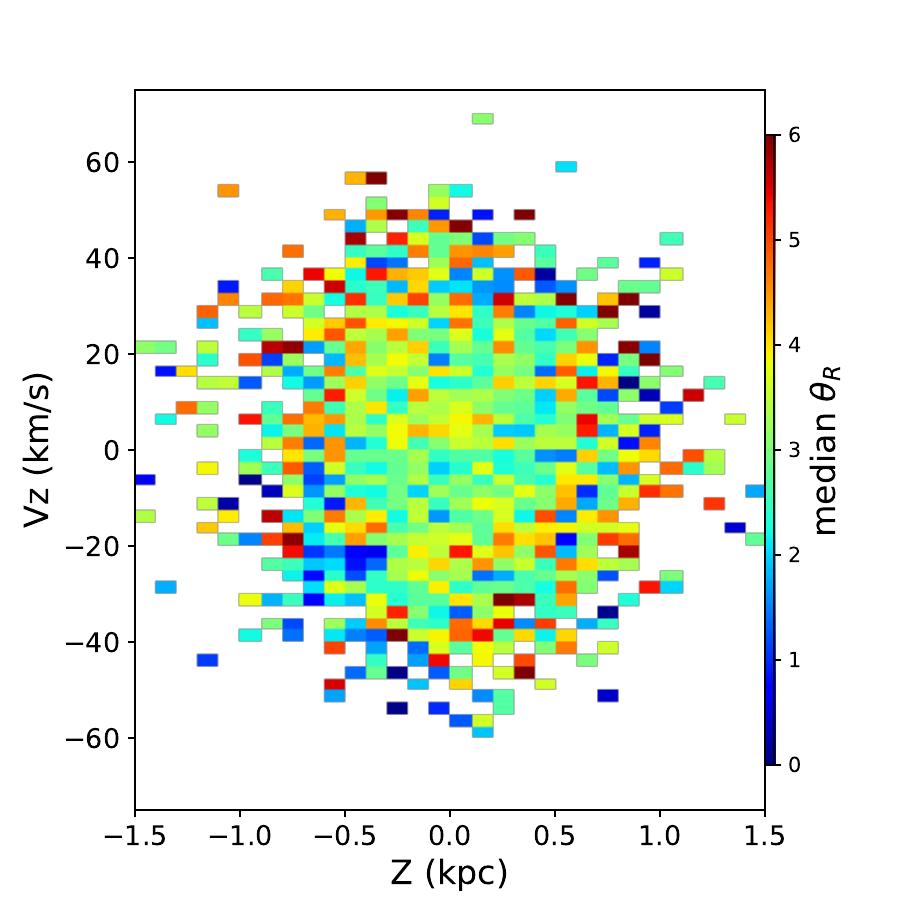}
\includegraphics[width=6cm]{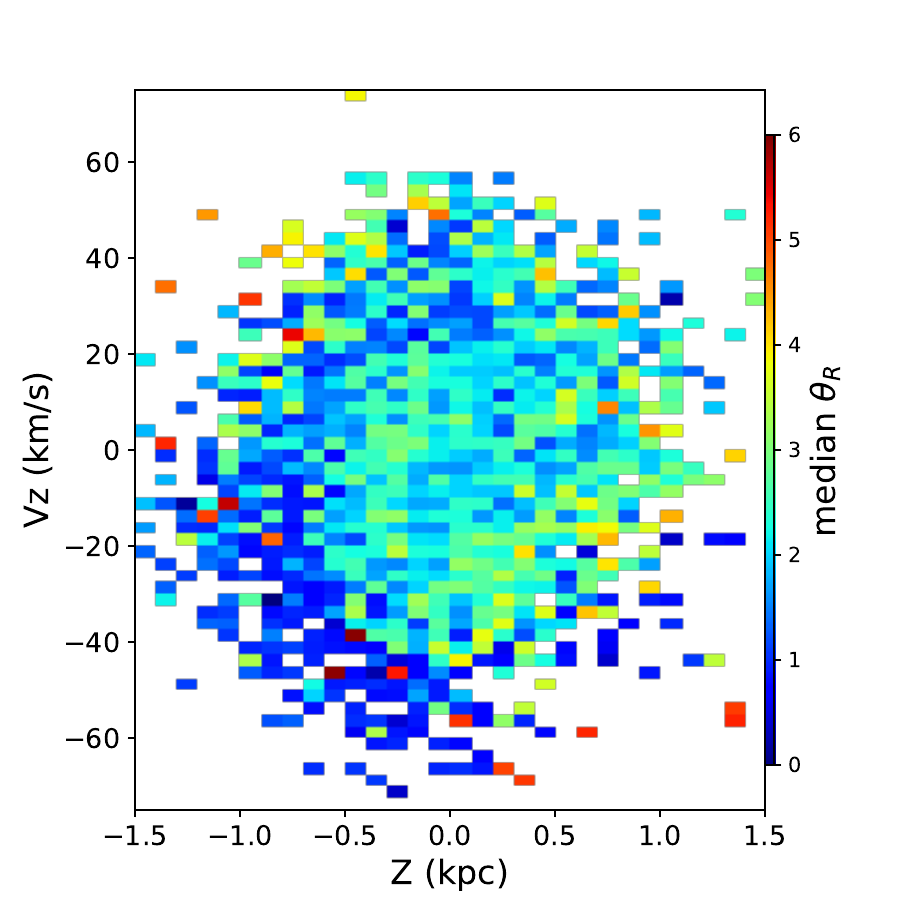}
\caption{Median $\theta_R$ distribution in Z-$V_Z$ phase space for stars of  ``SampleR8" from Figure~\ref{RZhess180_270}. The left panel is the median $\theta_R$  at the beginning of impact. The right panel is the median $\theta_R$  at 200 Myr after impact. Note that the stars on the phase spiral in the right panel have $\theta_R\sim1$ radian, 200 Myr after the impact.}
\label{AngleR8}
\end{figure}

From Section \ref{sec:asymmetric}, we know the features that are a projection of phase spirals in $Z-V_Z$ space are asymmetric around $Z=0$ kpc. We would like to know how a  change in the  integral invariants is related to the formation of phase spirals. The relationship between the phase spiral and the amount of change in the integral invariants is also studied in this subsection. The motion of stars  in $Z-V_Z$ phase space is determined by the integral invariants $J_Z$ and $\Omega_Z$ \citep{2018MNRAS.481.1501B,2019MNRAS.489.4962K}. The upper panels of Figure~\ref{ZVzR8} show the distribution of stars of ``SampleR8" in $Z-V_Z$ phase space at the moment of the beginning of the impact. The distribution is color coded by the amount of change in  $J_Z$, $\Omega_Z$ during the impact. The two upper panels of Figure~\ref{ZVzR8} show that the most disturbed stars whose $\Delta$ $J_Z$, $\Delta$ $\Omega_Z$ is largest concentrate in the quadrant with $Z>0$ kpc and $V_Z>0$ km/s. If the dwarf galaxy comes from the north side the Milky Way, the stars of $Z>0$ kpc are  influenced first. For the same reason, the influenced stars are pulled upwards, so the motion of stars with $V_Z>0$ km/s are strengthened. 

The two lower panels of Figure~\ref{ZVzR8} show the distribution of ``SampleR8" in  $Z-V_Z$ phase space at 200 Myr after the impact. The distribution is color-coded by the change in  $J_Z$ and  $\Omega_Z$.  Just like analysis around Figure 3 of \citet{2021MNRAS.504.3168B}, the stars with different $\Omega_Z$ have different orbital phase. The most disturbed stars are dragged into a spiral due to different $\Omega_Z$; the stars with higher $\Omega_Z$ wrap faster. This is evidenced by comparing the lower panels of Figure~\ref{ZVzR8} with the left panel of Figure~\ref{PS180_270};  we see strong similarities in the $V_\phi$ phase spiral and the spiral composed by the most disturbed stars that have largest $\Delta$$\Omega_Z$ and $\Delta$$J_Z$. The location of the high $V_\phi$ phase spiral is from ($Z,V_Z)=(0$ kpc, 40 km/s), to ($-0.7$ kpc, 0 km/s), then to (0 kpc, $-50$ km/s) in the left panel of Figure~\ref{PS180_270}. The phase spirals traced by the highest $\Delta \Omega$ and $\Delta J_Z$ are located at a similar place in $Z-V_Z$ space in the lower panels of Figure~\ref{ZVzR8}. 

The Section 6.1 shows that the disturbed stars may obtain identical $\theta_R$. Figure~\ref{ZVzR8} shows that the phase spiral in $Z-V_Z$ space composed of the most disturbed stars. We wonder if the stars on the phase spiral show identical $\theta_R$. Figure ~\ref{AngleR8} shows the median $\theta_R$ distribution in $Z-V_Z$ phase space at the beginning of the impact and 200 Myr after the impact. In the left panel of Figure~\ref{AngleR8}, there is no obvious structure in the distribution of median $\theta_R$ in Z-$V_Z$ phase space. From the right panel of Figure~\ref{AngleR8}, the stars on the high $V_\phi$ phase spiral have similar median $\theta_R$ of around 1 radian. From the distribution of median $\theta_R$, the stars on the phase spiral are closer to pericenter (where $\theta_R=0$ radian)  than the stars off the phase spiral. That is why the stars on the phase spiral show higher $V_\phi$ than the stars off the phase spiral. 

\begin{figure}
\centering
\includegraphics[width=18cm]{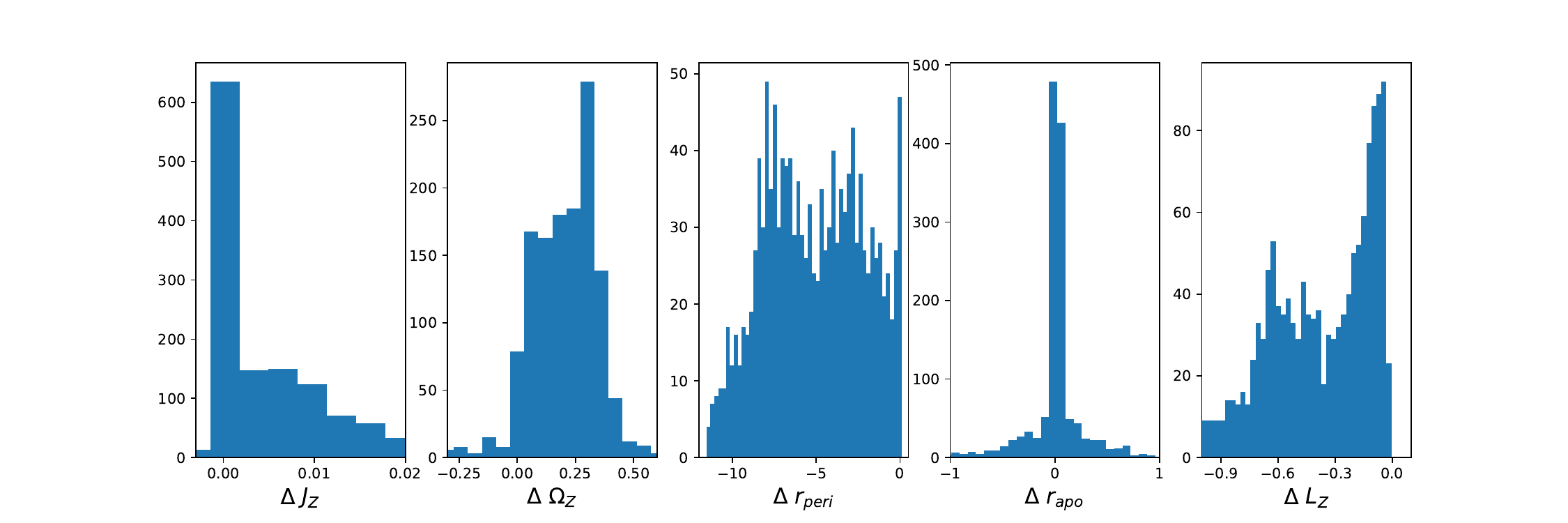}
\caption{Histogram of amount of change in integral invariants (vertical action $J_Z$, frequency of vertical action $\Omega_Z$, pericentric radius $r_\mathrm{peri}$, apocentric radius $r_\mathrm{apo}$, angular momentum $L_Z$) before and after the impact of stars  in ``SampleR13" from Figure~\ref{RZhess180_270}.}
\label{histR12}
\end{figure}

\begin{figure}
\centering
\includegraphics[width=8cm]{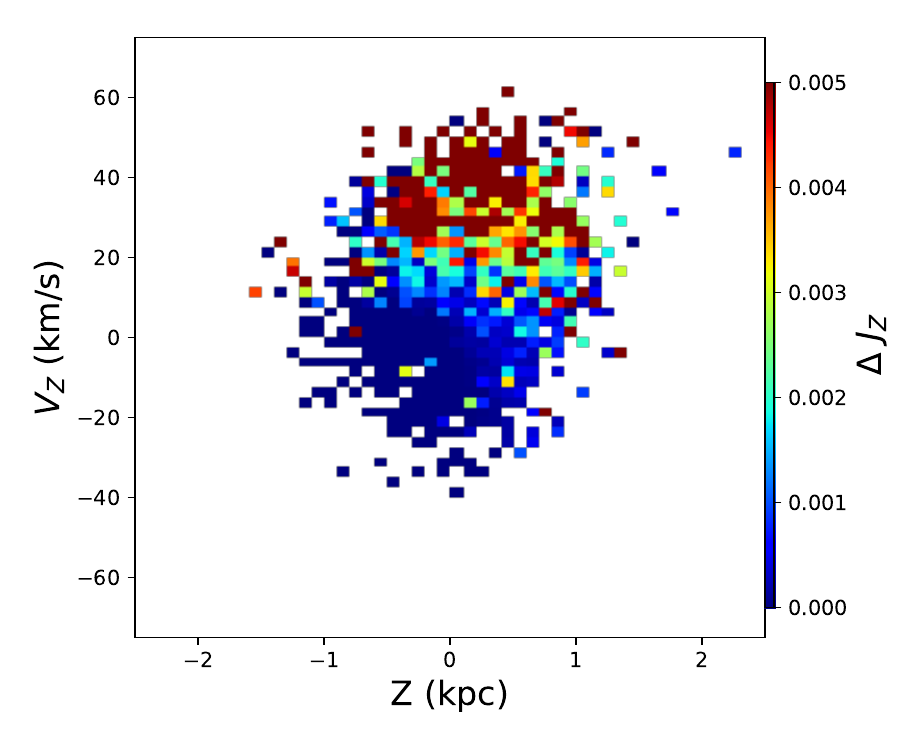}
\includegraphics[width=8cm]{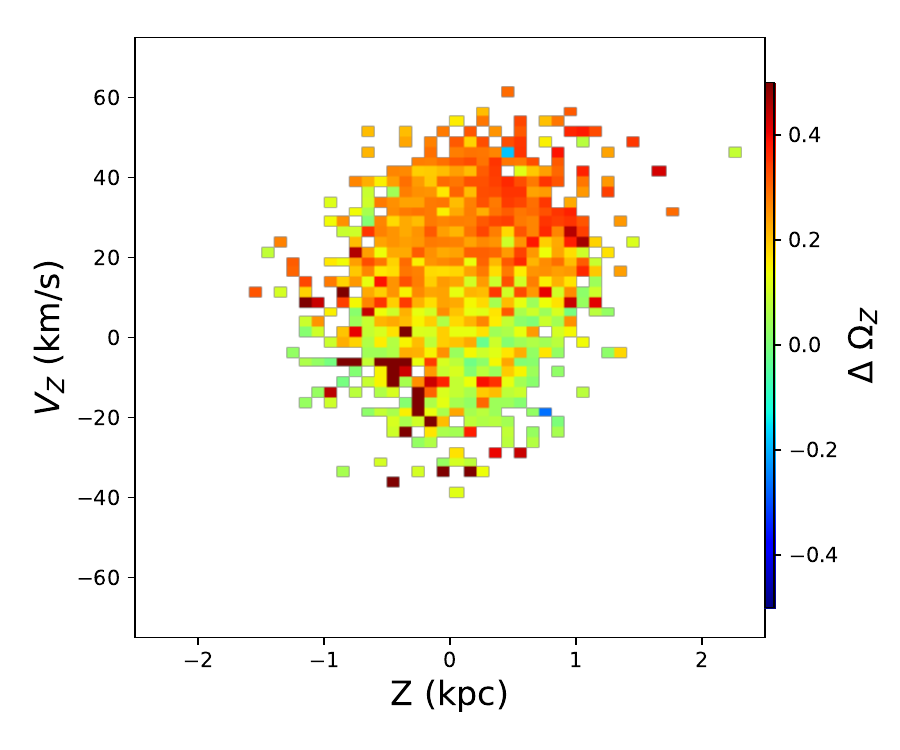}
\includegraphics[width=8cm]{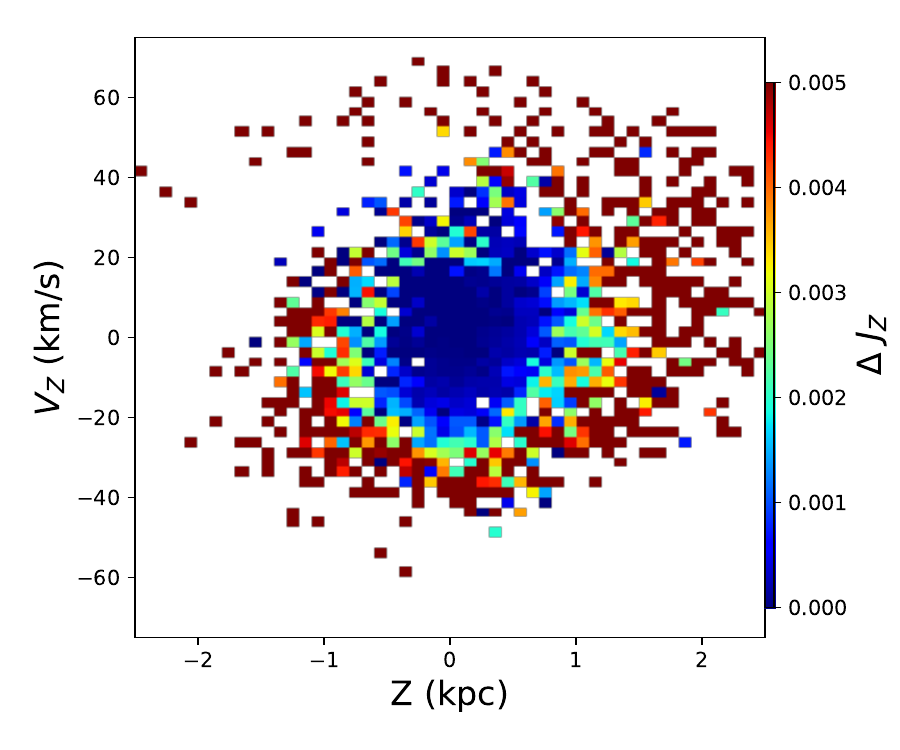}
\includegraphics[width=8cm]{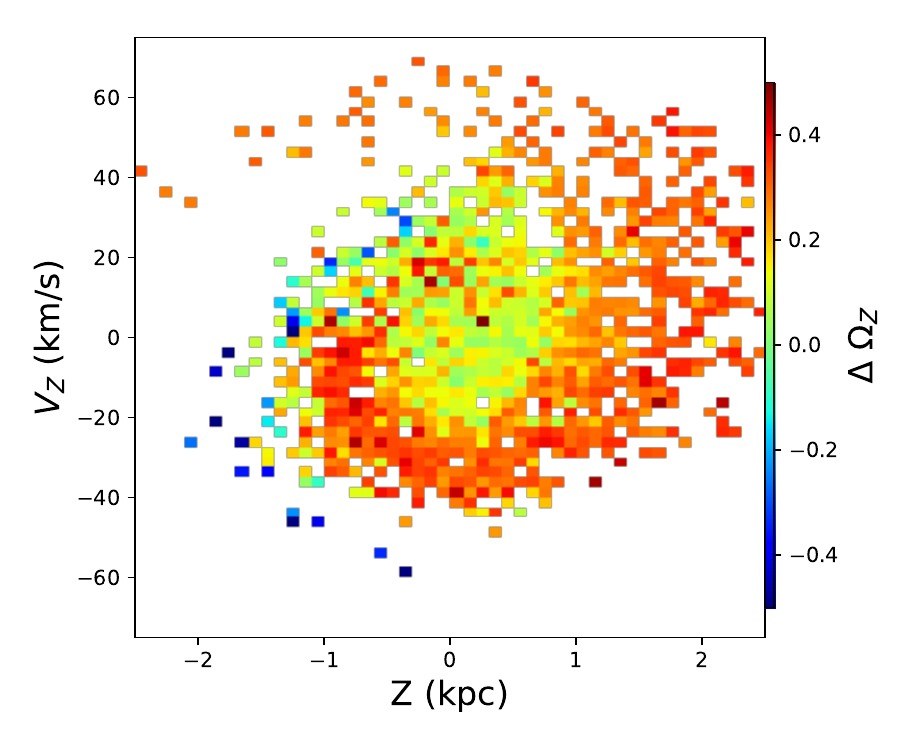}
\caption{The two upper panels show the distribution of stars in ``SampleR13" at the beginning of the impact in the $Z-V_Z$ phase space, color coded by the median of the amount of change in $J_Z$ and $\Omega_Z$ during the impact.  The two lower panels show the distribution of stars in ``SampleR13" at 200 Myr after  the impact in the $Z-V_Z$ phase space, color coded by median of amount of change in $J_Z$ and $\Omega_Z$ during the impact. The amount of change is defined similarly for $J_Z$ and $\Omega_Z$; $\Delta \Omega_Z=\Omega_Z$(at the beginning of the impact)-$\Omega_Z$(at the end of the impact). }
\label{ZVzR12}
\end{figure}

\begin{figure}
\centering
\includegraphics[width=6cm]{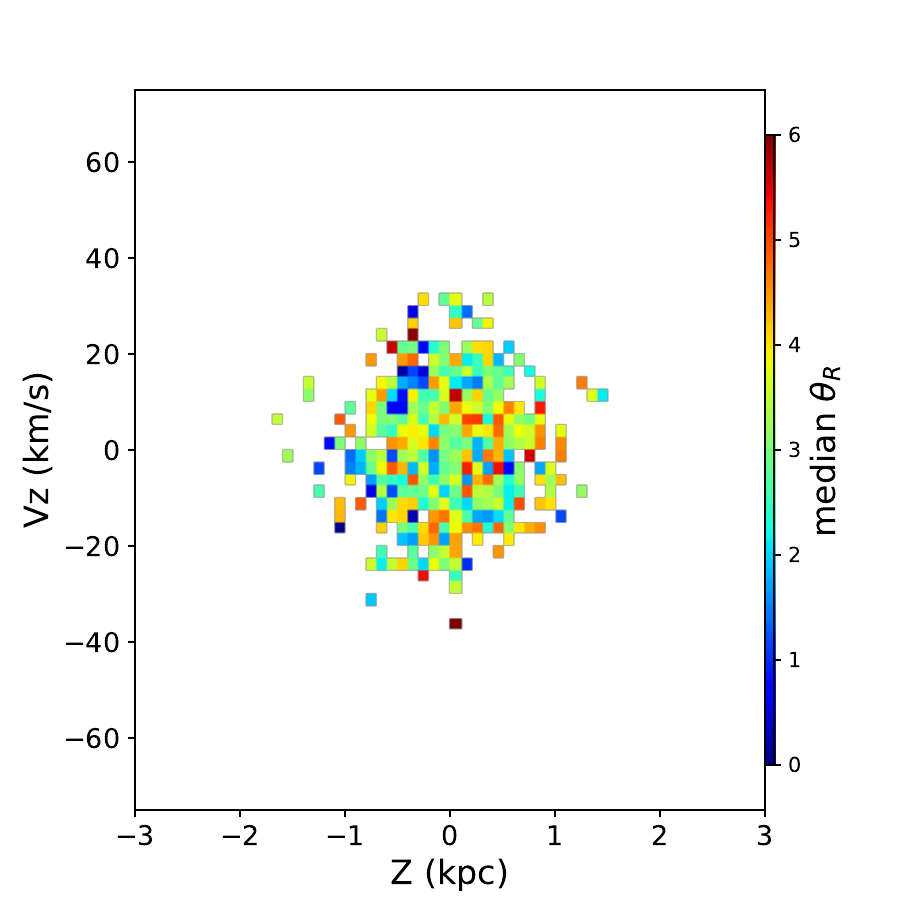}
\includegraphics[width=6cm]{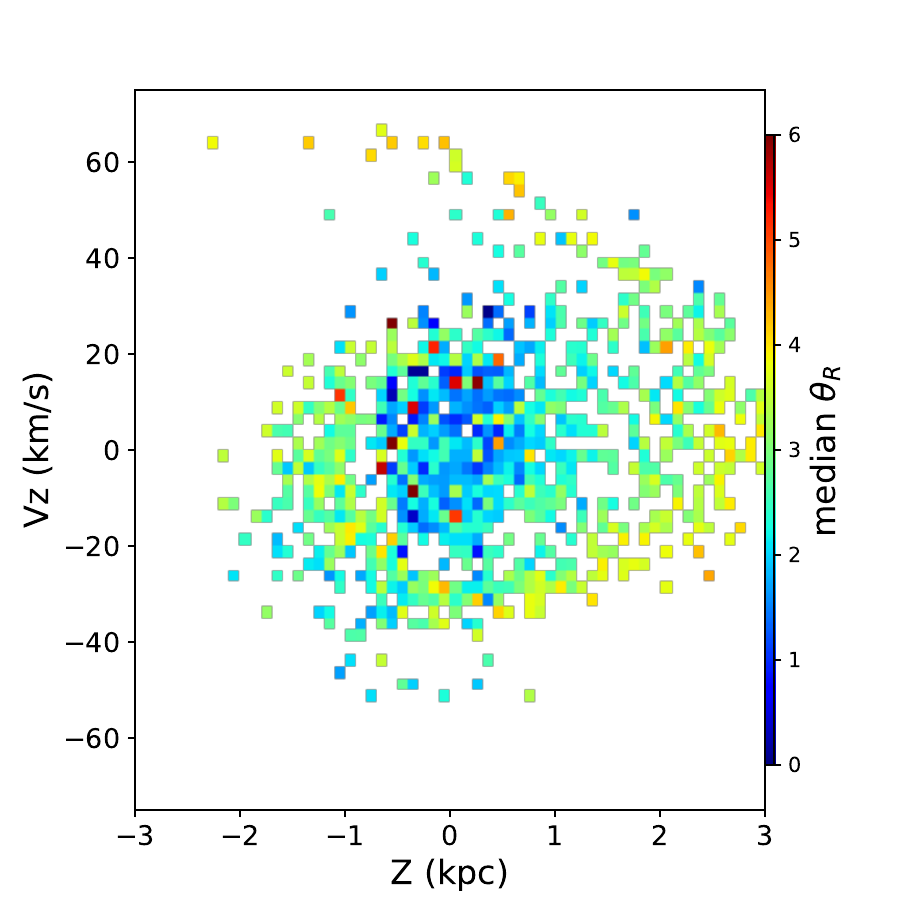}
\caption{Median $\theta_R$ distribution in $Z-V_Z$ phase space of stars  in ``SampleR13" of Figure~\ref{RZhess180_270}. The left panel is the median $\theta_R$ at the start of the impact. The right panel is the median $\theta_R$ at 200 Myr after the impact. }
\label{AngleR12}
\end{figure}

\subsubsection{The change in the integral invariants of ``SampleR13" during the impact}  
 Figure~\ref{histR12} shows the amount of change in integral invariants of ``SampleR13". From Figure~\ref{Rphi180_270_12}, the stars of ``SampleR13" are moving away from the location of impact at the beginning of the impact.  Correspondingly, Figure~\ref{histR12} shows that the angular momentum and pericentric radius of most of these stars  decreases. The  vertical frequency  of most of these stars is increased. 
 
Similar to the analysis of ``SampleR8" in subsection 6.2.3, we see that the $V_\phi$ phase spiral shows the most overlap with the spiral composed of the most disturbed stars, that have the largest $\Delta$$\Omega_Z$ and $\Delta$$J_Z$, compare the lower panels of Figure~\ref{ZVzR12} with the right panel of Figure~\ref{PS180_270}. In the right panel of Figure~\ref{PS180_270}, the low $V_\phi$ phase spiral wraps  from $(Z,V_Z)=(0$ kpc, 20 km/s) to (-1 kpc, 0 km/s), and then to (0 kpc, -30 km/s), and finally (2.5 kpc, 0 km/s). The phase spiral traced by the largest $\Delta$$\Omega_Z$ and $\Delta$$J_Z$ in the lower panels of Figure~\ref{ZVzR12} has a similar location.

Figure ~\ref{AngleR12} shows the median $\theta_R$ distribution in $Z-V_Z$ phase space before and 200 Myr after the impact.  The right panel of Figure~\ref{AngleR12} shows that the stars on the low $V_\phi$ phase spiral share more identical median $\theta_R$ that is around 3. From the distribution of median $\theta_R$, the stars on the phase spiral are closer to apocenter than the stars off the phase spiral. Therefore, the stars on the phase spiral show lower $V_\phi$ than the stars off the phase spiral. 

\section{Comparison of observational substructures with the test particle simulation }\label{comparison}

In this section we compare the results of the test particle simulation with the two obvious kinds of substructures in the observational data. The test particle simulation is a very simplified model. It does not include self-gravity or gas. We cannot match the results of the test particle simulation with the observational substructure in detail.  Instead, we compare similar properties of the observational data and test simulation  qualitatively. 

The analysis of Section \ref{sec:asymmetric} highlights that one kind of substructure is exemplified by the ``north branch" and the ``south branch", which have higher density than the spatially adjacent stars and are prefentially found in bins 1 and  2 in Figure~\ref{numAngleR}. The ``north branch" and ``south branch" are a projection of the high $V_\phi$ phase spiral on the $R-Z$ map  \citep{2020ApJ...905....6X}. In many recent works, the phase spirals are associated with the last impact of the Sgr dSph.  In this work we find that the last impact of a dwarf galaxy with similar properties to the Sgr dSph also causes clumps in $\theta_R$ space.  The second  kind of substructure in Figure~\ref{numAngleR} is at the position of  the ``Monoceros area",  a low $V_\phi$ structure located at larger Galactic radii. More than 50\% of the stars in the region of the ``Monoceros area" are in bin 4.  
 The substructure has a more narrow distribution in $\theta_R$ space than that of the ``north branch" or the ``south branch",  indicating that it is also undergoing phase mixing. 

From comparison, 
the ``north branch", and ``south branch" are analogous to  ``SampleR8"  of the test particle simulation in Figure~\ref{RZhess180_270}. They all show a phase spiral with high $V_\phi$. Analogous to the analysis of the test particle simulation, the ``north branch" and ``south branch" substructures may also be composed of stars disturbed by the passage of a dwarf galaxy; the disturbed stars obtain identical orbital phase (similar $\theta_R$), and they develop a high $V_\phi$ phase spiral when they are near  pericenter.

The ``Monoceros area" substructure is analogous to the substructure found in ``SampleR13"  of the test particle simulation in Figure~\ref{RZhess180_270}.  Both of them are low $V_\phi$ substructures compared to the spatially adjacent stars. Both  are concentrated in a narrow range of $\theta_R$ space. The substructure of ``SampleR13" is produced by the passage of the dwarf galaxy in the test particle simulation. Analogously, the ``Monoceros area" substructure may also be composed of stars disturbed by the passage of a dwarf galaxy that in response obtain identical orbital phase. In this case, ``Monoceros area" stars have low $V_\phi$ because  the orbital phase is near apocenter. 

There is also an obvious difference between ``SampleR13" and the ``Monoceros area". ``SampleR13" shows a low $V_\phi$ phase spiral. The stars of the ``Monoceros area" do not exhibit a low $V_\phi$  phase spiral in $Z-V_Z$ space but instead form a low $V_\phi$ substructure. This difference can be explained because in the real situation, there is a longer dynamical time scale at  large Galactic radius where the ``Monoceros area" is found; the phase spiral has not wrapped up yet. In contrast,  the test particle simulation forms phase spirals much more quickly than  real life, due to the lack of self-gravity. 
 
 \begin{figure}
\centering
\includegraphics[width=6cm]{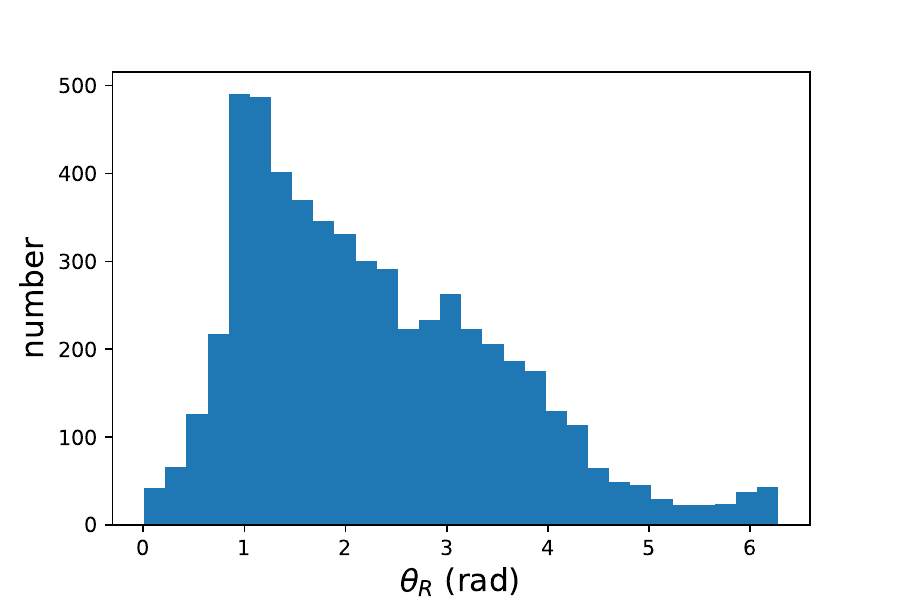}
\includegraphics[width=6cm]{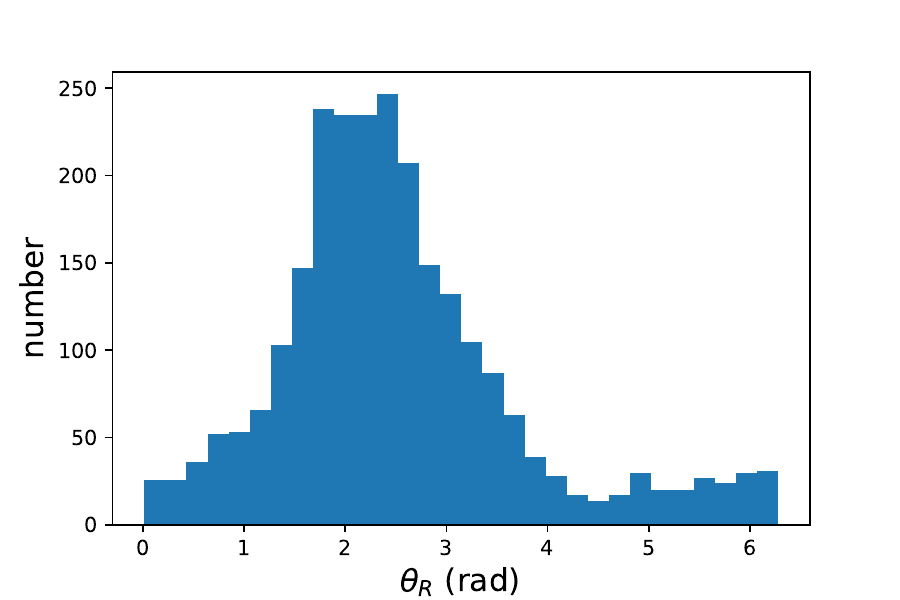}
\caption{Distribution in $\theta_R$ for ``SampleR8" (left panel) and ``SampleR13" (right panel). }
\label{histAngleR8and12}
\end{figure}

The  ``north branch" and ``south branch" substructures appear as  clumps in two $\theta_R$ bins (bin 1 and bin 2, respectively), while the ``Monoceros area" appears as a  clump only in one $\theta_R$ bin (bin 4).   Similarly, in the results of the test particle simulation, the stars in ``SampleR8"  have a wider $\theta_R$ distribution than that of ``SampleR13". In Figure~\ref{histAngleR8and12}, the standard deviation of the $\theta_R$ distribution of ``SampleR8" is 1.29 and that of ``SampleR13" is 0.95. We conjecture that these substructures are related to the presence of the different frequencies at different Galactic radii. 
The substructures at smaller $R$ have higher surface density thus have higher vertical frequency ($\Omega_Z$ describes the oscillation of the orbit in the vertical direction of the stars). 
Phase spirals can wrap up more quickly in the $Z-V_Z$ phase space with higher $\Omega_Z$.  At the same time, the stars with smaller $R$ have higher circular frequency ($\Omega_\phi$), so they mix more quickly due to differential rotation. The substructure at larger $R$ has lower $\Omega_Z$ and $\Omega_\phi$, and the wrapping up in $Z-V_Z$ and mixing due to differential rotation is slower, so it is still distributed narrowly in $\theta_R$. 
 
By comparing the observations and the results of the test particle simulation,  it is possible to explain the clumps in the $\theta_R$ space as a response to an interaction between the disk and dwarf galaxy.

\section{Discussion}\label{discussion}

\subsection{Importance of classifying orbital phase }
In this work, the data are divided by orbital phase represented by $\theta_R$. This work shows that classifying orbital phase is an important way to decipher the features of substructures. 

\citet{2021ApJ...911..107L} shows that the guiding radius ($R_g$), which is not influenced by the mixing of orbital phase, represents the phase spiral better than Galactocentric distance.  This indicates that the mixing of orbital phase can blur the features of the  kinematic structures. 

For example, orbital phase mixing may influence and introduce contamination to features detected in the distribution of $\alpha$ abundance within a given range of $R$. This is because in each subsample of a given $R$ and $\alpha$ abundance, the subsample includes stars near pericenter with larger $R_g$ and stars near apocenter with smaller $R_g$. Therefore, the subsample includes different features due to having stars at different guiding radii. We see this effect in the sixth row of Figure~\ref{velocityfeh}, where the scale heights of the $\alpha$ abundance contours are different at the same $R$ in different orbital phases. As a result, we cannot identify a  feature   by  $\alpha$ abundance alone; we also must consider the influence of orbital phase mixing, especially if the disk is not in equilibrium.  

Our method of examining substructure using orbital phase is suitable for \textit{Gaia} data. The identity of features can be made clearer and strengthened after taking into account the orbital phase mixing. Although these features might be made slightly less significant due to the division of the data into smaller subsets,  the vast quantity of data coming from \textit{Gaia} allows any slight loss of statistical significance to be more than offset by the increased purity of the sample achieved by considering orbital phase. 

\subsection{Comparison with results from previous simulations}
Our work provides new ways to observe phase mixing. We show from a test particle simulation that the Sgr dSph impact could cause  clumps in $\theta_R$ space similar to those that are observed. We then use the simulations to associate prominent clumps in $\theta_R$ space with their  dynamical origin.
From the test particle simulation analyzed in this work, we show that the Monoceros ring-like low $V_\phi$ substructure and the corresponding clump in the conjugate angle space of radial action can be produced by interaction between the disk and the dwarf galaxy.

Other simulations have also shown that the Monoceros-ring-like substructure can be excited by gravitational interaction \citep{2011MNRAS.414L...1M, 2011Natur.477..301P,2016MNRAS.456.2779G,2018MNRAS.481..286L}. 
 Using \textit{Gaia} DR2, DR3 and APOGEE data, \citet{2020MNRAS.492L..61L} show that the Anti-Center Stream included in the Monoceros ring have a narrow range of energies. 
From $N-$body simulations \citep{2018MNRAS.481..286L}, the position and kinematics of the Monoceros ring-like substructures can be reproduced by the cumulative interaction of multiple passes of the Sgr dSph. The tidally excited disk material at distance of TriAnd (25$-$30 kpc from the Galactic center) can stay coherent for several Gyrs \citep{2018MNRAS.481..286L}. 

Recently, \citet{2020arXiv201112323G}  found a Monoceros-ring-like substructure in the low-inclination, late-type galaxy VV304a. They are able to build a reasonable VV304a-like model in a fully self-consistent cosmological simulation. In the model, the mass of the host galaxy is $M_\mathrm{tot}=1.25\times10^{12}M_\odot$, and the mass of the passage of dwarf galaxy is $M_\mathrm{sat}=3\times10^{11}M_{\odot}$. The dwarf galaxy passes inside the virial radius at $R=40$ kpc. A vertical displacement that is similar to the substructure of the Monoceros ring is detected in the snapshots that are 0.73 Gyr and 0.9 Gyr after the impact.  

The most significant difference between the full $N-$body results and our test particle simulation is time scale.  In our work, the ``Monoceros area" is analogous to ``SampleR13". The corresponding clump in $\theta_R$ space of ``SampleR13" is a transient structure in the test particle simulation. It can be observed in a snapshot that is 200 Myr after the impact and it can exist for 100 Myr.  In the $N$-body simulations, Monoceros-ring-like structure survives for a  much longer time period \citep{2018MNRAS.481..286L}. The difference in the survival rate can be traced to the lack of self-gravity in our test particle simulation.

\subsection{Effect of the LAMOST footprint on the ``Monoceros area"}
We showed that the ``Monoceros area" is dominated by stars that are receding from the apocenter. However, this could be affected by the fact that the observational data is limited by the LAMOST footprint. The range of $\phi$  observed in the ``Monceros area" is $-20^\circ<\phi<10^\circ$ (see Figure 2 of \citealt{2020ApJ...905....6X}). Since the LAMOST data samples only a part of the Monoceros ring, we cannot determine the complete orbital phase distribution along the Monoceros ring.
  
\begin{figure}
\centering
\includegraphics[width=18cm]{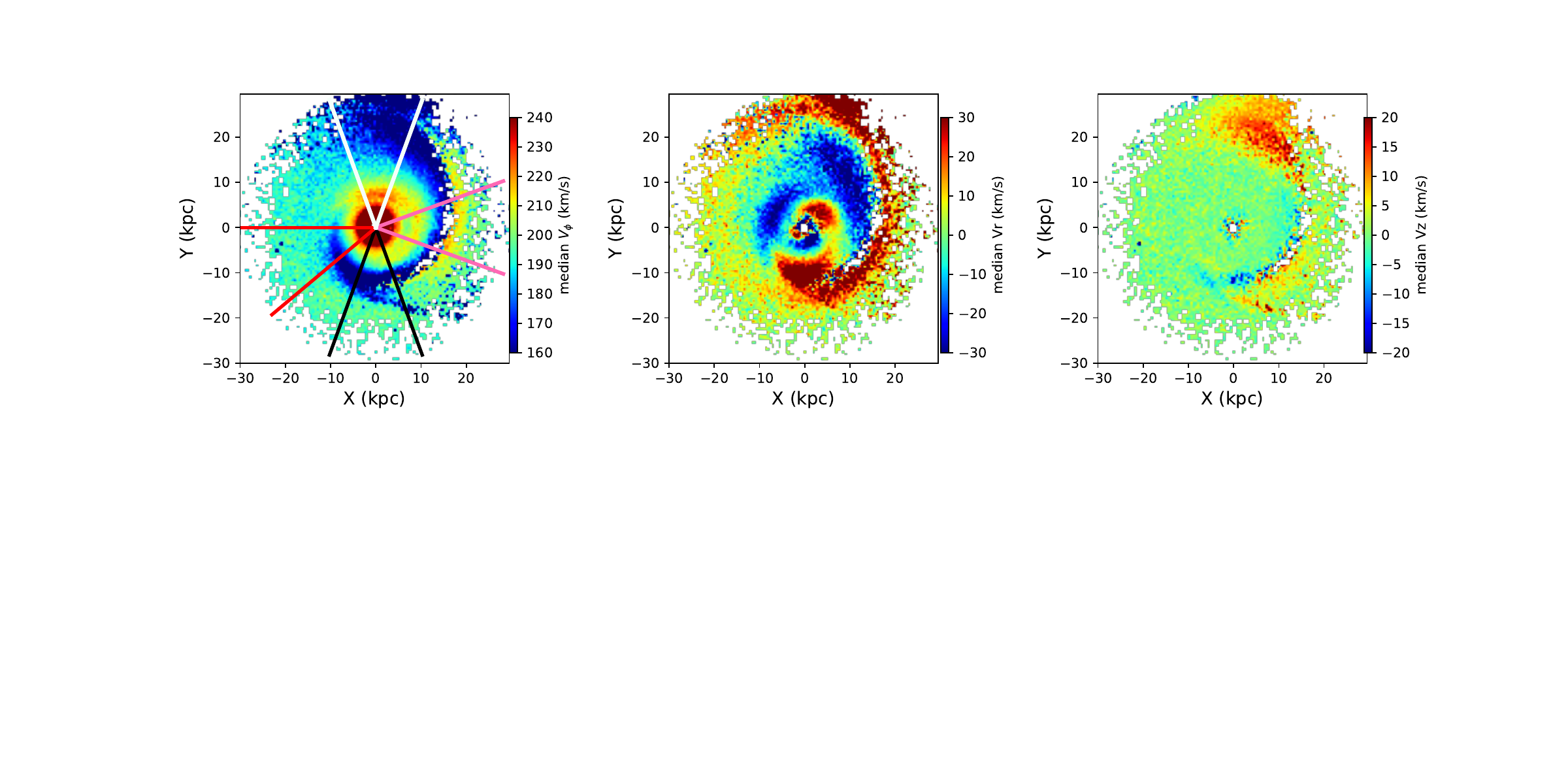}
\caption{The panels show the $V_\phi$ (the left panel), $V_r$ (the middle panel) and $V_Z$ (the right panel) distribution at 200 Myr after the impact. In the left panel, the locations of wedges with $340^\circ-20^\circ<\phi<340^\circ+20^\circ$ (the boundary lines of the wedge colored by red), $270^\circ-20^\circ<\phi<270^\circ+20^\circ$ (black boundary lines, the wedge selected in Figure~\ref{RZhess180_270}), $180^\circ-20^\circ<\phi<180^\circ+20^\circ$ (pink boundary lines), $90^\circ-20^\circ<\phi<90^\circ+20^\circ$ (white boundary lines) are shown.}
\label{map194}
\end{figure}

\begin{figure}
\centering
\includegraphics[width=5.5cm]{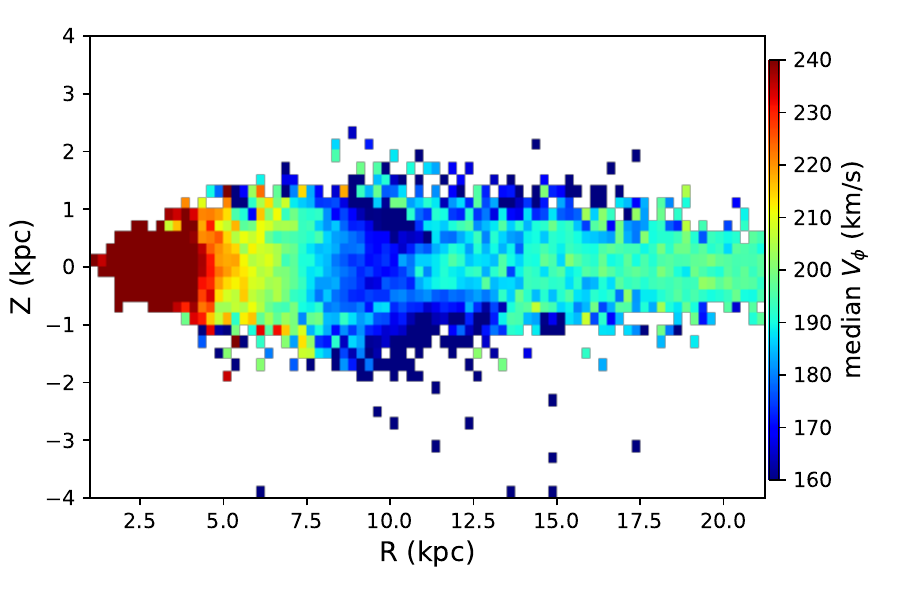}
\includegraphics[width=5.5cm]{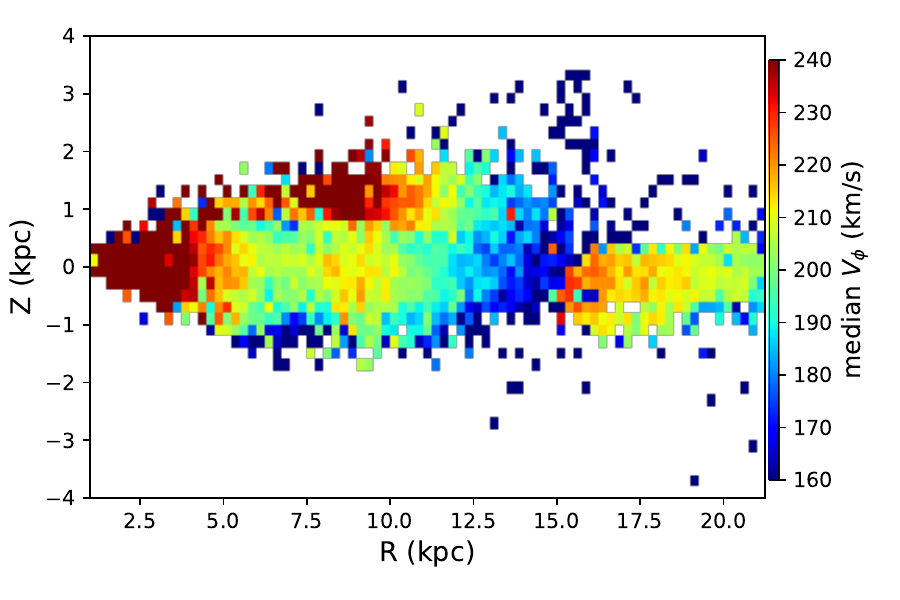}
\includegraphics[width=5.5cm]{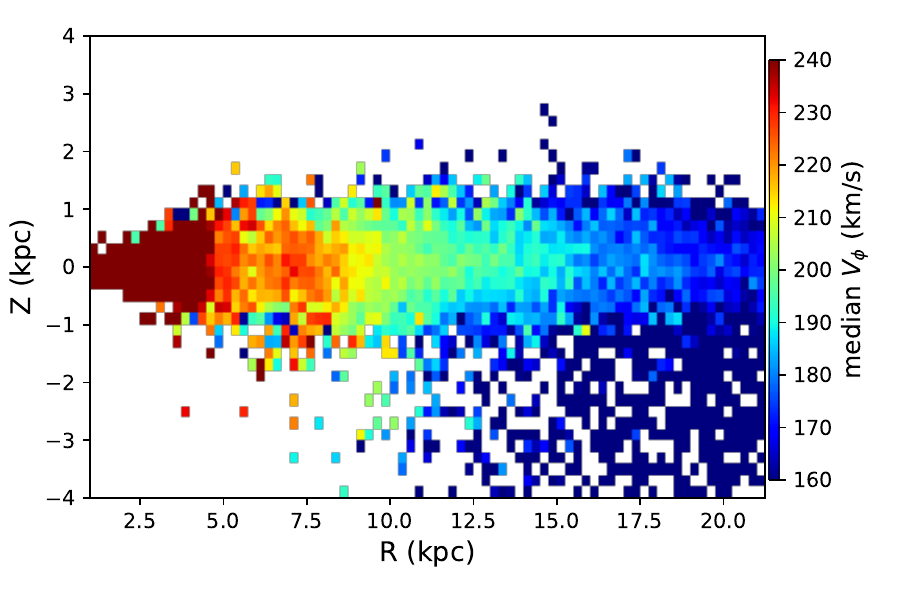}
\caption{The panels shows the median $V_\phi$ distribution in the $R - Z$ map for the bodies with $340^\circ-20^\circ<\phi<340^\circ+20^\circ$ (the left panel), $180^\circ-20^\circ<\phi<180^\circ+20^\circ$ (the middle panel), $90^\circ-20^\circ<\phi<90^\circ+20^\circ$ (the right panel) at 200 Myr after the impact.}
\label{RZhesswedges}
\end{figure}

\begin{figure}
\centering
\includegraphics[width=13cm]{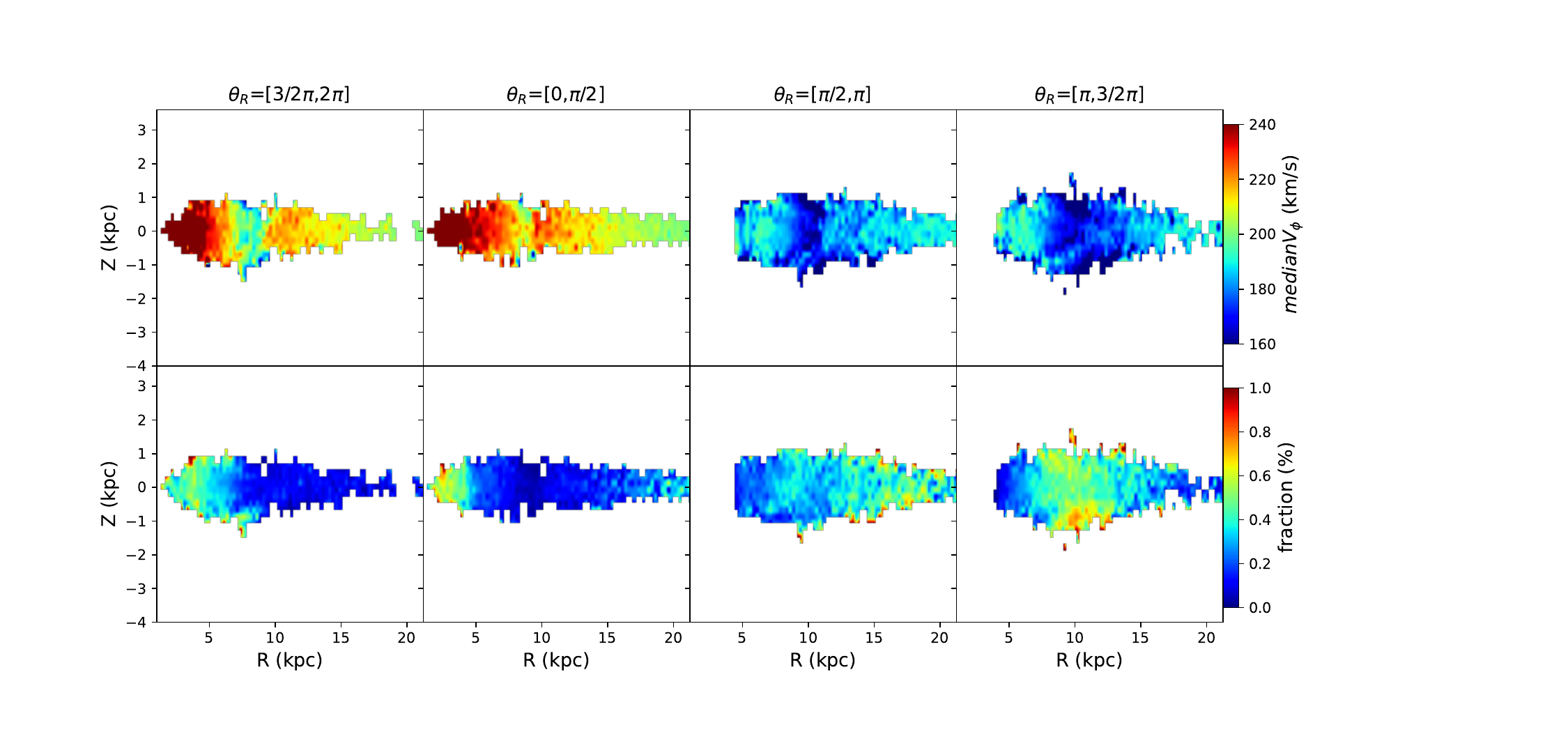}
\includegraphics[width=13cm]{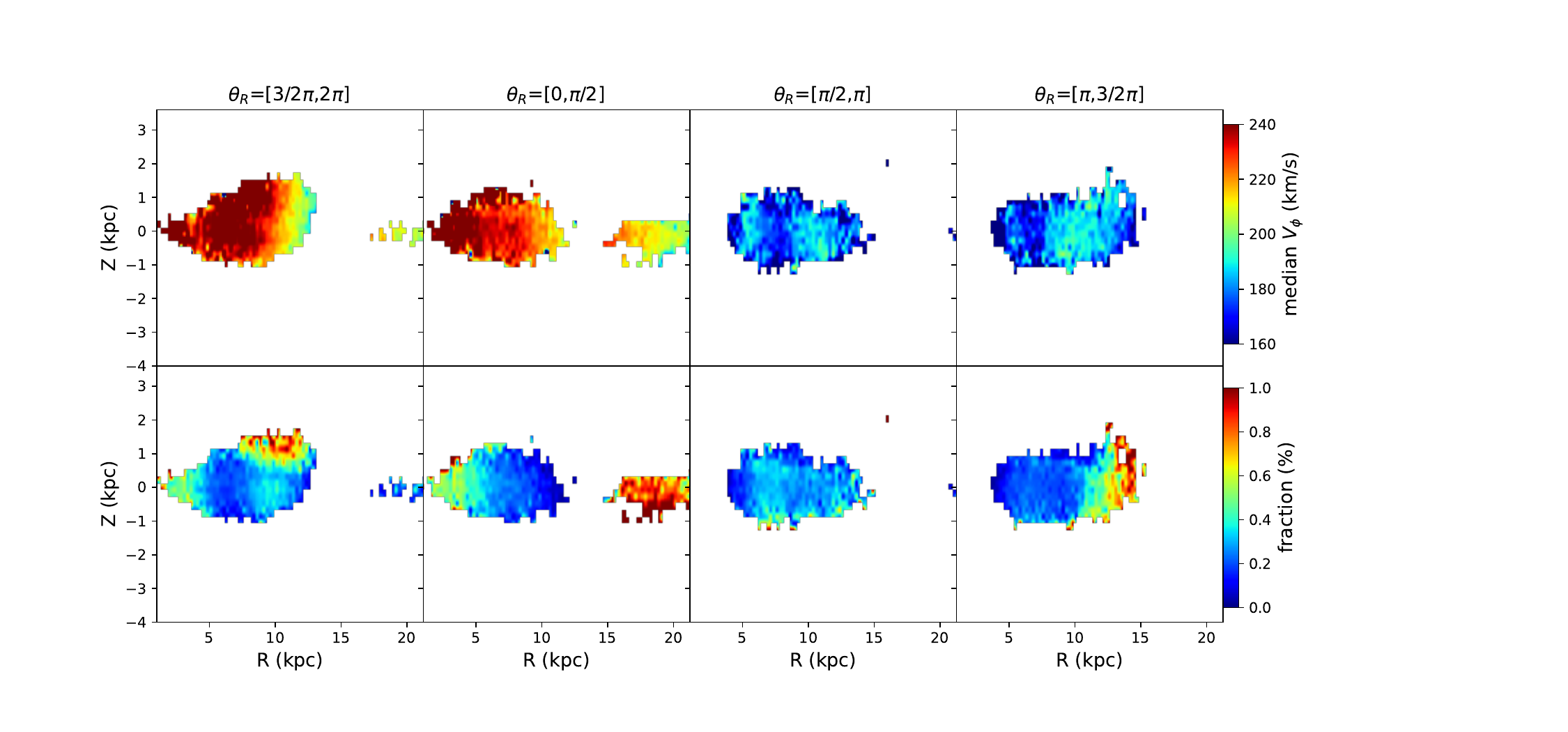}
\includegraphics[width=13cm]{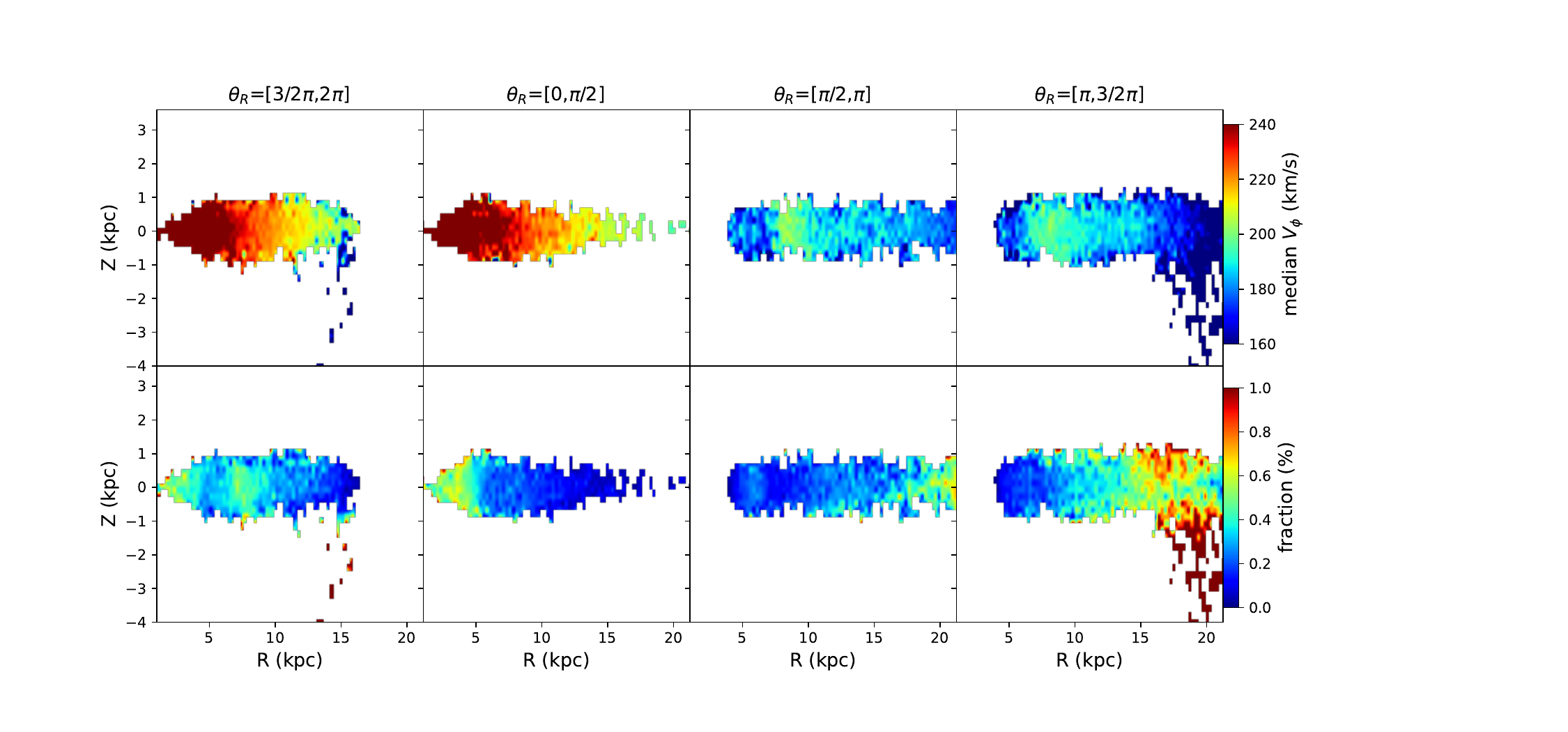}
\caption{The panels show the $V_\phi$ distribution and number fraction of the wedges with $340^\circ-20^\circ<\phi<340^\circ+20^\circ$ (the top panel), $180^\circ-20^\circ<\phi<180^\circ+20^\circ$ (the middle panel), $90^\circ-20^\circ<\phi<90^\circ+20^\circ$ (the bottom panel) respectively, now divided into the 4 $\theta_R$ bins.}
\label{RZhessAngleR}
\end{figure}

 In  Section 6.2, we selecte a wedge of the simulation with $270^\circ-20^\circ<\phi<270^\circ+20^\circ$ and find that the low $V_\phi$ features are analogous to the ``Monoceros area".  Figure~\ref{map194} shows the full map of  the velocity distribution in the $X-Y$ plane at 200 Myr after the impact. We see the low $V_\phi$ stream  at $12<R<15$ kpc, wrapping from $\phi=70^\circ$ to $\phi=360^\circ$. 
 Figure~\ref{RZhesswedges} shows the kinematics in the $R-Z$ plane for the the simulation wedges identified by the red, black and white radial lines in the left panel of Figure~\ref{map194}. Figure~\ref{RZhessAngleR} shows the  same data  split into four ranges of $\theta_R$. Note the change in orbital phase along the stream. 
  Stars in the low $V_\phi$ stream in wedges $180^\circ-20^\circ<\phi<180^\circ+20^\circ$ and $90^\circ-20^\circ<\phi<90^\circ+20^\circ$ show clumps in the $\pi<\theta_R<3\pi/2$ bin (see Figure~\ref{RZhessAngleR}).  Stars in the low $V_\phi$ stream in the wedge with $270^\circ-20^\circ<\phi<270^\circ+20^\circ$ shows clumps in $\pi/2<\theta_R<\pi$ bin (see Figure~\ref{RZhess180_270_AngleR}). The wedge with $340^\circ-20^\circ<\phi<340^\circ+20^\circ$ is near the end of the low $V_\phi$ stream. The stars in the low $V_\phi$ stream in the wedge $340^\circ-20^\circ<\phi<340^\circ+20^\circ$ show more clumps in $\pi/2<\theta_R<\pi$ and $\pi<\theta_R<3\pi/2$ bins (see Figure~\ref{RZhessAngleR}). 
 
The change of orbital phase of the low $V_\phi$ stream in the test particle simulation in Figure~\ref{map194} is due to the different relative position between the disturbed stars and the passage of the dwarf galaxy. 
However we do not know if they are similar to the Monoceros ring due to the limitation of the narrow azimuthal range of the observational data.  Determining whether the Monoceros ring stars populate the whole orbital phase, or whether they are very clumpy in orbital phase space, could constrain parameters  such as the location of the impact and the time since the dwarf galaxy passed the disk. Further observations are required to learn more about the origin of the Monoceros ring. 


\subsection{Effects of different potential models }
One might wonder how dependent our results are on our choice of Galactic potential. The radial phase ($\theta_R$) can shift by a large amount, especially for stars with small radial action, if a different Milky-Way-like potential is used.   In this subsection, we calculate $\theta_R$ with the galpy potential $\tt McMillan17$ (\citealt{2017MNRAS.465...76M},  kindly recommended by Jo, Bovy.) instead of $\tt MWPotential2014$. The density profiles of components of $\tt MWPotential2014$ and $\tt McMillan17$ are listed in Table 3.  In addition to different profile shapes and parameter, the $\tt McMillan17$ model includes gas disk. Both the $\tt MWPotential2014$ and $\tt McMillan17$ potentials fit  the observational data well. The $\tt MWpotential2014$ is constrained by SEGUE data \citep{2013ApJ...779..115B, 2015ApJS..216...29B}. The $\tt McMillan17$ potential is constrained by RAVE data \citep{2017MNRAS.465...76M}.

\begin{figure}
\centering
\includegraphics[width=20cm]{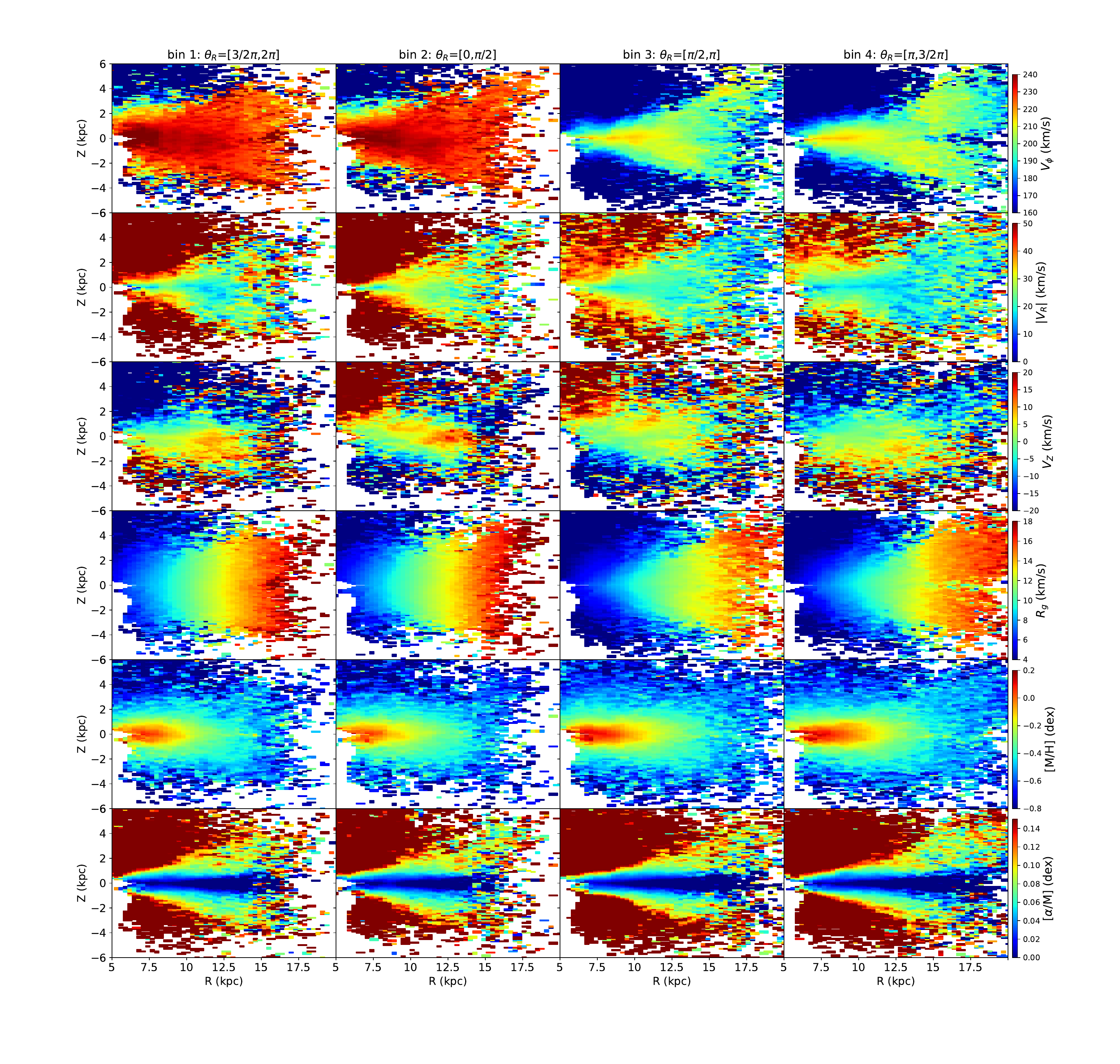}
\caption{$R-Z$ distributions of median $V_\phi$, median $|V_R|$, median $V_Z$, median of guiding center radii (median $R_g$), median [M/H] and median [$\alpha$/M]  in four $\theta_R$ bins of LAMOST K giants. In the second row, the median $V_R$ is positive in the second and third bins, and it is negative in the first and the fourth bins; in each panel of the second row, the absolute value of median $V_R$ distribution is presented to make it easier to compare stars in adjacent bins. $\theta_R$ in this section is calculated with galpy $\tt McMillan17$ potential. }
\label{velocityfeh2}
\end{figure}

\begin{figure}
\centering
\includegraphics[width=8cm]{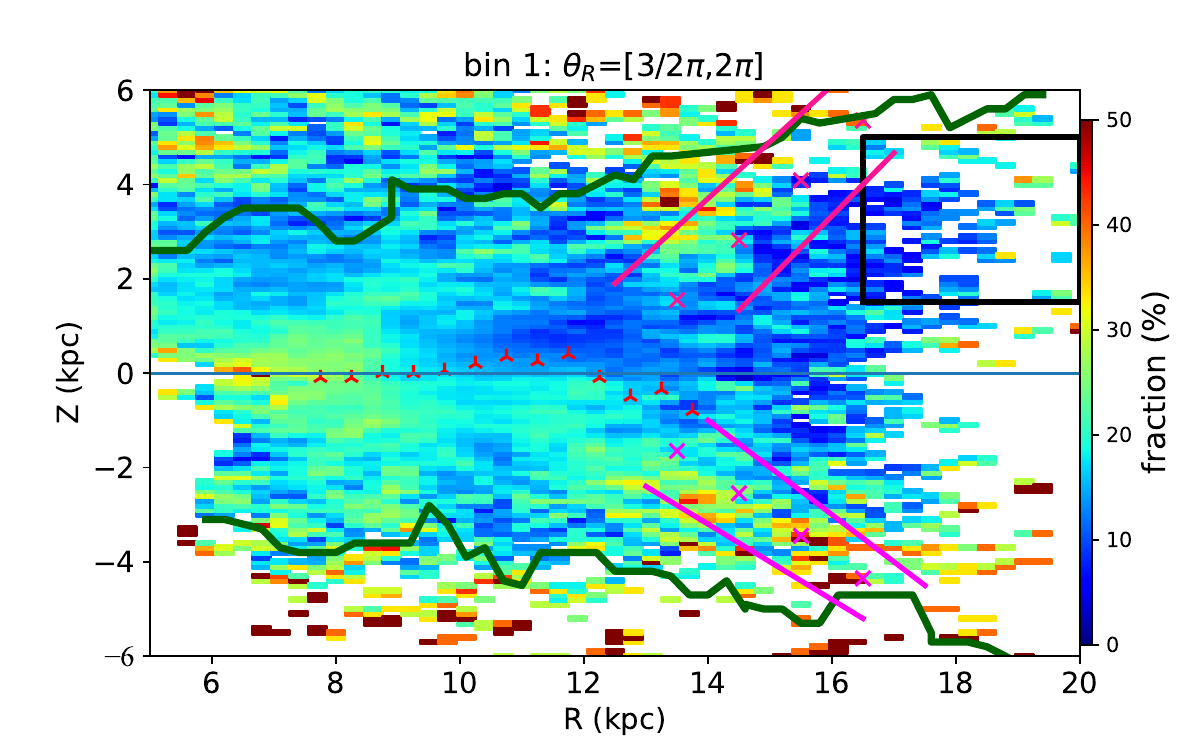}
\includegraphics[width=8cm]{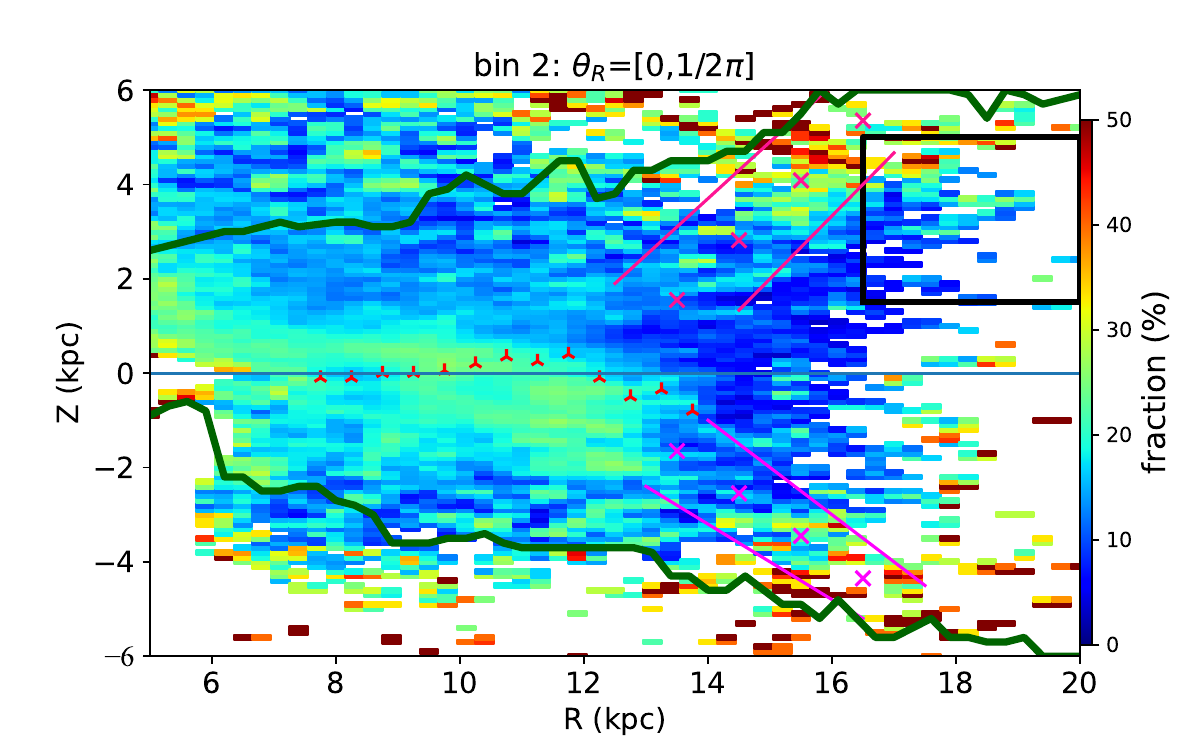}\\

\includegraphics[width=8cm]{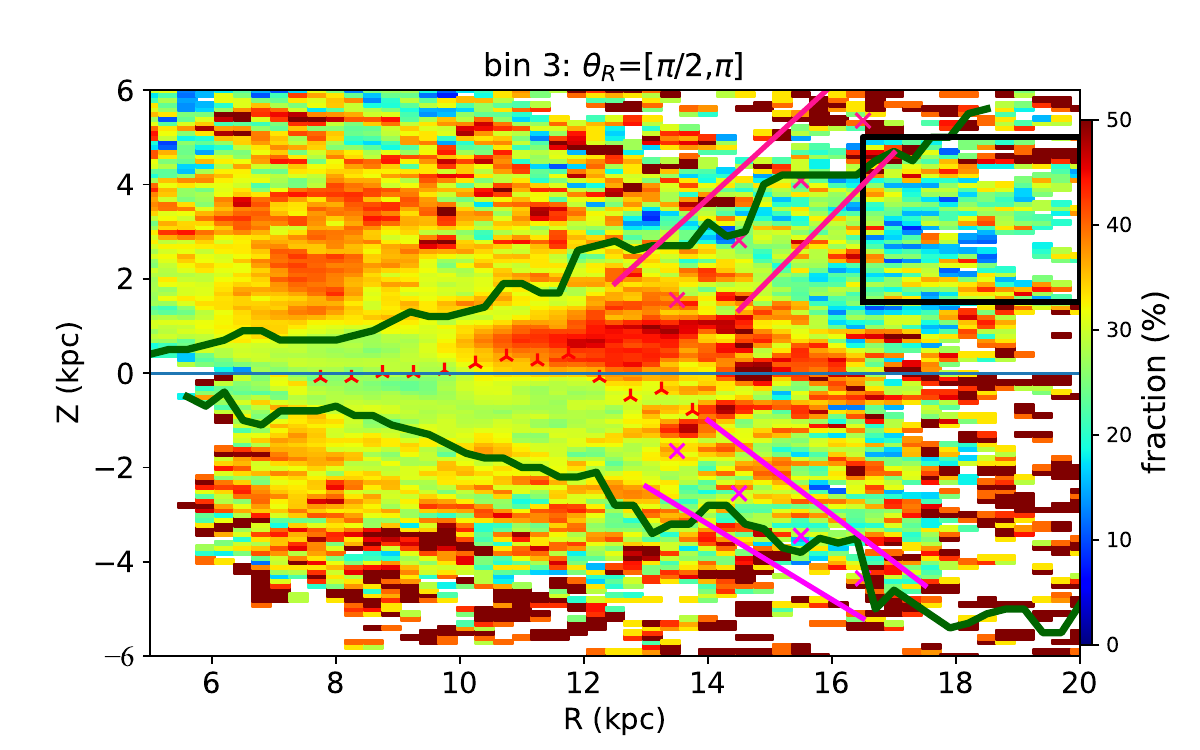}
\includegraphics[width=8cm]{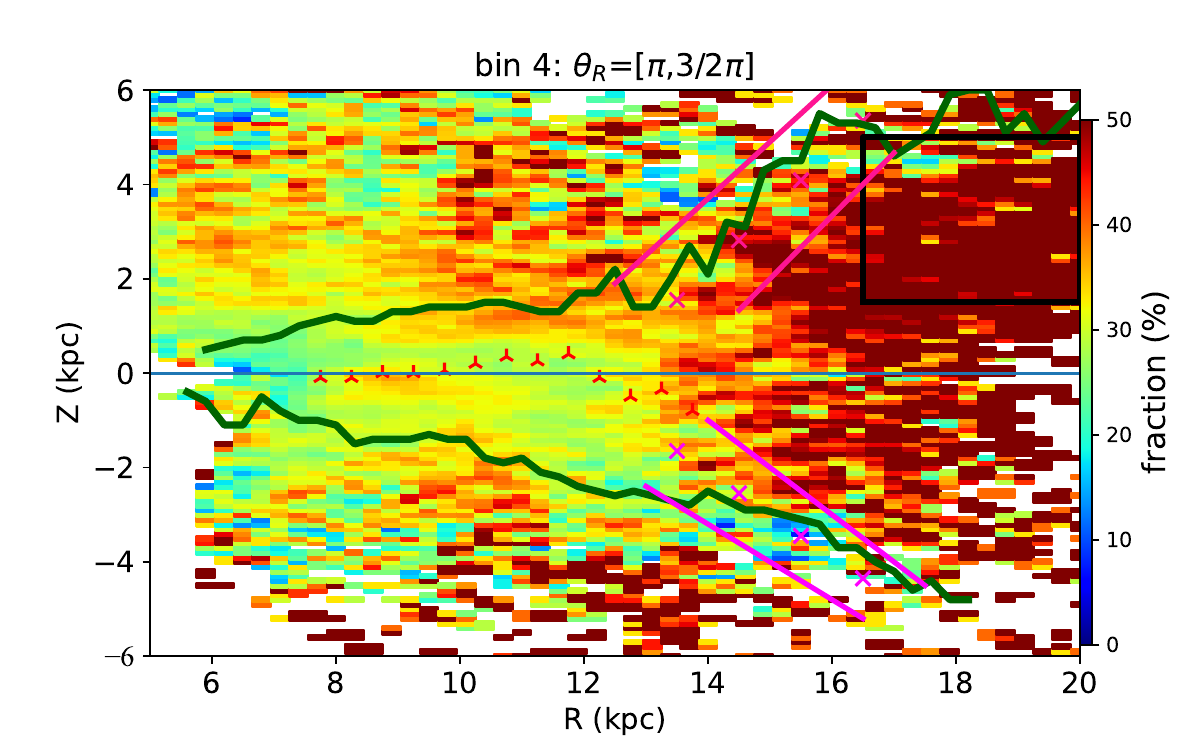}

\caption{This picture shows the relative fraction of stars in each $\theta_R$ bin, where the relative fraction is defined as the number of stars in a particular  $\theta_R$ bin divided by the total number of stars at a particular region of $(R, Z)$.  The boundaries and peaks of the ``north branch" and ``south branch" are labeled  with pink lines and crosses, respectively. The oscillation of the ``mid-plane" is labeled with triangles. The position of  ``Monoceros area" is labeled  by the black rectangle \citep[see Figure 8 of][]{2020ApJ...905....6X}. The wavy dark green curves label the boundary area where median($V_\phi)=190$ km/s.  $\theta_R$ in this section is calculated with galpy $\tt McMillan17$ potential.}
\label{numAngleR2}
\end{figure}

 The LAMOST K giants are reassigned to the four $\theta_R$  bins based on  $\theta_R$ calculated from the $\tt McMillan17$ potential. Figure~\ref{velocityfeh2} and Figure~\ref{numAngleR2} show the kinematic properties of Milky Way stars and fraction of stars in each $\theta_R$ bin as function of spacial position for the $\tt McMillan17$ potential; they are analogous to Figure~\ref{velocityfeh} and Figure~\ref{numAngleR} for the $\tt MWpotential2014$ potential.  

\begin{table*} \label{table:number}
\scriptsize
\centering
\caption{Summary of $\tt MWPotential2014$ and $\tt McMillan17$ potentials}
\begin{threeparttable}
\begin{tabular}{lll}
   \hline
   & density profile of $\tt MWPotential2014$& density profile of $\tt McMillan17$\\
   \hline
bulge&$\mathrm{amp}(r_1/r)^\alpha \exp(-(r/r_c)^2)$ \qquad  (1) & $\rho_0/(1+r'/r_0)^\alpha \exp(-(r'/r_\mathrm{cut})^2)$  \qquad (2)\\
disk & $-\mathrm{amp}/\sqrt{R^2+(a+\sqrt{z^2+b^2})^2}$ \qquad (3)  & $\Sigma_0 /(2z_d) \exp(-|z|/z_d-R/R_d)$ \qquad (4)\\ 
dark matter halo &$\mathrm{amp}/(4\pi a^3 (r/a) (1+r/a)^2)$ \qquad (5)& $\rho_0/(\chi^\gamma(1+\chi)^{3-\gamma})$ \qquad (6)\\
gas disk &&$\Sigma_0/(4z_d)\exp(-R_m/R-R/R_d)\mathrm{sech}^2(z/2z_d)$ \qquad (8)\\
   \hline
\end{tabular}

 \begin{tablenotes}
        \footnotesize
        \item[] (1) amp (amplitude to be applied to the potential) , $\alpha=1.8$, $r_c=1.9$ kpc (cut-off radius), $r_1$ (reference radius for amplitude), $\mathrm{normalize}=0.05$, normalize means that the force is this fraction of the force necessary to make the circular velocity at $R_0$ and $Z_0$ equal to one, $V_C(R_0,Z_0)=1$
                \item[] (2) $r'=\sqrt{R^2+(z/q)^2}$, $\alpha=1.8$, $r_0=0.075$ kpc, $r_\mathrm{cut}=2.1$ kpc, $\rho_0=9.93\times10^{10}$ M$_\odot \mathrm{kpc}^{-3}$
        \item[] (3) MiyamotoNagai Potential is provided instead of density profile, $a=3$ kpc, $b=0.28$ kpc, $\mathrm{normalize}=0.6$
        \item[] (4) $z_d=300$ pc (thin disk scale height), $z_d=900$ pc (thick disk scale height), $R_d=2.6$ kpc (thin disk scale length), $R_d=3.6$ kpc (thick disk scale height), $M_d=2\pi \Sigma_0 R^2_d$ (total disk mass) 
     \item[] (5) density profile of NFW potential,  $a=16$ kpc, $\mathrm{normalize}=0.35$
      \item[] (6) $\chi=r/r_h$, $r_h$ (the scale radius), $\gamma=1$ 
      \item[] (8) $R_d=7$ kpc, $R_m=4$ kpc, $z_d=0.085$ kpc, $\Sigma_0=53.1$ M$_\odot \mathrm{pc}^{-2}$ (HI)\\
                     $R_d=1.5$ kpc, $R_m=12$ kpc, $z_d=0.045$ kpc, $\Sigma_0=2180$ M$_\odot \mathrm{pc}^{-2}$ (H$_2$)
             
      \end{tablenotes}
    \end{threeparttable}
\label{Table:potential}
\end{table*}

 Similar to Figure~\ref{velocityfeh}, the stars in bins 1 and  2  also have higher $V_\phi$ than  stars in bins 3 and 4 at the same $R$, as shown in the first row of Figure~\ref{velocityfeh2}. We also see the discontinuous distribution of $V_\phi$ at $R=15$ kpc, $Z>1$ kpc that separates the ``Monoceros area" and the other flare area in bin 4 in the first row of Figure~\ref{velocityfeh2}. 

 Table 4 shows the median and standard deviation of the fraction of stars within each bin. 
 In Figure~\ref{numAngleR2}, the higher median fractions in bins 3 and 4 compared to bins 1 and  2 might be caused by selection effects.
 
In Figure~\ref{numAngleR2},  we also see asymmetric substructures that are similar with those in Figure~\ref{numAngleR}. First, in Figure~\ref{numAngleR2}, in bins 1 and  2, the relative fraction of stars in the area of the ``mid-plane'' (25\%) is larger than that in the adjacent area (15\%). The high fraction area near the ``mid-plane'' also follows the trend of shifting south after $R>12$ kpc. Second, in the area defined by the ``north branch" and ``south branch", we see high fraction clumps in bins 1 and 2, with fraction higher than 30\%. Though they are not as significant as those of Figure~\ref{numAngleR}. In the region of the ``north branch" and ``south branch", labeled by the pink oblique lines, there are also low fraction pixels. There is no obvious high median fraction in the ``north branch" and ``south branch" in bins 1 and 2 in Table 4. Third, the clump at $10<R<14$ kpc, $Z>0$ kpc in bin 3 that is described in Section 5.2 also shows a high relative fraction of stars (higher than 40\%). Last but not least, the clump in the ``Monoceros area " in bin 4 is very strong. The relative fraction is higher than 50\%. 

To summarize, using either potential we find similar asymmetric clumps of stars. However, the significance of these asymmetric substructures is dependent on the adopted potential, i.e., $\tt MWPotential2014$ and $\tt McMillan17$ as tested in this work.

\begin{deluxetable*}{l*{7}{c}}
\tablewidth{0pt}
\tiny
\tabletypesize{\footnotesize}
\renewcommand{\arraystretch}{1.1}
\tablecaption{Median and standard deviation (std) of the fraction of  stars in each $\theta_R$ bin for the full sample and kinematic substructures of \citet{2020ApJ...905....6X}. $\theta_R$ in this section is calculated with the $\tt galpy$ $\tt McMillan17$ potential.}
\tablehead{
\colhead{} &
\colhead{median (std) of } &
\colhead{median (std) of } & 
\colhead{median (std) of } &
\colhead{median (std) of } &
\colhead{} &
\colhead{} \\
\colhead{} &
\colhead{ fraction (\%) of } &
\colhead{ fraction (\%) of } & 
\colhead{ fraction (\%) of } &
\colhead{ fraction (\%) of } &
\colhead{} &
\colhead{} \\\colhead{} &
\colhead{total sample of this $\theta_R$ bin} &
\colhead{ north branch} & 
\colhead{south branch} &
\colhead{ Monoceros area} &
\colhead{} &
\colhead{} }

\startdata
$\theta_R=[3/2\pi,2\pi]$ (bin 1)  & 18.1 (6.3)  & 14.28 (7.35) & 20.0 (8.4) & 11.1 (5.9) \\
$\theta_R=[0,1/2\pi]$ (bin 2) & 16.8 (6.2)  & 16.7 (12.2)  & 16.6 (7.7) & 12.5 (6.7) \\
$\theta_R=[\pi/2,\pi]$ (bin 3) & 33.3 (8.6) & 30.0 (11.4) & 32.2 (10.7) & 26.9 (12.2) \\
$\theta_R=[\pi,3/2\pi]$ (bin 4) & 36.1 (13.2) & 37.5 (12) & 31.0 (10.8) & 60 (19.9) \\
 \enddata
\label{Table:fraction2}
\end{deluxetable*} 

\subsection{Effect of the LAMOST selection function}
The LAMOST selection function for the main survey is  described in \citet{2017RAA....17...96L}, and the selection function for the LAMOST Spectroscopic Survey of the Galactic Anti-centre (LSS-GAC) is described by \citet{2015MNRAS.448..855Y}. The LAMOST targets are separated into different spectroscopic plates according to their apparent magnitude: the VB plates ($9<r<14$), the B plates ($14<r<16.8$), the M plates ($r<17.8$) and the F plates ($r<18.5$). For the VB and B plates in the main survey region and also LSS-GAC, the target stars are randomly selected from color magnitude diagrams.  For the M and F plates in the main survey region, the selection function depends only on $r$ magnitude.

Although selection effects can dramatically change the apparent distribution of stars in phase space \citep{2023MNRAS.521.5917F}, such effects, as introduced by the LAMOST selection function, do not influence the kinematic groups when separated by $\theta_R$. Thus, our main conclusion on the usefulness of studying kinematic groups through the means of $\theta_R$ remains robust.

\subsubsection{Effect of the LAMOST selection function during disk equilibrium}

We generate a mock star catalog using $\tt {Galaxia}$  \citep{2011ApJ...730....3S} in order investigate the $\theta_R$ distribution of Milky Way like disk stars in equilibrium.  We select a 1000 square degree area in the anticenter direction. Because our observed data are exclusively LAMOST K giants, we only select mock stars  with surface gravity $\log{g}<3.5$.  We obtain 1,509,633 mock data.  The action and angle are calculated with the software package $\tt {galpy}$ \citep{2015ApJS..216...29B}, which uses the $\tt {galpy}$ potential $\tt {MWPotential2014}$ and Stackel approximation \citep{2012MNRAS.426.1324B}, in the same way we calculated these quantities for our observed Milky Way K giant sample.

\begin{figure}
\centering
\includegraphics[width=9cm]{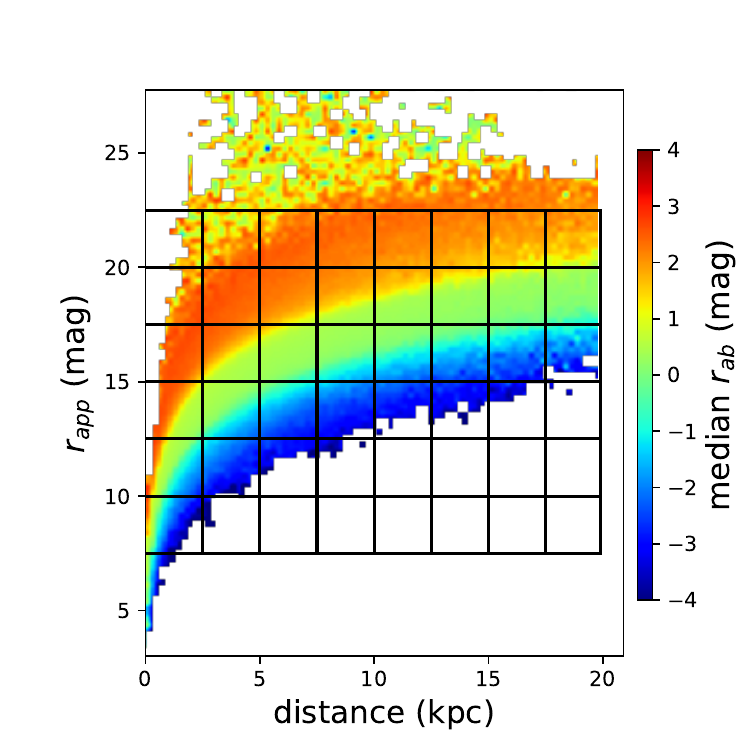}
\caption{The median absolute r magnitude distribution ($r_\mathrm{ab}$) of mock stars with $log(g)<3.5$ as a function of apparent magnitude r ($r_\mathrm{app}$) and distance $D$ from the Sun. }
\label{R_rmag_rab}
\end{figure}

 Figure~\ref{R_rmag_rab}  shows the median absolute $r_\mathrm{ab}$ magnitude distribution of the mock stars with $log(g)<3.5$ as a function of apparent $r_\mathrm{app}$ magnitude and distance $D$ from the Sun. As shown in Figure 4 of  \citet{2023MNRAS.521.5917F},  stars at a particular apparent magnitude and distance range have a different median absolute magnitude. If the observational limit is $r_{app}<17.8^m$, then $median(r_{ab})<2^m$ at $D=5$ kpc, $median(r_{ab})<0^m$ at $D=10$ kpc and $median(r_{ab})<-0.5^m$ at $D=10$ kpc. D is distance from the Sun's location in {\tt Galaxia}.

\begin{figure}
\centering
\includegraphics[width=9cm]{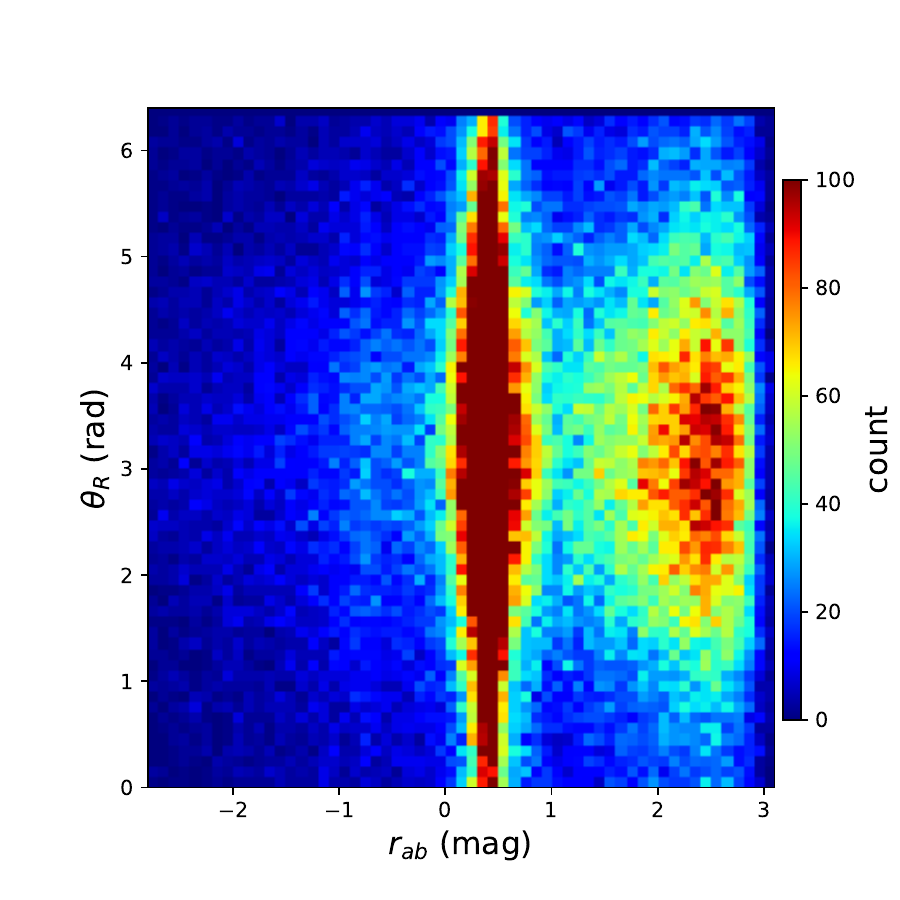}
\caption{Counts of the mock stars within $5<D<6$ kpc of Figure~\ref{R_rmag_rab} in $\theta_R$ vs.  $r_{ab}$ space. }
\label{rmag_theta_R_count}
\end{figure}

\begin{figure}
\centering
\includegraphics[width=9cm]{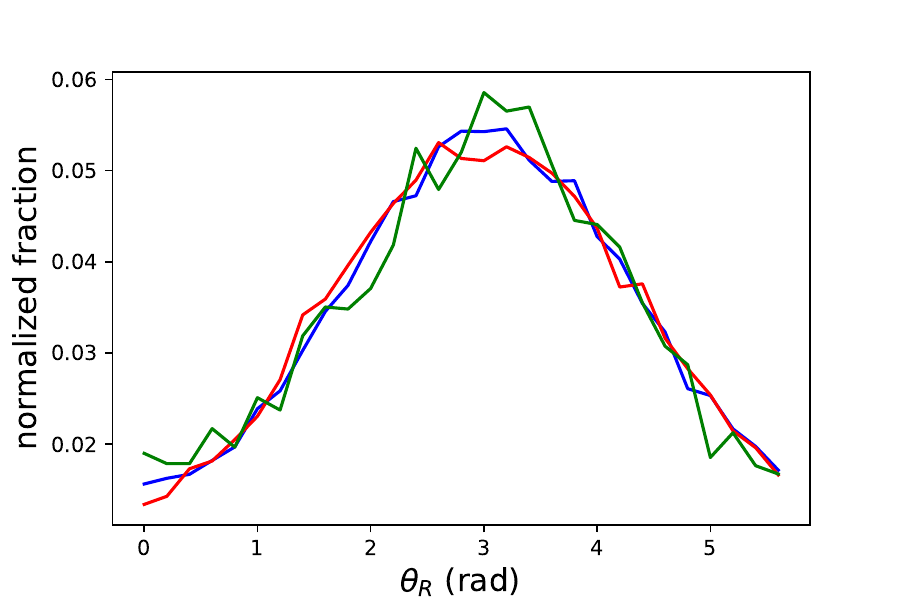}
\caption{The $\theta_R$ distribution of stars with different apparent magnitudes in Figure~\ref{rmag_theta_R_count}. The blue, red, green curves represent the $\theta_R$ distribution of stars in $0^m<r_{ab}<1^m$, $2^m<r_{ab}<3^m$ and $-2^m<r_{ab}<-1^m$, respectively.}
\label{histfraction}
\end{figure}

We select a narrow distance range ($5<D<6$ kpc) of the sample in Figure~\ref{R_rmag_rab} in order to study the $\theta_R$ distribution as a function of absolute $r$ magnitude. 
In Figure~\ref{rmag_theta_R_count}, the star counts within $5<D<6$ kpc of Figure~\ref{R_rmag_rab} are shown as a function of $\theta_R$ vs. $r_{ab}$.  There are vertical overdensities at in $r_{ab}=0.5^m$, $2.5^m$ not a function of $\theta_R$.   Figure~\ref{histfraction}  shows histogram of $\theta_R$ in the range of  $0^m<r_{ab}<1^m$, $2^m<r_{ab}<3^m$ and $-2^m<r_{ab}<-1^m$ for stars in Figure~\ref{rmag_theta_R_count}. In Figure~\ref{numAngleR}, the fraction of bins 3 and 4 are higher than that of bins 1 and 2, due to the intrinsic selection effect that the stars near apocenter have higher sampling rate than stars near pericenter, just like we explained in Section \ref{sec:asymmetric}. We see the same situation in Figure~\ref{histfraction}, such that the stars near apocenter ($\theta_R\sim \pi$) have a higher fraction than the stars near pericenter ($\theta_R\sim0,2\pi$).  From Figure~\ref{histfraction}, $\theta_R$ has the same distribution in all three  $r_{ab}$ bins.

To verify this phenomenon is the same for all distance ranges, we select subsamples with $2<D<3$, $7<D<8$, $10<D<11$ kpc from the sample in Figure~\ref{R_rmag_rab}. We find that $\theta_R$ still has the same distribution in different absolute magnitude bins, just as we find in Figure~\ref{histfraction} for the stars with $5<D<6$ kpc. 

The mock data from $\tt Galaxia$  support the conclusion that the $\theta_R$ distribution is not influenced by  the selection effects  in the LAMOST data set for the case of disk equilibrium. If the $\theta_R$ has the same distribution at each Galactic radius in different $r_{ab}$ bins, then the $\theta_R$ distribution of stars extracted from different $r_{ab}$ is same as the $\theta_R$ distribution of the total sample. 

\subsubsection{Effect of the LAMOST selection function  during disk nonequilibrium}

 In a nonequilibrium state, such as the Sgr dSph impact, the selection function which is applied to the apparent magnitude does not obviously influence the kinematics grouped by $\theta_R$. This is because the stars of different mass (or absolute magnitude) at  the same location have the same acceleration toward a Sgr dSph like satellite. The amount of change of $\theta_R$ is related to the initial velocity and location of the  disk stars and the mass and velocity of the Sgr dSph, but not related to the mass (or absolute magnitude) of the stars. 

\begin{figure}
\centering
\includegraphics[width=9cm]{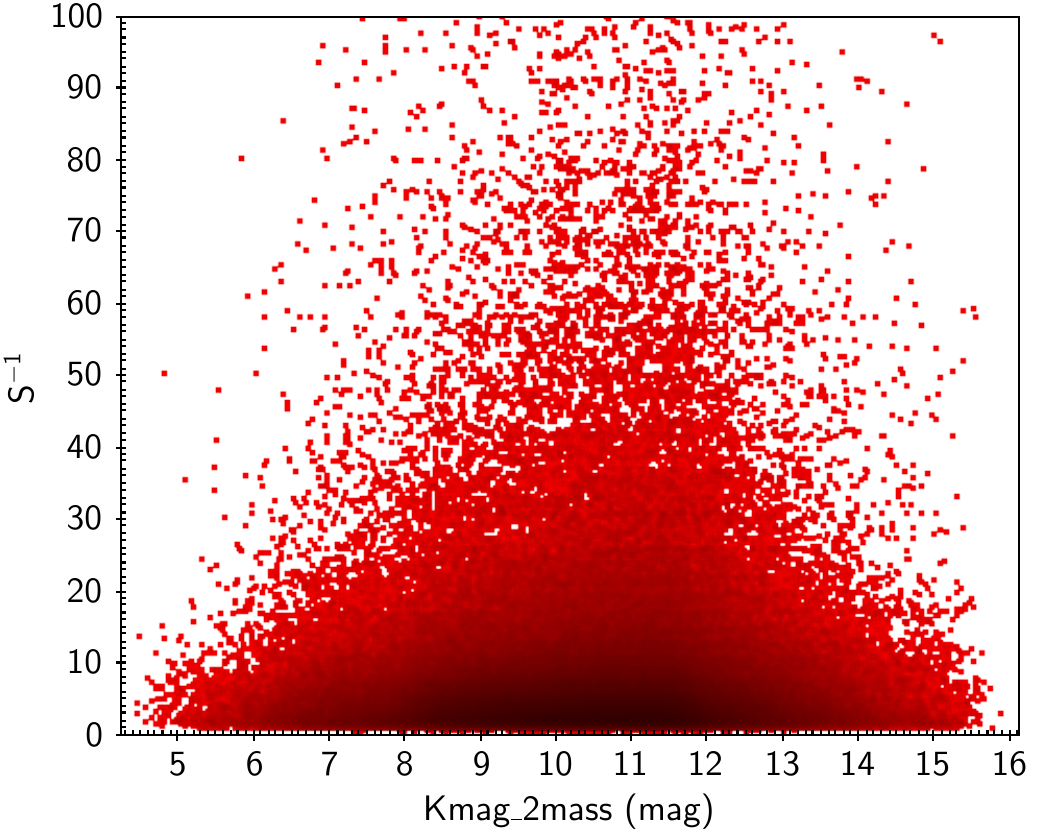}
\caption{The plot shows the inverse of the selection function ($S^{-1}$) applied to each  LAMOST K giant vs. their 2MASS $K$ magnitude.}
\label{sf}
\end{figure}

Because the disturbance of stars is unknown and model dependent, we cannot correct for the selection effect by estimating the distribution of $\theta_R$ for the unobserved stars by using our observed LAMOST K giant sample. Despite this, we can still  partly show the effect of the selection function.  From Figure 1 of \citet{2017RAA....17...96L}, the selection function is inclined to bright stars, indicating bright stars have a higher  chance of being selected. We  resample the data based on the selection function ($S$) defined by equation 16 of  \citet{2017RAA....17...96L} to erase the inclination to the bright stars. In \citet{2017RAA....17...96L}, $S$ is calculated by the number of spectroscopic and photometric stars at a given color, apparent magnitude, l and b. In practice, the inverse of the selection function $S^{-1}$ is used instead of S. Figure~\ref{sf} shows the inverse of the selection function $S^{-1}$ applied to each star in our sample.  A total of 89\% of our sample stars have $S^{-1}\le10$. We choose the stars within $S^{-1}\le10$ to make resampling and the very rare stars with $S^{-1}>10$ are ignored. We select this criterion because if we choose an $S^{-1}$ with a lower value, the data set which is used to be resampled cannot include most of our sample. If we choose a $S^{-1}$ with a higher value, the resampled data will be too few to make a thorough statistical analysis.

\begin{figure}
\centering
\includegraphics[width=8cm]{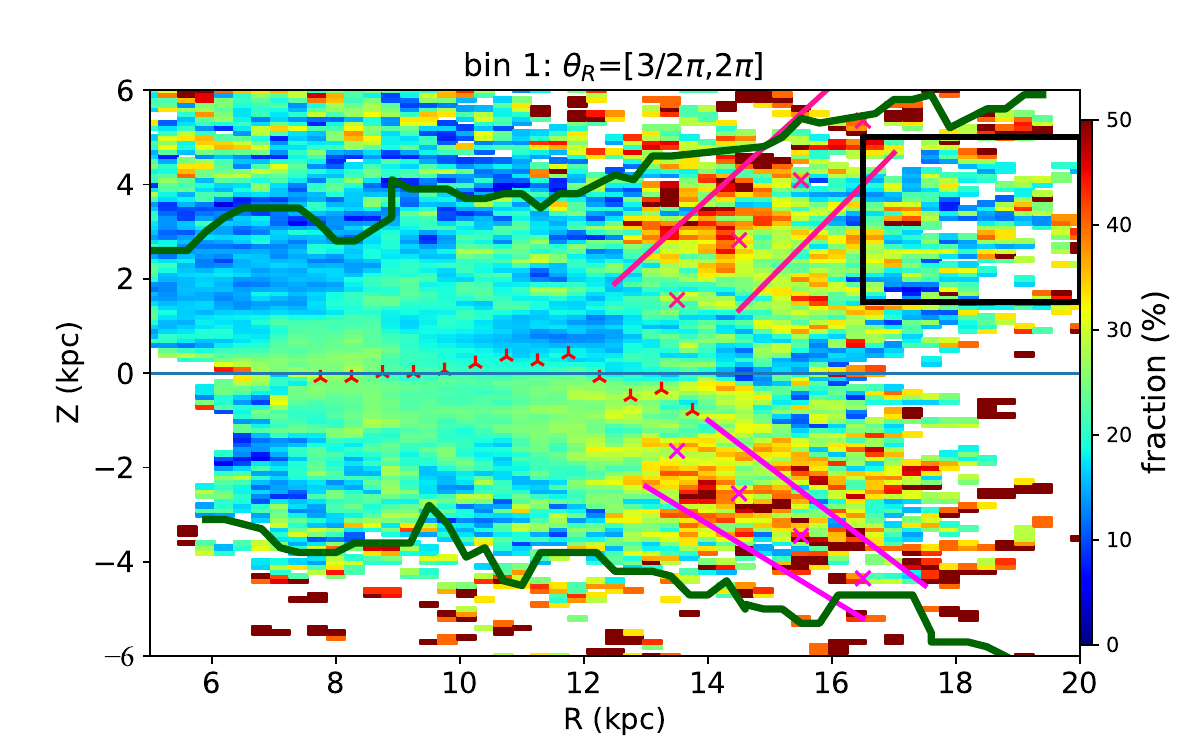}
\includegraphics[width=8cm]{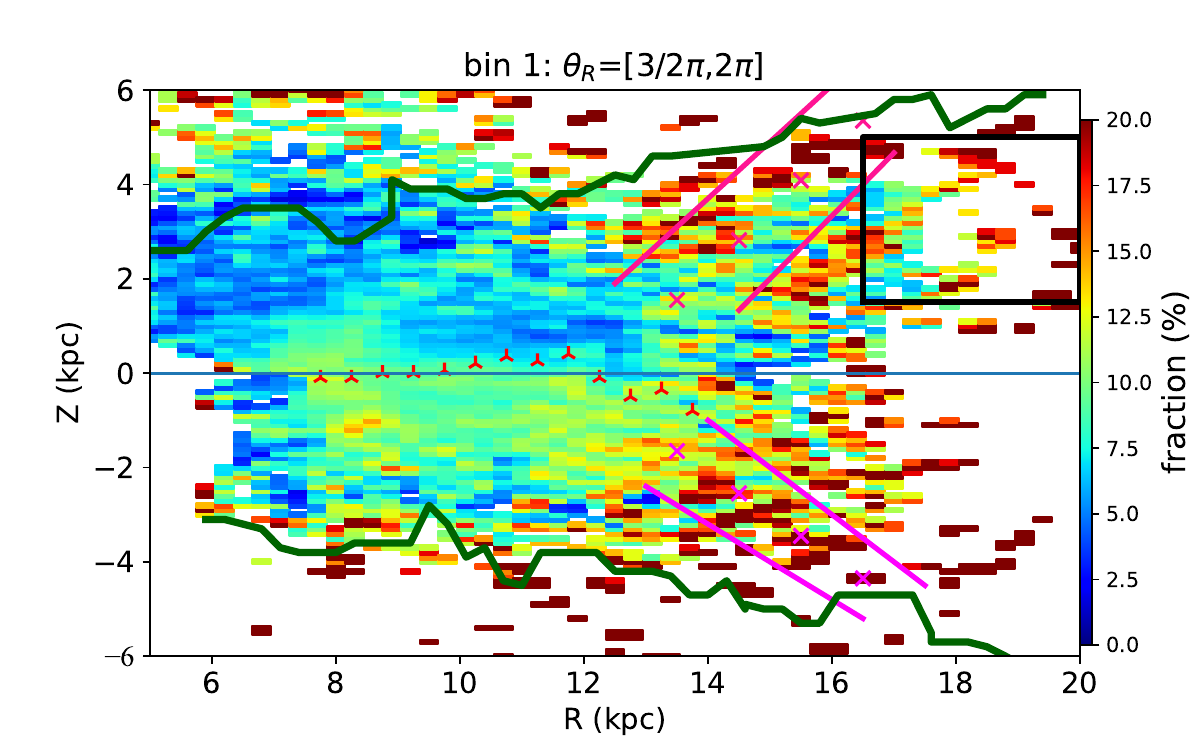}\\
\includegraphics[width=8cm]{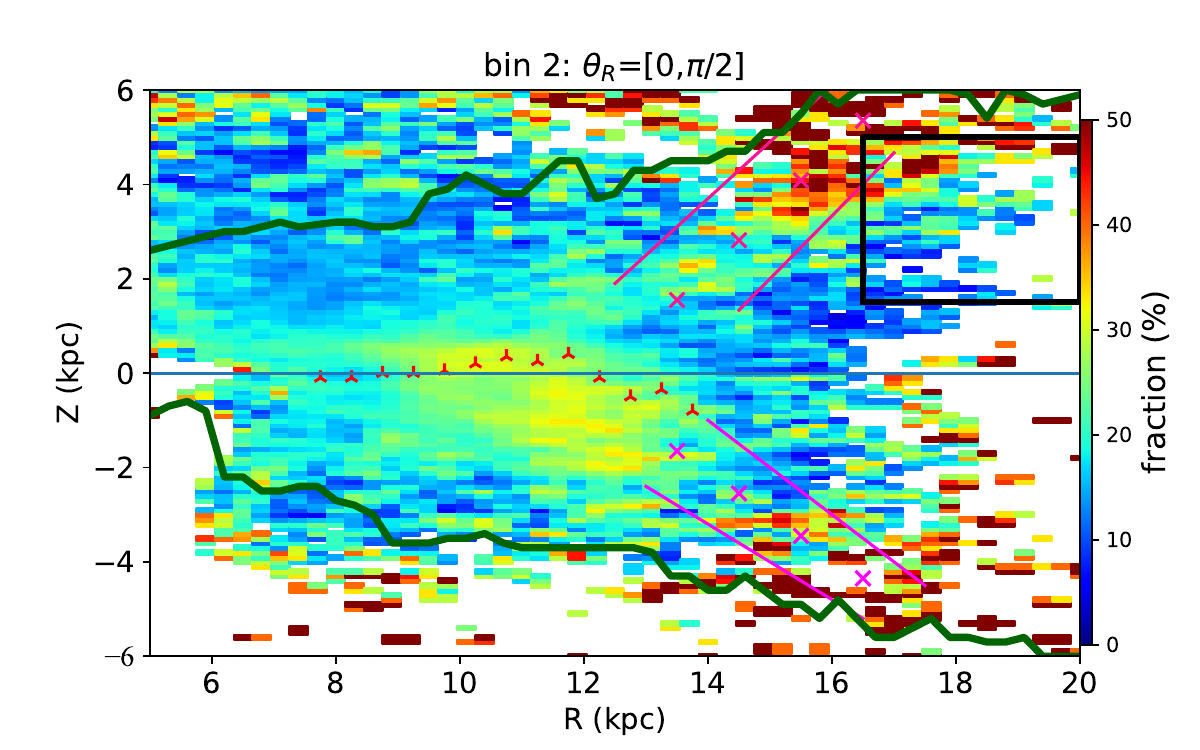}
\includegraphics[width=8cm]{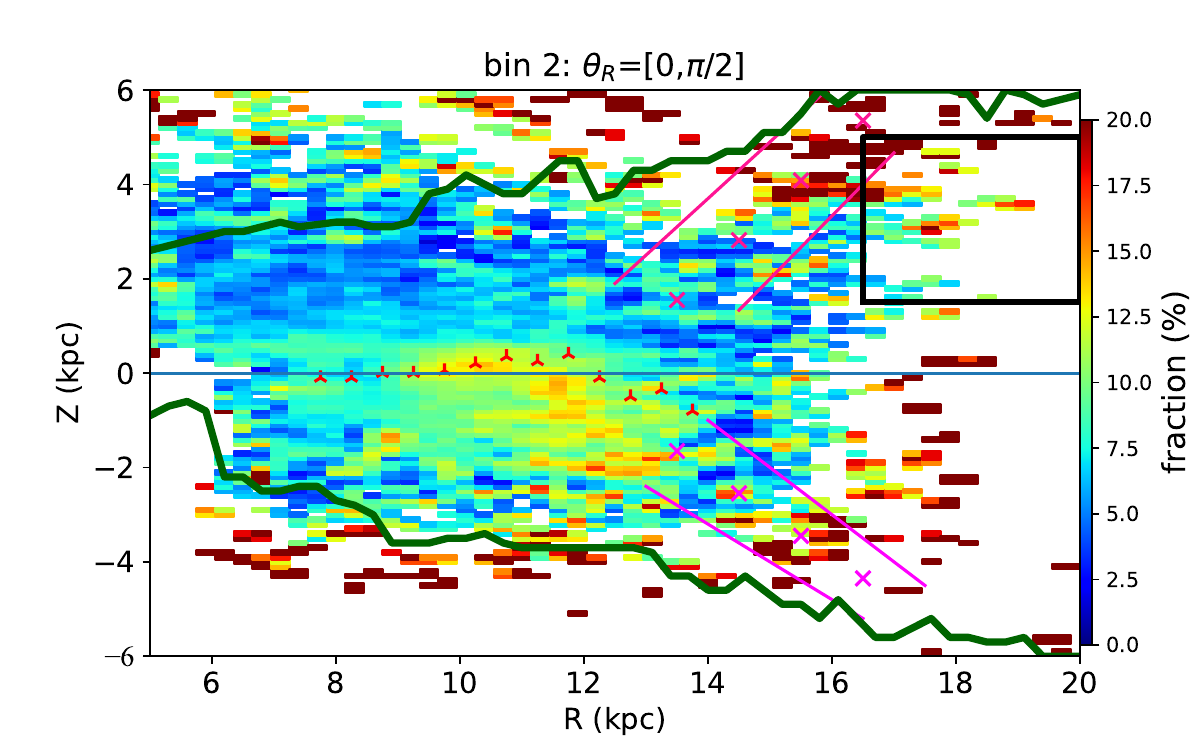}\\
\includegraphics[width=8cm]{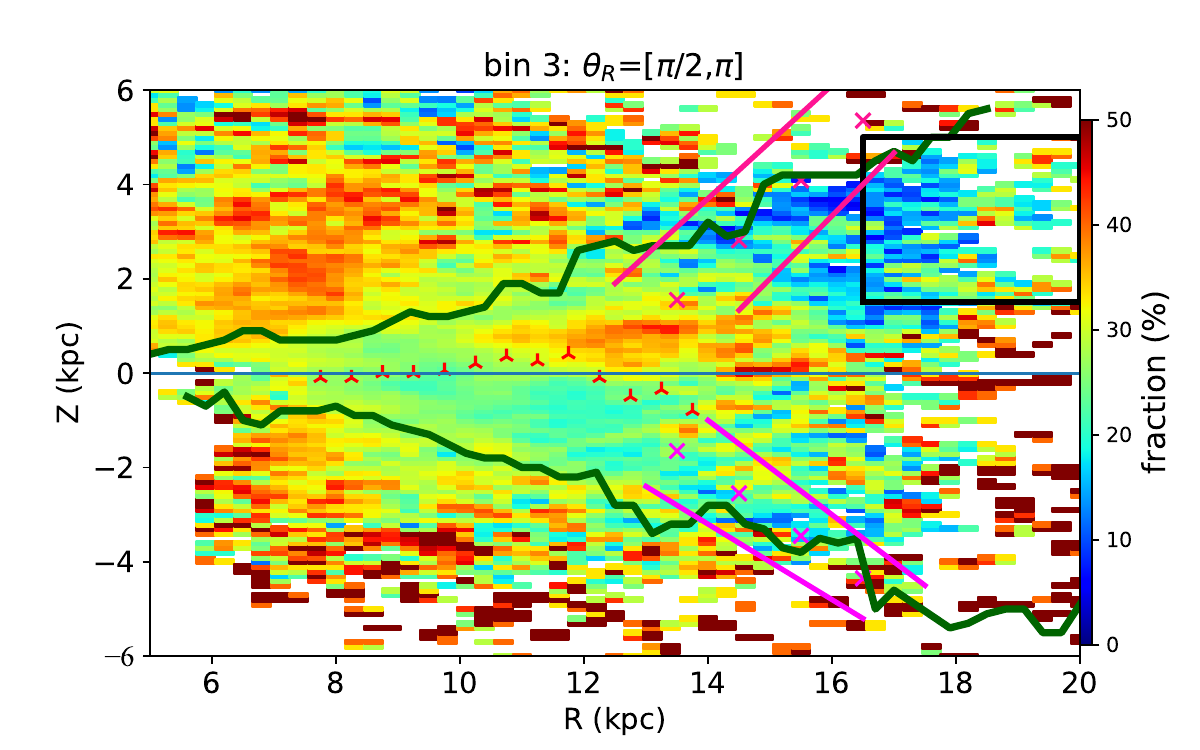}
\includegraphics[width=8cm]{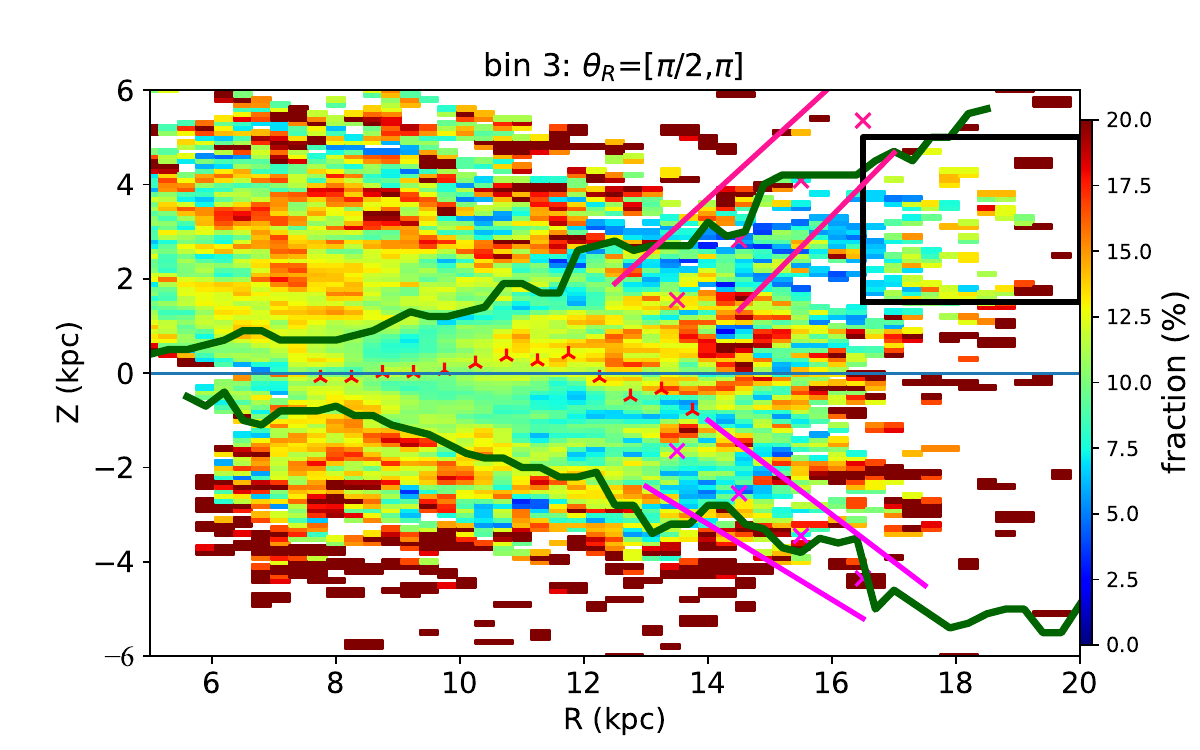}\\
\includegraphics[width=8cm]{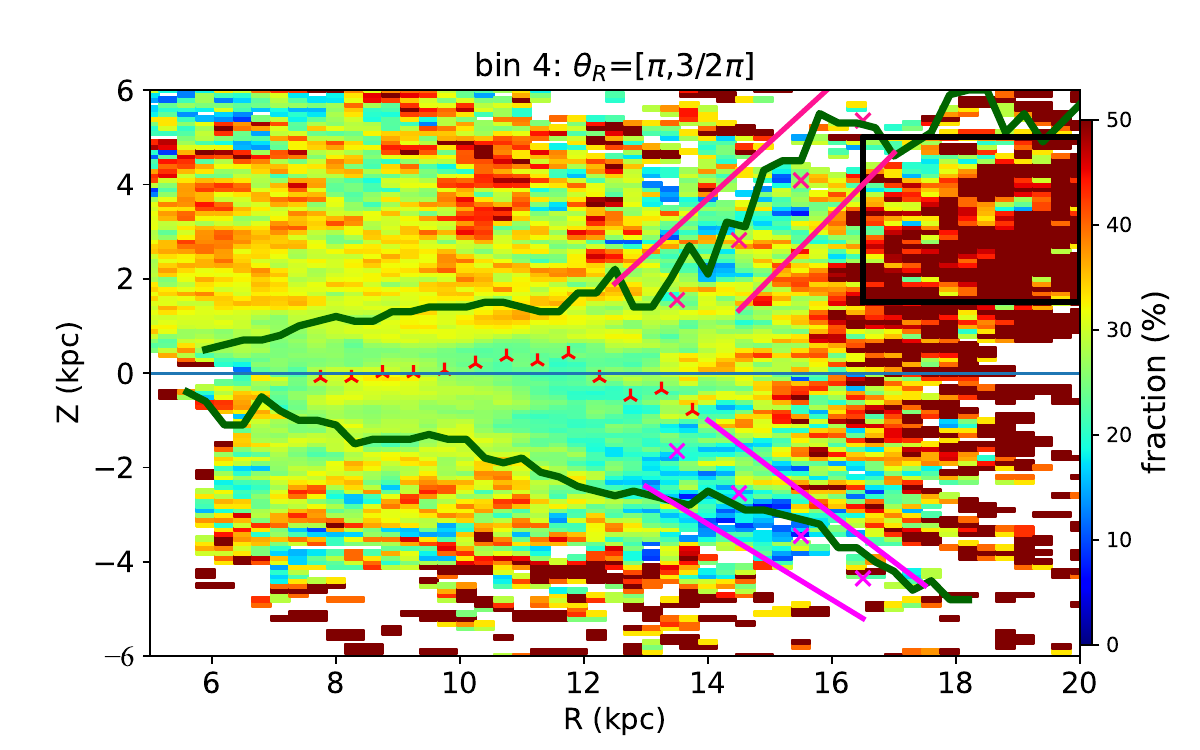}
\includegraphics[width=8cm]{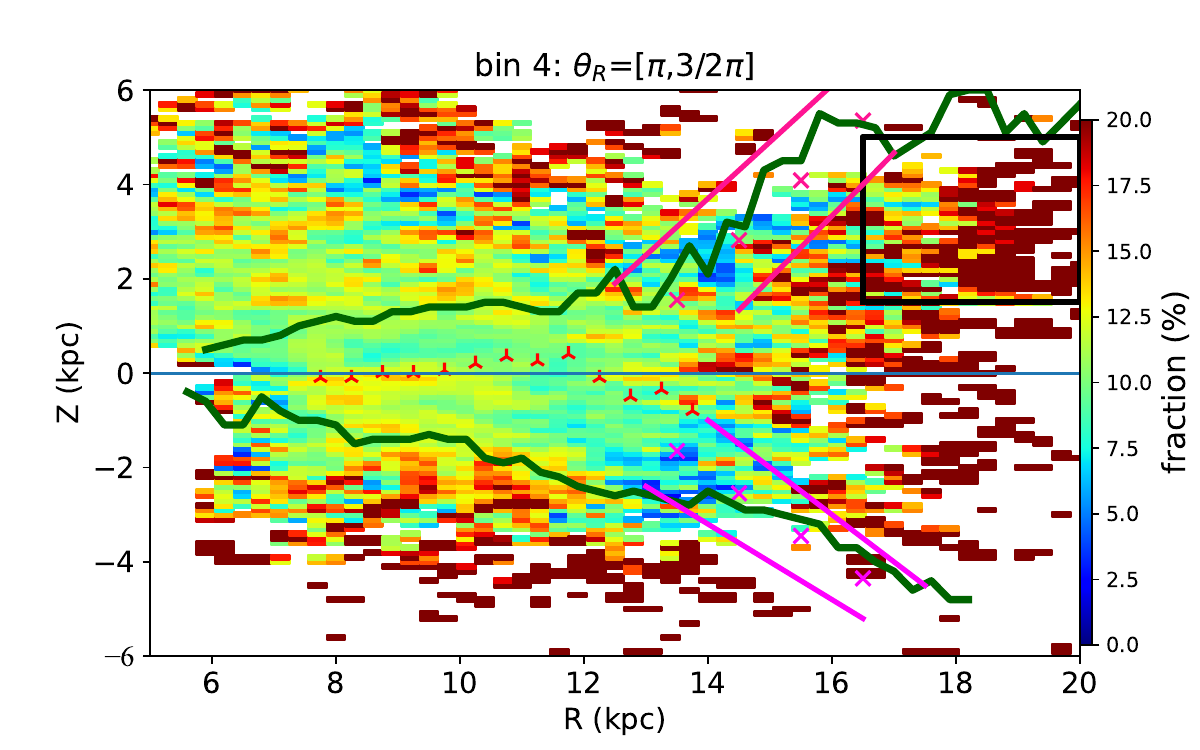}\\
\caption{Left panels:  relative fraction of stars in each $\theta_R$ bin whose $S^{-1}\le10$.  Right panels:  relative fraction of stars in each $\theta_R$ bin resampled with criterion $S^{-1}=10$.}
\label{RZ_num_sf}
\end{figure}

 The left panels of Figure~\ref{RZ_num_sf} show the relative fraction of each $\theta_R$ bin of stars with $S^{-1}\le10$.  The right panels of  Figure~\ref{RZ_num_sf} show one of the results for the relative fraction for each $\theta_R$ bin of stars resampled with the criterion of $S^{-1}=10$.  As a result, the stars in the resampled data now have the same chance to be selected. Although the noise increases in the right panels of Figure~\ref{RZ_num_sf} due to the reduced number of stars, we still are able to detect the high relative fraction of the relevant kinematic features, namely the `south branch' in bin 1, the `mid-plane' shifting to the south in bin 2, the $V_R$ related feature described in Section 5.2 in bin 3, and the `Monoceros area'  in bin 4.

These results are in harmony with the results of  the test particle simulation.  In the test particle simulation, the kinematic features are composed of the most disturbed stars of the Sgr dSph impact. Because  the Sgr dSph impact  very strongly influences  stars near the projection point of the Sgr dSph, these stars are highly influenced in the similar way. Following this assumption, we expect that in a particular portion of the sky, we might observe a portion of the 
nonequilibrium substructure that is preferentially in one or more $\theta_R$ phases.
We conclude that the appearance of the asymmetric substructures of stars grouped by $\theta_R$ is not seriously influenced by the LAMOST selection function.

 The $\tt {Galaxia}$ mocks do not include nonequilibrium substructures.  $N$-body simulations are needed to completely explore the effect of the selection function on the $\theta_R$ distribution in the nonequilibrium case.

\section{Summary}\label{summary}

We separate our sample of LAMOST K giants into 4 bins according to conjugate angle space of radial action that represents the orbital phase. The 4 bins roughly correspond to orbital phases approaching (bin 1: $3\pi/2<\theta_R<2\pi$), or receding from (bin 2: $0<\theta_R<\pi/2$) pericenter and approaching (bin 3: $\pi/2<\theta_R<\pi$), or receding from (bin 4: $\pi<\theta_R<3\pi/2$) apocenter. The kinematic, chemical and number fraction distribution of the data of the 4 bins on $R-Z$ plane are analysed. 

(i) From the analysis of LAMOST K giants in conjugate angle space of radial action, the kinematic substructures in $R-Z$ space and $Z-V_Z$ phase space that are considered to be possibly associated with the last impact of Sgr dSph  \citep{2020ApJ...905....6X}  also show clumps in $\theta_R$ space. 

 From Figure~\ref{numAngleR}, the location of the oscillation of ``mid-plane" that has been associated with the centroid of phase spirals \citep{2020ApJ...905....6X}  shows high number fraction in the range of $10<R<14$ kpc, $-2.5<Z<0.5$ kpc in bin 2 ($0<\theta_R<1/2\pi$), where the stars are moving from pericenter towards the guiding center radius. 

The ``north structure" and ``south structure" that are projections of the $V_\phi$ phase spirals in the $R-Z$ plane show high number fractions in bins 1 and 2 ($3/2\pi<\theta_R<2\pi$, $0<\theta_R<1/2\pi$),  where the stars are approaching and receding from pericenter, respectively.   

(ii) The substructure ``Monoceros area" shows significantly high number fraction in bin 4 where the stars are receding from the apocenter. We find that the stars that are clumped in $\theta_R$ space are analogous to the known kinematic substructures that are considered to be possibly associated with the Sgr dSph impact \citep{2020ApJ...905....6X}. The high overdensity  associated with the ``Monoceros area" in $\theta_R$ space indicates a possibility that this substructure may also be associated with a gravitational interaction such as the Sgr dSph impact. Further investigation is needed to understand any other possible causes of the $\theta_R$ concentration found for the stars of ``Monoceros area". The discontinuity from $(R, Z)=(14, 1.5)$ to (16, 0.5) kpc of the $V_\phi$ distribution  in the $R-Z$ map (bin 4 of the first row of Figure~\ref{velocityfeh}) shows extra evidence that the flared disk is not smooth.  

(iii) The kinematic substructures are analogous with the results from our test particle simulation. From the test particle simulation, the last impact of Sgr dSph can reproduce the characteristicly narrow range in $\theta_R$ space of substructures like the ``north branch", ``south branch" and ``Monoceros area". In the test particle simulation results, the $\theta_R$ space clumping of the ``Monoceros area" is more highly concentrated than that of the ``north branch" and ``south branch". This may be related to the difference in the frequencies  $\Omega_\phi$ and $\Omega_z$ at different $R$. 

By using different potential models to divide our observational data into bins of $\theta_R$, we find that the detection of asymmetric substructures is robust, although their exact relative fractional contribution varies depending on the adopted potential.

Shortcomings of the test particle simulation include the lack of self-gravity, influence of  the Large Magellanic Cloud, gas, and bar buckling. Investigations using more realistic cosmological simulation  are expected in future work.

\acknowledgments
This work is supported by the National Key R\&D Program of 
China No. 2019YFA0405500, the National Natural Science Foundation of China (NSFC) with grant No. 12173046, and science research grants from the China Manned Space Project with NO. CMS-CSST-2021-B03.
This work is also supported by the National Natural Science Foundation of China under grant No. 12122301, by a Shanghai Natural Science Research Grant (21ZR1430600), by the Cultivation Project for LAMOST Scientific Payoff and Research Achievement of CAMS-CAS, by the ``111'' project of the Ministry of Education under grant No. B20019, and by the China Manned Space Project with No. CMS-CSST-2021-B03.
HJN is supported by US
National Science Foundation grant AST19-08653.
Guoshoujing Telescope (the Large Sky Area Multi-Object Fiber Spectroscopic Telescope 
LAMOST) is a National Major Scientific Project built by the Chinese Academy of Sciences.
 Funding for the project has been provided by the National Development and Reform 
 Commission. LAMOST is operated and managed by the National Astronomical Observatories, 
 Chinese Academy of Sciences.
 This work has made use of data from the European Space Agency (ESA)
mission {\it Gaia} (\url{https://www.cosmos.esa.int/gaia}), processed by
the {\it Gaia} Data Processing and Analysis Consortium (DPAC,
\url{https://www.cosmos.esa.int/web/gaia/dpac/consortium}). Funding
for the DPAC has been provided by national institutions, in particular
the institutions participating in the {\it Gaia} Multilateral Agreement. Thanks for the very helpful discussion with Frankel.

\setcounter{figure}{0}    
\renewcommand{\thefigure}{A\arabic{figure}}
 \appendix
\section{influence of the passage of the Sgr dSph} 
\begin{figure}
    \includegraphics[width=18cm]{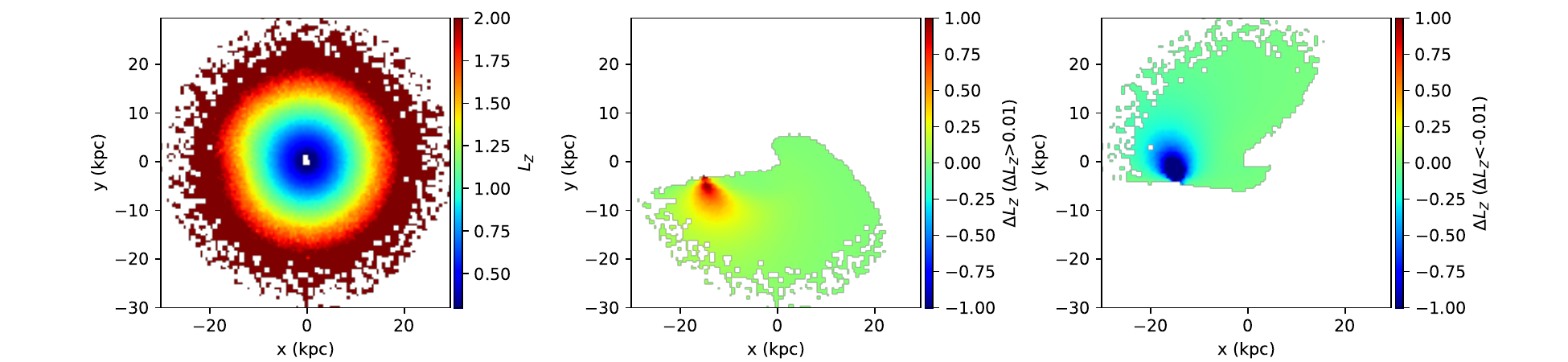}
    \caption{The left panel shows the distribution of angular momentum ($L_Z$) at the start of the impact. The middle panel shows the change in $L_Z$ ($\Delta L_Z>0.01$) during the impact. The right panel shows the change in $L_Z$ ($\Delta L_Z<-0.01$) during the impact.}
    \label{Lz}
\end{figure}

\begin{figure}
    \includegraphics[width=18cm]{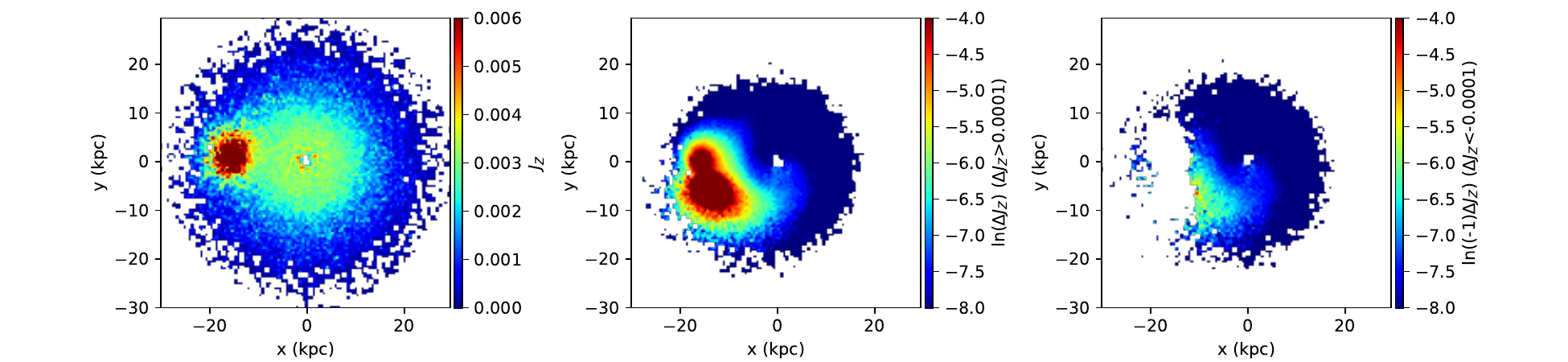}
    \caption{The left panel shows the distribution of vertical action ($J_Z$) at the beginning of the impact. The middle panel shows the amount of change in $J_Z$ ($\Delta J_Z>0.0001$) during the impact. The right panel shows the amount of change in $J_Z$ ($\Delta J_Z<-0.0001$. ) during the impact. Note the change in $J_Z$ is shown in log scale.}
    \label{Jz}
\end{figure}

\begin{figure}
    \includegraphics[width=18cm]{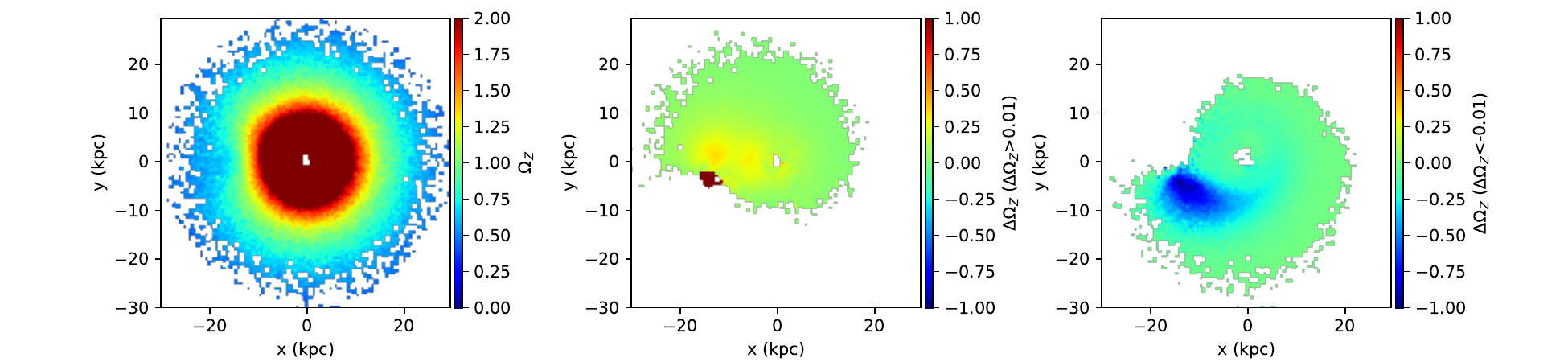}
    \caption{The left panel shows the distribution of vertical frequency ($\Omega_Z$) at the beginning of the impact. The middle panel shows the amount of change in $\Omega_Z$ ($\Delta \Omega_Z>0.0001$) during the impact. The right panel shows the amount of change in $\Omega_Z$ ($\Delta \Omega_Z<-0.0001$) during the impact.}
    \label{Omegaz}
\end{figure}

\begin{figure}
    \includegraphics[width=18cm]{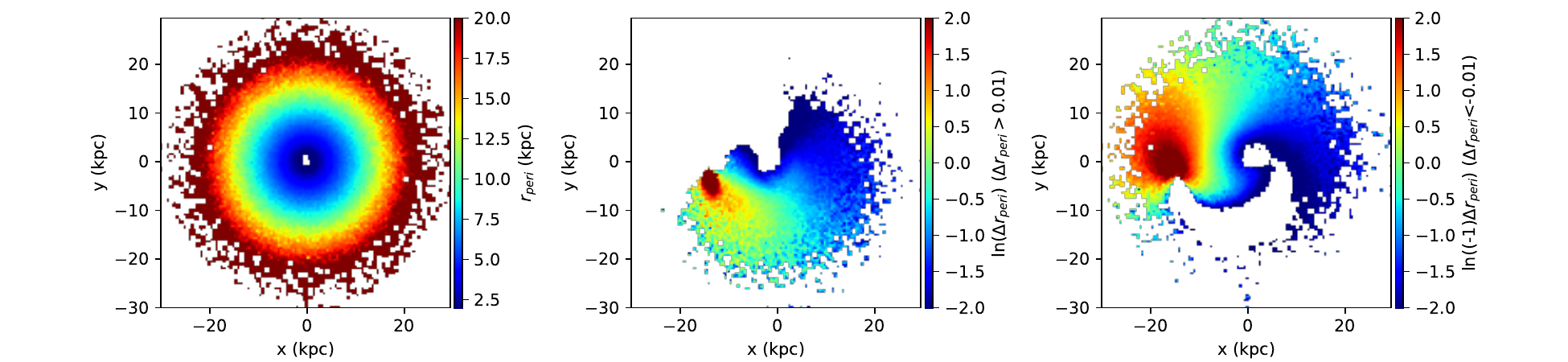}
    \caption{The left panel shows the distribution of pericenter radius ($r_\mathrm{peri}$) at the beginning of the impact. The middle panel shows the amount of change in pericenter radius ($\Delta r_\mathrm{peri}>0.01$) during the impact. The right panel shows the amount of change in pericenter radius ($\Delta r_\mathrm{peri}<-0.01$) during the impact.}
    \label{rperi}
\end{figure}

\begin{figure}
    \includegraphics[width=18cm]{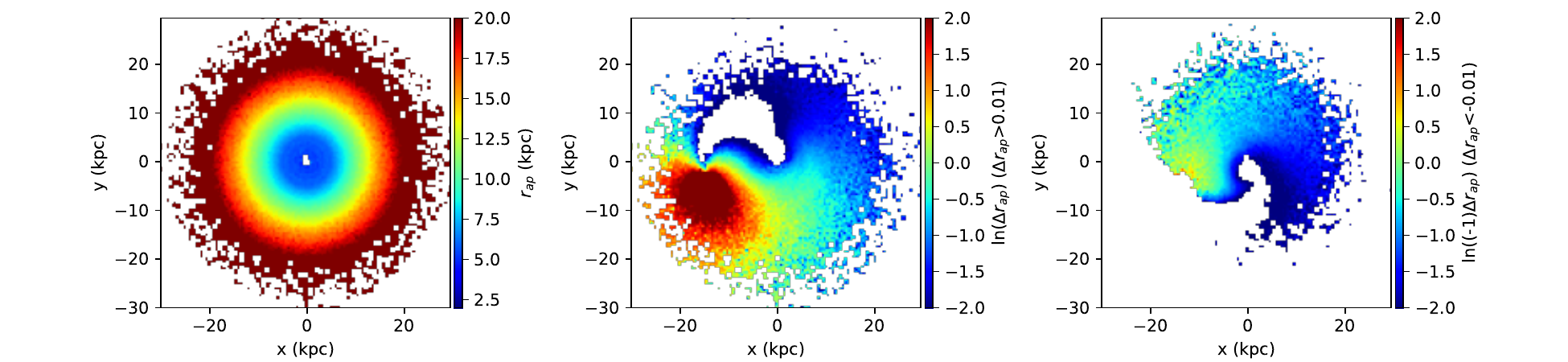}
    \caption{The left panel shows the distribution of apocenter radius ($r_\mathrm{apo}$) at the start moment of the impact. The middle panel shows the amount of change in apocenter radius ($\Delta r_\mathrm{apo}>0.01$) during the impact. The right panel shows the amount of change in apocenter radius ($\Delta r_\mathrm{apo}<-0.01$) during the impact.}
    \label{rap}
\end{figure}

 Figures~\ref{Lz} to ~\ref{rap} show the distribution of integral invariants ($L_Z$, $J_Z$, $\Omega_Z$, $r_\mathrm{peri}$, $r_\mathrm{apo}$) at the start of the impact and the amount of  change in the integral invariants during the impact  for all particles in the test particle simulations described in Section 6. 

\bibliographystyle{aasjournal}

\end{document}